%% file: manuscript.tex
\documentclass[twocolumn,twocolappendix]{aastex63}

\usepackage{natbib}
\usepackage{hyperref}
\usepackage{booktabs}
\usepackage{amsmath}
\usepackage{subfigure}

\newcommand{\hh}{H$_2$}

\graphicspath{{./}{figures/}}

\accepted{October 4, 2024}
\submitjournal{ApJ}

\shorttitle{H$_2$ in Flare Spectra}
\shortauthors{Jaeggli \& Daw}

\begin{document}

\title{Molecular Hydrogen Line Identifications in Solar Flares Observed by IRIS: Lower Atmospheric Structure from Radiometric Analysis}

\correspondingauthor{Sarah Jaeggli}
\email{sjaeggli@nso.edu}

\author[0000-0001-5459-2628]{Sarah A. Jaeggli}
\affiliation{National Solar Observatory \\ 
22 Ohia Ku Street \\
Pukalani, HI 96768, USA}

\author{Adrian N. Daw}
\affiliation{NASA Goddard Space Flight Center \\
Heliophysics Science Division, Code 671 \\
Greenbelt, MD 20771, USA}

\begin{abstract}
A rich spectrum of molecular hydrogen (\hh{}) emission lines is seen in sensitive observations from the far ultraviolet (FUV) channels of the Interface Region Imaging Spectrograph (IRIS) during flare activity in solar active region NOAA 11861.  
Based on this observation we have determined 37 new line identifications by comparing synthetic spectra produced using 1D modeling of \hh{} fluorescence. 
To avoid misidentification of the \hh{} lines, we have also compiled a complete list of atomic line identifications for the IRIS FUV bandpasses from previous work.  
We carry out analysis of the spatially resolved \hh{} emission that occurs during the flares and find that: 
(1) in spatially resolved observations the \hh{} line ratios may show optically thick line formation, contrary to previous results;
(2) comparison of the spatial distribution of \hh{} Doppler velocities with those measured from other species reveals that \hh{} remote sensing probes an intermediate depth in the atmosphere between the photosphere and chromosphere, consistent with expectations from modeling;
(3) the relationship between \hh{} line intensity and the observed intensity of its exciter is related to the atmospheric stratification; however,
(4) \hh{} fluorescence can sometimes occur in response to radiation from distant sources many Mm away across the solar surface.
\end{abstract}


\section{Introduction} \label{sec:intro}
Molecules are good indicators of cool gas in the Sun's atmosphere.
Simple diatomic molecules form in regions in the Sun's atmosphere where the gas is relatively cool and dense, primarily at the temperature minimum that occurs between the photosphere and the chromosphere.
Molecular hydrogen (\hh{}) constitutes the largest molecular population at photospheric heights, and in very cool sunspots it may make up a significant portion of the gas population, up to 10\% \citep{maltby86}.
The absorption lines of the carbon monoxide (CO) fundamental band provide compelling evidence that very cool gas ($<$ 4000 K) may be present at chromospheric heights alongside much warmer material \citep[e.g.][]{ayres81, solanki94}.
Recent observations show that cool bubbles of CO may form in situations where acoustic heating of the chromosphere is suppressed \citep{stauffer22}.
Simulations suggest that \hh{} and CO can form in the wake of magneto-acoustic shocks propagating upward through the chromosphere \citep{leenaarts11}.
The abundance of molecules is likely the best evidence for a thermally inhomogeneous chromosphere, and investigating molecular diagnostics that can occur at chromospheric heights can provide important evidence for solving this mystery.

Molecular hydrogen should be the most abundant molecule in the solar atmosphere, but it is also one of the most difficult to observe directly.  
\hh{} is a simple homonuclear molecule, and it does not produce a strong photospheric absorption spectrum typical of other molecular species in the solar atmosphere such as CO, OH, CH, and CN.  
Rotational-vibrational transitions of \hh{} which occur in the infrared are forbidden, requiring long collisional timescales. While these transitions are commonly seen in low-density astrophysical regimes \citep[e.g.][]{spinrad66}, they are absent in the high density solar environment.
  
Although the rotational-vibrational states of \hh{} do not produce easily observable transitions in the Sun, transitions in the ultraviolet from exited electron configurations can be seen in emission above the weak solar continuum.
The Lyman and Werner bands of molecular hydrogen are electronic transitions that occur from 850 - 1850 \AA\ in the ultraviolet between the $B^1\Sigma^+_u$ (Lyman) and $C^1\Pi_u$ (Werner) upper levels and the $X^1\Sigma^+_g$ ground state.  
These transitions of \hh{} were first investigated for the solar regime in theoretical work by \citet{krishna75}, who predicted small signatures from the \hh{} lines in emission and absorption on top of the UV continuum assuming thermal population of the energy levels.

Emission from the Lyman and Werner bands was first identified in the solar atmosphere by \citet{jordan77} using observations of the High Resolution Telescope and Spectrograph (HRTS) sounding rocket, which covered wavelengths 1175 - 1714 \AA.  
They found that the upper levels of the transitions were not thermally populated, but instead they were excited by strong transition region emission lines with similar wavelengths to the \hh{} lines.

Additional line identifications of \hh{} were made by \cite{jordan78} and \cite{bartoe79}, where it was identified in a variety of features such as active regions, above sunspots, in light bridges, and in flares.  
They carried out simple calculations of the \hh{} column density based on a 1D atmospheric model.  
Additional line identifications were made by \citet{sandlin86} using spectra from HRTS and the Apollo Telescope Mount (ATM) on {\it Skylab}, covering wavelengths 1175 - 1710 \AA.

The SUMER (Solar Ultraviolet Measurements of Emitted Radiation) spectrograph on the Solar and Heliospheric Observatory (SOHO) again made it possible to investigate \hh{}, covering wavelengths 670-1609 \AA\ in the first order of the spectrograph.  
\citet{curdt01,curdt22} provide comprehensive line identifications of \hh{} for the SUMER spectrum drawing on the identifications in \citet{bartoe79}.
The brightest \hh{} lines can be seen in the quiet Sun spectrum, but they appear most prominently in the sunspot atlas spectrum and the HRTS atlas ``light bridge'' spectrum.

Other authors made use of the same SUMER sunspot observation used in the \citet{curdt01} atlas for further analysis of the \hh{} lines.  
\citet{schuhle99} found that the relative line strengths from the Werner $J=3, v=1$ upper level (where $J$ is the rotational quantum number and $v$ is the vibrational quantum number) excited by \ion{O}{6} at 1030 \AA\ were in good agreement with the expected transition probabilities.  
\citet{morgan05} investigated these same lines and found that the enhanced strength of the \ion{O}{6} 1032 \AA\ line relative to the doublet line at 1038 \AA\ in sunspots may be due to resonance with the \hh{} lines.  
\citet{kuhn06} found that the \hh{} lines show enhanced emission over the sunspot penumbra relative to the umbra in this sunspot.  
They inferred that a diffusive process must be responsible for draining the umbra of neutral species, causing a greater density of \hh{} to accumulate in the penumbra.  
However, the work of \citet{labrosse07} shows that the same enhanced emission is present in both \ion{O}{6} lines at 1032 and 1038 \AA, indicating that it is likely the spatial enhancement of the \ion{O}{6} emission which causes the enhanced \hh{} emission over the penumbra.

Given that the \hh{} Lyman and Werner bands are excited by bright UV emission lines, it is unsurprising that \hh{} emission can be seen prominently in flares when heating in the upper atmosphere causes enhanced emission from bright lines and also continuum emission.
\hh{} emission was seen to correspond with strong UV and x-ray emission of microflares by \citet{innes08}, caught in repeated scans of an active region by the SUMER spectrograph.  
While \hh{} emission was also bright over a nearby sunspot, the \hh{} emission during the microflares did not correspond to cool photospheric features.

The UV lines of \hh{} can now be investigated with the highest spatial and temporal resolution yet achieved using the Interface Region Imaging Spectrograph \citep[IRIS,][]{depontieu14}, which covers the wavelength ranges 1332 - 1358 \AA\ and 1389 - 1406 \AA\ in the far ultraviolet (FUV) and includes many lines of the \hh{} Lyman band.    
\citet{mulay21} and \citet{mulay23} have used IRIS FUV observations to examine the spatially and temporally resolved properties of \hh{} emission lines excited by \ion{Si}{4} in flare ribbons.
Highly unusual behavior has already been observed by \citet{schmit14} during a series of C-class flares over an active region.  
A rich absorption spectrum, which appears to be due to \hh{} and possibly CO, appears on top of continuum emission and the line emission of the transition region.

For select \hh{} lines in the IRIS bandpass, IRIS sees both the line and its major excitation source (\ion{C}{2} or \ion{Si}{4}), so that there is the potential to diagnose the condition of the atmosphere between those two regions.
The UV transitions of \hh{} are largely unexplored for solar studies, but they have the potential to provide crucial information on the molecular content and the physical conditions of the low chromosphere. 
For some \hh{} lines, the IRIS data may contain all the information necessary to understand their formation.

There are approximately 40 \hh{} transitions per Angstrom in the FUV channels of IRIS, and it would be a difficult task to positively identify the lines based on the wavelength alone, however, only particular lines are bright due to the fluorescence mechanism.  
Lines of the Werner and Lyman bands of molecular hydrogen are produced when bright emission lines similar in wavelength to the \hh{} transitions excite electrons into the upper energy levels within the molecule.  
The downward transitions occur to multiple lower levels with differing probabilities, producing a distinctive fluorescence spectrum where only lines from strongly pumped upper levels show emission.

The average stratified structure of the solar atmosphere provides a small region of overlap in the low chromosphere where UV photons can reach \hh{} molecules.  
The pumping photons originate high in the transition region and travel downward into regions of rapidly increasing optical depth.  
Extinction at 1400 \AA, the wavelength of interest for the IRIS FUV spectra, is mostly due to photoionization of Si and C \citep[e.g.][fig. 7]{socasnavarro15}.  
Although cool conditions are important for the formation of \hh{}, high densities are also critical because \hh{} forms in a 3-body process.  
An additional particle must escape with the excess energy in order to leave the hydrogen atoms bound together.  
Therefore, \hh{} fluorescent emission can originate no lower and no higher than the low chromosphere which is cool and dense enough for the formation of molecules and still optically thin to UV photons.  
The greatest contribution to the \hh{} fluorescence should come from the greatest depth that significant pumping radiation can reach, i.e. where $\tau=1$ for those wavelengths.  
This does not necessarily correspond to the temperature minimum region, rather it depends on the stratification of atmospheric parameters.

The separation of the pumping source and fluorescing regions, the location of \hh{} in the low chromosphere, and the small amount of absorption and radiation produced by \hh{} allowed \citet{jordan78}, \citet{bartoe79}, and \citet{jaeggli18} to approach the synthesis using a stratified 1D atmosphere without directly including the large number of \hh{} transitions in the radiative transfer calculation.  
The work of \citet{jaeggli18} uses extensive new \hh{} line lists from \citet{abgrall00} as well as an updated partition function for \hh{} from \citet{barklem16}.  
Implementation in the Rybiki-Hummer Radiative Transfer and Chemical Equilibrium code \citep[RH,][]{uitenbroek00} allows for the treatment of complicated, even 3D, atmospheres, and the wavelength dependence of the radiation field can be arbitrarily complex.

It is necessary to understand the basic process of fluorescence, and explore the available lines, before \hh{} lines can be used as diagnostic tools.  
New work by \citet{jaeggli18} provides a tool for synthesizing the spectral signature resulting from fluorescent processes for model atmospheres, and in the current work we take the first steps to develop the diagnostic potential of the fluorescent \hh{} lines in the IRIS FUV bandpass.  
We investigate the rich spectrum of narrow emission lines produced by cool molecular and atomic species during IRIS observations of flare activity above an active region described in Section \ref{sec:dataset}.  
Parameters crucial to the interpretation of the observations are derived from spectral synthesis of one dimensional model atmospheres in Section \ref{sec:synth}.  
In Section \ref{sec:lineids}, the spectral synthesis results are compared with an average flare spectrum from the observation to confirm past \hh{} and atomic line identifications and reveal many previously unidentified \hh{} lines.  
An analysis of the spatially resolved observations is carried out in Section \ref{sec:spec2d}, in order to get properties of the strongest \hh{} lines in the IRIS FUV bandpasses and compare them to their excitation sources and diagnostics representative of the photosphere and chromosphere.
Results of this analysis are presented in Section \ref{sec:results}.
We carry out a discussion of the results, comparing with past work, and theorizing on future research in Section \ref{sec:discussion}.
Finally, we summarize our work and provide conclusions in Section \ref{sec:conclusions}.

\section{Dataset\label{sec:dataset}}

\subsection{Instrument}\label{sec:instrument}
IRIS \citep{depontieu14} is an innovative UV spectrograph and imager that saw first light on July 17, 2013.  
It combines high resolution spectroscopy in two far UV passbands (FUV 1: 1333 - 1358 \AA, FUV 2: 1389 - 1407 \AA) and one near UV passband (NUV: 2783 - 2834 \AA) with high spatial resolution imaging of the region around the slit (the slit-jaw imager, or SJI) in four selected passbands.  
The wavelengths of IRIS were selected to focus on bright spectral lines that form in the temperature range from 10,000 K to 100,000 K, corresponding to the chromosphere and transition region in the Sun's atmosphere.  
The NUV spectrograph of IRIS covers the \ion{Mg}{2} k and h line pair (at 2796 and 2803 \AA\ respectively), the broad wings of these lines, as well as continuum and photospheric absorption lines.  
The FUV 1 channel covers the \ion{C}{2} line pair (1334.5 and 1335.7 \AA), \ion{Fe}{12} (1349.4 \AA), and \ion{Fe}{21} (1354.1 \AA).  
The FUV 2 channel covers the \ion{Si}{4} lines (1393.8 and 1402.8 \AA), \ion{O}{4} (1399.8, 1401.2 \AA), and \ion{S}{4} (1404.8, 1406.1 \AA).  
Both FUV channels also include emission from many neutral and singly ionized species and a low level continuum formed by the hydrogen Balmer continuum emission.  
The three spectrograph channels operate simultaneously with the SJI, which provides context imaging around the entrance slit of the spectrograph.  
By using filters, the SJI is capable of imaging in four narrow spectral bandpasses centered around the \ion{C}{2} lines (1330 \AA), \ion{Si}{4} lines (1400 \AA), core of the \ion{Mg}{2} k line (2796 \AA), or far out in the wing of the \ion{Mg}{2} lines (2830 \AA) which mainly contains photospheric continuum.  
The spectrograph is capable of observing a $175''$ tall region, while the SJIs cover a field of approximately $175''\times175''$, with $0.3''$ resolution sampled at $0.167''$ pixel$^{-1}$.  
The FUV and NUV channels have a spectral resolution of 26 and 53 m\AA\ sampled at 12.8 and 25.6 m\AA\ pixel$^{-1}$ respectively.

\begin{figure}
    \begin{interactive}{animation}{sji_movie.mp4}
	   \begin{center}
		  \includegraphics[width=3.55in]{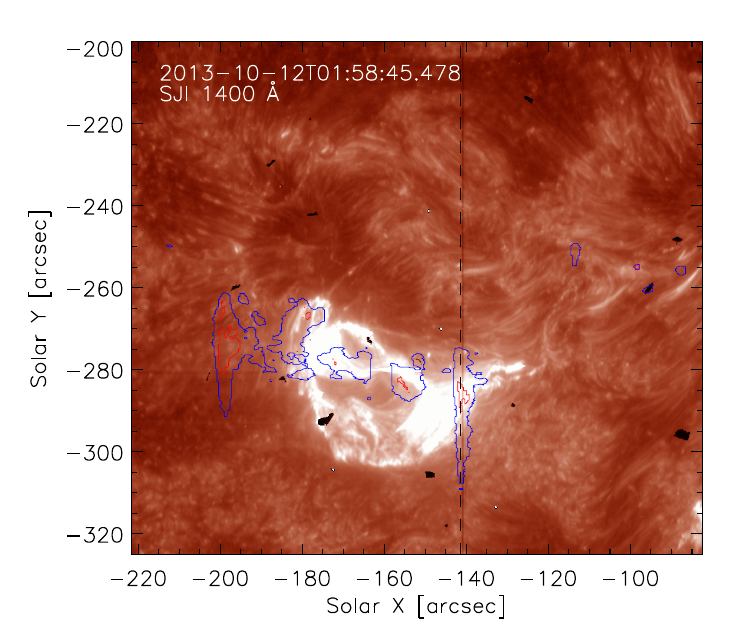}
	   \end{center}
    \end{interactive}
	\caption{This single frame from the 1400 \AA\ IRIS slit-jaw imager is representative of the complex structure of the emission from the \ion{Si}{4} lines which are major sources for \hh{} excitation.  In the animated version of this figure, found in the online material, there are several such instances of bright dynamic structures produced by flare activity throughout the 3.6 hour observation as the spectrograph slit (black dashed line) scans over them from left to right.  The contours from Figure \ref{fig:spec_map} are overplotted on the SJI image, although it should be noted that at any given time the contours are only aligned with the features at the slit location due to solar rotation during the scan.}
	\label{fig:sji_frame}
\end{figure}

\subsection{Observation}\label{sec:observation}
The IRIS observation that we have selected for analysis of \hh{} emission occurred early in the mission and contains flare activity over a sunspot umbra, penumbra, and the surrounding quiet Sun.
The C1.5 (00:46 UT) and C5.2 (01:54 UT) flares of October 12, 2013 occurred between the main sunspot pair in NOAA 11861, a typical $\beta$-type active region in the Hale classification system with fully developed sunspots located approximately $120''$ east and $250''$ south of Sun center.  
IRIS observed this region from 23:54 UT on the 11th to 03:29 UT on the 12th with a very large, dense raster of 400, $0.5''$ steps, resulting in an observed area of $166''\times175''$ seen with the spectrograph.  
During the raster the spectrograph slit followed the flare activity as it moved from below the east sunspot toward the west sunspot. 
Only the east sunspot is fully visible within the spectrograph field of view.
The SJI obtained images using only the \ion{Si}{4} 1400 \AA\ channel at the same cadence as the spectrograph steps.
Figure \ref{fig:sji_frame} shows the SJI image at one time step during the raster, and an overview of the activity can be seen in the animated version of this figure in the online material.

\begin{figure*}
	\begin{center}
		\includegraphics[width=7.25in]{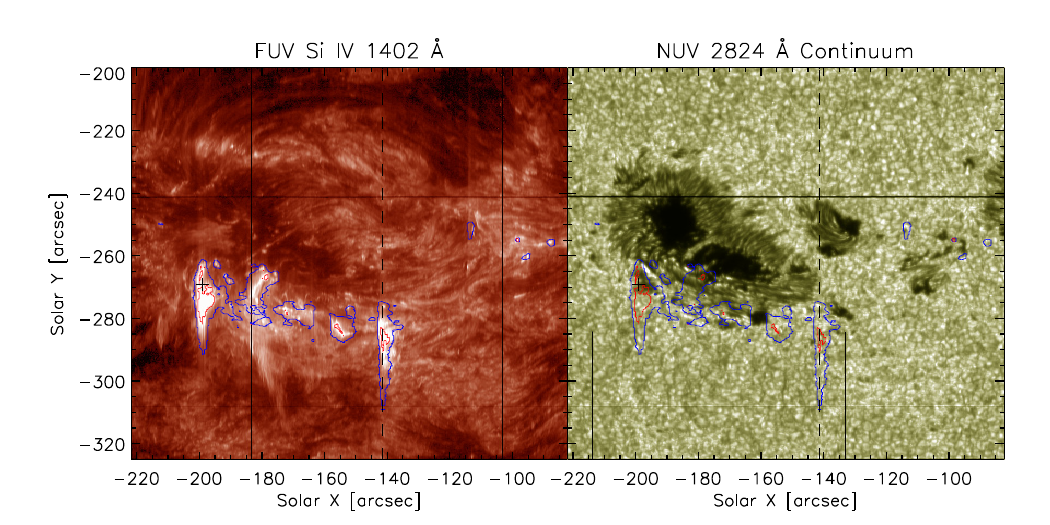}
	\end{center}
	\caption{Spectroheliogram images constructed from the spectrograph raster for the FUV \ion{Si}{4} 1402 \AA\ line (left) and the photospheric continuum far out in the NUV \ion{Mg}{2} line wing (right).  The dashed black line shows the location of the slit position for the image shown in Figure \ref{fig:sji_frame}.  The blue contours show regions of high \hh{} intensity and the red contours show regions where there is saturation in the bright \ion{C}{2} or \ion{Si}{4} lines.  The dark horizontal line at Solar Y=-242$''$ comes from the alignment fiducial at the top of the spectrograph slit, and missing data occurs at two different time steps in each of the channels, shown by the black vertical lines.}
	\label{fig:spec_map}
\end{figure*}

During this observation, 30 sec exposures were used to provide higher signal levels in weak lines.  
Contrary to typical operating procedure for active regions, the automatic exposure control was not enabled.  
As a result the brightest transition region lines became saturated during the flare activity, but the weaker lines and continuum radiation can be seen more easily above the background noise level.  
In addition, no binning or windowing of the spatial and spectral dimensions was used, the observations contain the full spatial and spectral range possible with the FUV and NUV channels of the spectrograph.  

Figure \ref{fig:spec_map} shows heliogram maps reconstructed from the spectrograph observations of the \ion{Si}{4} 1402 \AA\ line in the FUV 2 passband and the continuum near 2824 \AA\ in the NUV passband.  
The full field of view has been reduced vertically to show the region of interest.

\subsection{Data Reduction}\label{sec:reduction}
For this analysis we have used the standard IRIS Level 2 data products for the SJI images and NUV spectra which were downloaded from the IRIS website at \url{http://iris.lmsal.com/}.  
The FUV spectra used in this work were processed from the Level 1 (unprocessed, uncalibrated) data product using some of the \verb+iris_prep+ routines in {\it SolarSoft IDL}, however, our processing departs from the standard calibration techniques described by \citet{wuelser18}.  
We use non-standard techniques in order to control the noise more effectively, which is critical for the analysis of the signals from weak spectral lines.  
The standard dark calibration images used in the IRIS processing pipeline are built from short exposures and are scaled by the exposure time to apply to longer exposures.
They therefore have a much higher ratio of read and photon counting noise to signal level than is reached in long exposure data.  
For the FUV data, only the time-dependent, average dark level for each readout of the CCD was subtracted.  
The FUV stray light background was subtracted, and the flat field was applied as in the usual data pipeline.  
Despiking, or removal of discrete bright pixels from the detector, cosmic rays, and particle strikes, was carried out prior to the geometric dewarping to avoid smearing discrete spikes over multiple pixels.

The dewarping for the IRIS spectra makes the spectral and spatial dimensions rectilinear with respect to the pixel dimensions and is applied by cubic interpolation of the data to the unwarped coordinate frame.  
Typical IRIS Level 2 data is resampled once during the geometric unwarping, and again to apply the wavelength and spatial drift corrections.  
There is inherently some information loss or smoothing that occurs during interpolation.  
We have combined the geometric and drift corrections in this data reduction in order to resample the spectra only once.  
The wavelength drift correction was determined by fitting the average position of the \ion{Cl}{1} 1351.7 \AA\ line in the FUV 1 channel and the \ion{S}{1} 1401.5 \AA\ line in the FUV 2 channel at each scan step.  
The average position of the lines at each scan step was fitted with a harmonic function and the wavelength shift was removed from the data simultaneously with the dewarping.  
The fine wavelength adjustment to the local standard of rest was determined during line fitting and identification discussed in Section \ref{sec:lineids}.

Having skipped the coalignment steps which are part of the standard data reduction from Level 1 to Level 2, the FUV spectra were coaligned with respect to the NUV Level 2 spectra based on the position of the slit fiducials.  
The FUV spectra were shifted spatially to the nearest pixel, and then the coordinate metadata from the NUV data was adopted for the FUV data.

Counts were converted to absolute intensity units using the routine \verb+iris_get_response.pro+ which gives the wavelength- and time-dependent instrument sensitivity \citep{wuelser18}.
The wavelength-dependent component is based on the pre-launch measured effective areas, while the time-dependent component is determined based on trending of IRIS's daily throughput observations at disk center and cross-calibration with observations from the Solar Stellar Irradiance Comparison Experiment (SOLSTICE) on the Solar Radiation and Climate Experiment (SORCE) sounding rocket.  
This correction does not account for the detailed dosage burn-in for each detector, however this is only severe for the core of the \ion{C}{2} lines and should not be large for data from this early in the mission.

\section{Spectral Synthesis}\label{sec:synth}
To interpret the IRIS FUV spectrograph observations we model \hh{} fluorescent emission using the computational framework developed in \citet{jaeggli18}.
For 1D model atmospheres, the RH code was used to solve the equations of radiative transfer and chemical equilibrium given different cases for the radiation injected into the top of the atmosphere to simulate different flare intensities.
The \hh{} abundance, radiation, and opacity as a function of height were taken from the RH results and used in a separate IDL code to calculate the fluorescent emission from molecular hydrogen.
The resulting line intensities were used to generate synthetic spectra and other observable quantities to assist with identification of \hh{} lines and provide insight into the behavior of \hh{} seen in the IRIS observations.

We used the version of the RH code downloaded from \url{https://github.com/han-uitenbroek/RH} on July 7, 2023.
The RH calculations were performed using the same three semi-empirical 1D model atmospheres used in \citet{jaeggli18}, two of these atmospheres are distributed with the RH code.
RH includes \verb+FALC_82.atmos+, a version of the model ``C'' from \cite{fontenla93} with 82 gridpoints (called here FALC).
FALC was made to reproduce the average UV emission spectrum from the photosphere, chromosphere, transition region, and corona of quiet Sun regions.
RH also includes the model \verb+FALXCO_80.atmos+ (called here XCO).  
This model is a version of the ``M$_{\text{CO}}$'' atmosphere presented in \citet{avrett95} with 80 gridpoints.
It is similar to FALC but with a lower minimum temperature that is reached higher in the atmosphere in order to explain the average behavior of the strongest lines of the fundamental band of the infrared spectrum of carbon monoxide. 
In addition, we have transcribed the 32 gridpoint ``F2'' model atmosphere of \citet{machado80} into the RH atmosphere format (called here F2).
F2 was made to reproduce various spectral properties, including the intensities of the hydrogen Lyman series lines and Lyman and \ion{C}{1} continuua, of spatially averaged spectra from ``bright flares.''
The essential properties of all three atmospheres are shown in Figure 3 of \citet{jaeggli18}.

RH was used to solve the non-LTE radiative transfer equations for 5 atoms included in ``active'' mode:  a hydrogen 6-level atom, a 16-level silicon atom and 14-level oxygen atom to account for the continuum opacity in the IRIS bandpass, an 11-level carbon atom to account for the optical density of the \ion{C}{2} 1330 \AA\ lines, and a 16-level iron atom.
Other atomic species were calculated in ``passive'' mode (Al, N, S, Na, Mg, Ca, Ni) to account for their populations in LTE.  
All molecules (\hh{}, \hh{}+, C$_2$, N$_2$, O$_2$, CH, CO, CN, NH, NO, OH, and \hh{}O) were included in ``passive'' mode, assuming LTE and instantaneous chemical equilibrium to get their abundances.

For each atmosphere, RH was run to convergence to produce a starting solution with no additional radiation, then ad hoc radiation was injected into the top of the atmosphere, and the calculation was run again, using the starting solution, to produce a self-consistent atmosphere.
For the down-going radiation we took the quiet Sun line intensities for the brightest transition region emission lines important for the fluorescence of the \hh{} Lyman band, including Ly $\alpha$, \ion{C}{2}, \ion{O}{5}, \ion{Si}{4}, and \ion{O}{4}, and applied a uniform multiplier to the intensities to simulate the impact of different flare intensities.
See Table 1 in \citet{jaeggli18} for a list of the exciters and their assumed intensities and line widths.
For intensity multipliers larger than 10 the line width was increased by a factor of 2.
In this work, we used radiation multipliers of 1, 3, 10, 30, 100, 300, 1000, 3000, and 10,000.

The radiative impact of \hh{} is relatively minor compared to other sources of opacity and emission in the upper photosphere and chromosphere.
This fact allows us to treat the detailed problem of excitation and emission of the thousands of possible transitions in a separate calculation which we have implemented in {\it IDL}.
As in the column density estimations of \citet{jordan78} and \cite{bartoe79}, \hh{} is treated in partial NLTE, where the ground state populations are in LTE and given by the partition function, the upper levels are populated only by radiative excitation, and emission is instantaneous, that is, there is no significant collisional de-excitation of the upper levels.
The total \hh{} LTE abundance, radiation field, and opacities from RH were used as inputs to this calculation, along with the \hh{} line list of \citet{abgrall00} and the \hh{} partition function of \citet{barklem16}.
For each grid point in the atmosphere, the lower level population was determined based on the partition function.
The radiation from RH was used to determine the upper level population based on photo-excitation for each transition in the line list. 
The downward transitions from the upper levels were then determined by the transition and dissociation probabilities from the line list.
From this point we carried out two cases of the calculation.
For the first case, we assumed that \hh{} does not affect the opacity and the lines are formed in an optically thin regime.
For the second case, we assumed the lines are optically thick and can be treated approximately with escape probabilities.
The \hh{} line emission was propagated back out of the atmosphere using the wavelength-dependent opacities from RH.
We take the synthetic spectrum output from RH for a viewing angle of $\mu$=1 along with the \hh{} emission as our resulting synthetic spectrum.
For more details on the calculation of \hh{}, see Section 3.2 of \citet{jaeggli18}.

\hh{} is pumped most strongly when particular lines are coincident in wavelength with bright emission from other sources to populate particular upper levels.
The downward transitions from a particular upper level are closely related in intensity and form a ``line family'' where properties of the line formation region should be identical and the lines differ only in their intensity determined by their relative transition probability, or branching ratio, and optical depth effects.
In the optically thin case they ought to be related by the branching ratio.  
In the optically thick case, this branching ratio is slightly modified by the escape probabilities.
In the rest of the paper we will identify the upper and lower energy levels of the Lyman band using the notation $(J,v)$, where $J$ is the rotational quantum number and $v$ is the vibrational quantum number for the level.  
For the energy levels within the Lyman band, the upper level configuration is always $B^1\Sigma^+$, abbreviated to $B$, and the lower level configuration is always $X^1\Sigma^+$, abbreviated to $X$.
The Lyman band transitions that we list with the same $(J,v)$ upper level are part of the same ``line family.''

\begin{figure}
	\begin{center}
		\includegraphics[width=3.5in]{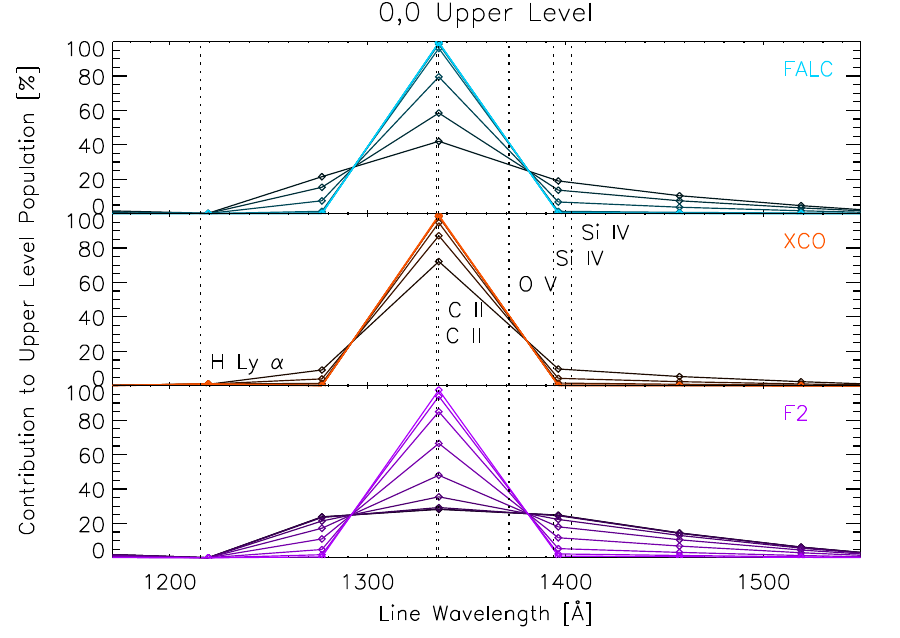}
        \includegraphics[width=3.5in]{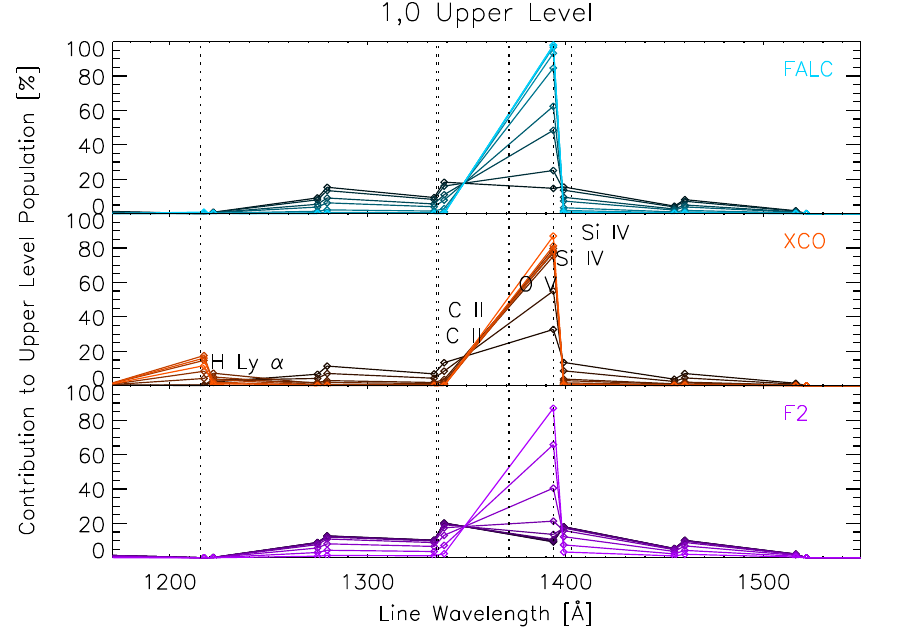}
        \includegraphics[width=3.5in]{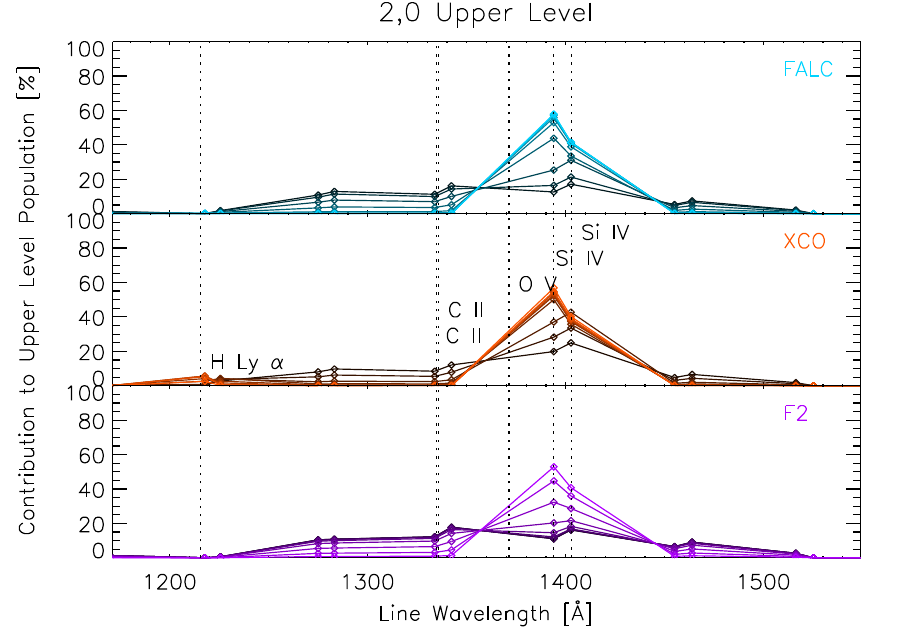}
	\end{center}
	\caption{The relative contribution to the upper level populations for the first three upper levels in the Lyman band which produce the strongest lines \hh{} lines seen in the IRIS FUV spectra.  For each upper level, The contribution of each line is shown for each baseline model (FALC, XCO, F2) indicated by the hue, and intensity multiplier indicated by the hue saturation.  The highest multipliers have the brightest hue.  Bright emission lines in this wavelength range are shown by the dotted lines.}
	\label{fig:pump_contrib}
\end{figure}

It is instructive to look at the intermediate results from the synthesis calculation to determine the major pumping sources and the relative contribution of various pumps to the upper level \hh{} populations.
The (0,0), (1,0), and (2,0) upper levels are of particular interest because they produce the brightest \hh{} lines in the IRIS FUV channels.  
Figure \ref{fig:pump_contrib} shows the contribution to the upper level population from lines in the 1170 - 1550 \AA\ range for these three upper levels for each model and intensity multiplier.
The (0,0) upper level is excited by the \ion{C}{2} 1335 \AA\ line, the (1,0) upper level is excited by the \ion{Si}{4} 1393 \AA\ line, and the (2,0) has significant contributions from both of the \ion{Si}{4} lines at 1393 and 1402 \AA.
It is important to note that the contribution from these pumping sources does change as a function of multiplier and is different for each of the models.
For small multipliers and warmer atmospheric models, the \hh{} upper level is excited broadly through many transitions of the line family by continuum emission.
In cases where the multiplier is very large, or the atmosphere very cool, the line emission is a more dominant source, exciting the upper level through particular \hh{} transitions at wavelengths corresponding to the transition region emission lines.

\begin{figure}
    \centering
    \includegraphics[width=3.3in]{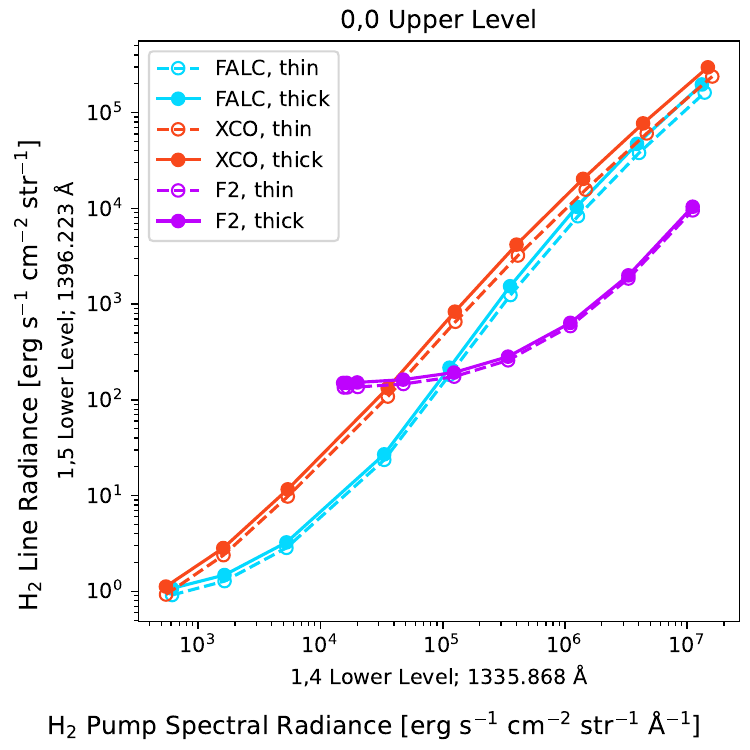}
    \caption{Radiance of the 1396.223 \AA\ \hh{} line as a function of the spectral radiance at the wavelength of the 1335.969 \AA\ \hh{} line pumped by \ion{C}{2} to populate the common (0,0) upper level.  The individual curves include all multipliers for a particular baseline model.  The dashed and solid lines show the results from the optically thin and optically thick calculation for the \hh{} line emission, respectively.  The ratio of the \hh{} line intensity to its pump provides a way to distinguish between different atmospheric models.}
    \label{fig:synth_pump_ratio}
\end{figure}

To understand how well these models match to observations in an absolute sense, we have generated some observable quantities from the spectral synthesis results.
To give an example, Figure \ref{fig:synth_pump_ratio} shows the relation between the total radiance of the 1396.223 \AA\ \hh{} line and the spectral radiance at the wavelength of the main pumped line responsible for populating the (0,0) upper level at 1335.868 \AA\ in the wing of the \ion{C}{2} line at 1335.708 \AA.
In log-log scale we show the line radiance versus the pump spectral radiance for each of the models and for the optically thin and thick cases.
The different multipliers for the ad hoc radiation are indicated by the data points along the line for each model.
The hot F2 model is well separated from the cooler FALC and XCO models, so this comparison of quantities provides fairly good discrimination between different atmospheric models and flare conditions.

\section{Line Identifications}\label{sec:lineids}
There are about 1030 transitions of the \hh{} Lyman band that fall within the two IRIS FUV bandpasses.
While only certain transitions are stimulated through the mechanism of fluorescence, the results of spectral synthesis described in the previous section only provide a first guess of the \hh{} line intensities.
In order to positively identify \hh{} lines in the IRIS FUV bandpass, we have compiled an extensive list of both atomic and molecular lines, constructed a reference spectrum from the IRIS data, performed fitting of line profiles in the reference spectrum, and applied criteria to determine if a potential line might be part of the \hh{} Lyman band.
This work is discussed in the following subsections alongside the resulting detailed reference spectrum, with line identifications shown in Figure \ref{fig:FUV_spec}, and the corresponding line parameters listed in Table \ref{tbl:lines}.

\subsection{Average Flare Spectrum}
As our observational reference for carrying out the line identification, we constructed a 1D spectrum from the FUV channels of the IRIS observation.  
We selected spatial regions of the raster where the intensity of the bright \hh{} line at 1342.256 \AA\ was larger than a certain threshold and easily distinguishable from the background noise in the spectrum.  
The selected positions are indicated by the blue contours in Figures \ref{fig:sji_frame} and \ref{fig:spec_map}, and they are well correlated with the high intensity regions in the \ion{Si}{4} map.  
We excluded regions within the blue contours that showed saturated line profiles for the \ion{C}{2} or \ion{Si}{4} lines, this region is indicated by the red contours.  
The ``average flare spectrum'' combines 7,457 individual spectra from the positions between the blue and red contours, and contains only valid intensity values.  
The resulting average flare spectrum is shown by the black line in Figure \ref{fig:FUV_spec}.
For reference, a 2D spectrum from one step in the IRIS raster is shown as the background image in the plot.  This image has a reversed log scale so that the faint emission from weak lines can be seen clearly.
This spectrum comes from the slit position indicated by the dotted line in Figures \ref{fig:sji_frame} and \ref{fig:spec_map}.
The average flare spectrum was compared directly to the synthetic \hh{} spectrum from the FALC$\times100$ optically thick calculation, shown by the cyan line in Figure \ref{fig:FUV_spec}.

\subsection{Line Lists}
In order to determine new \hh{} line identifications, we have compiled and cross-referenced lists of \hh{} and atomic lines from previous references for the IRIS FUV bandpasses.
Using the results of the spectral synthesis described in Section \ref{sec:synth} we constructed a complete \hh{} line list for comparison with the IRIS FUV bandpasses.
We cross-referenced this with existing \hh{} line identifications from \citet{jordan77}, \citet{jordan78}, \citet{bartoe79}, \citet{sandlin86}, and \citet{jaeggli18}.
All theoretical rest wavelengths for \hh{} listed here come from \citet{abgrall00}.

\begin{figure*}
	\begin{center}
		\includegraphics[width=7in]{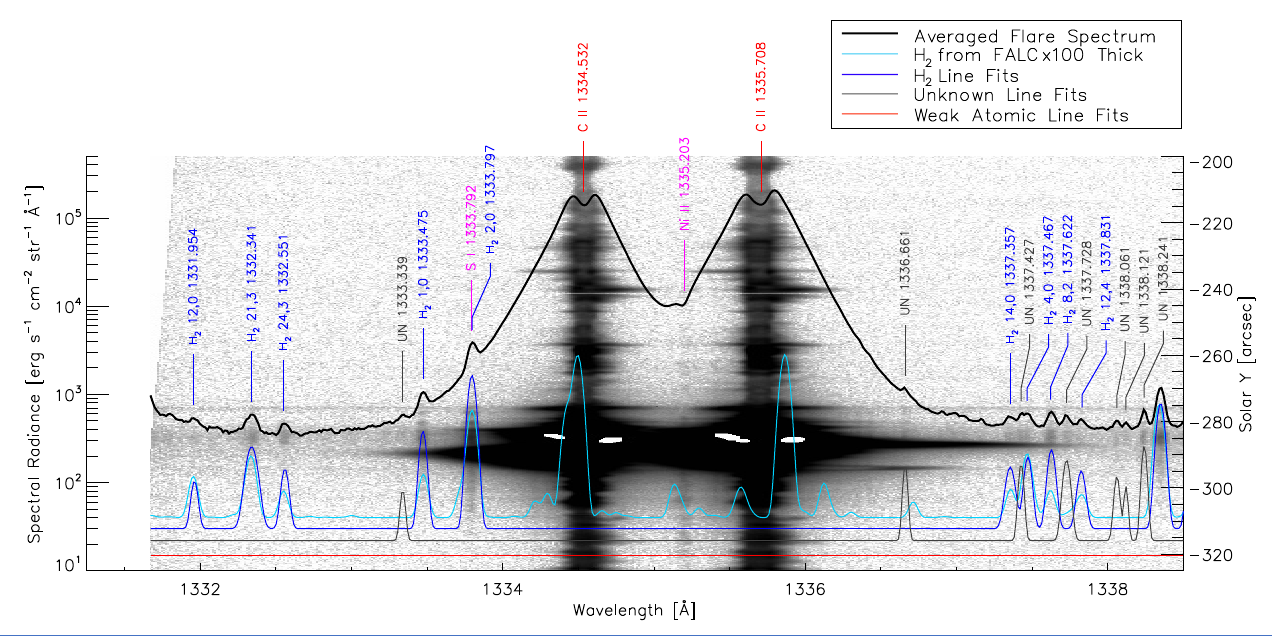}
		\includegraphics[width=7in]{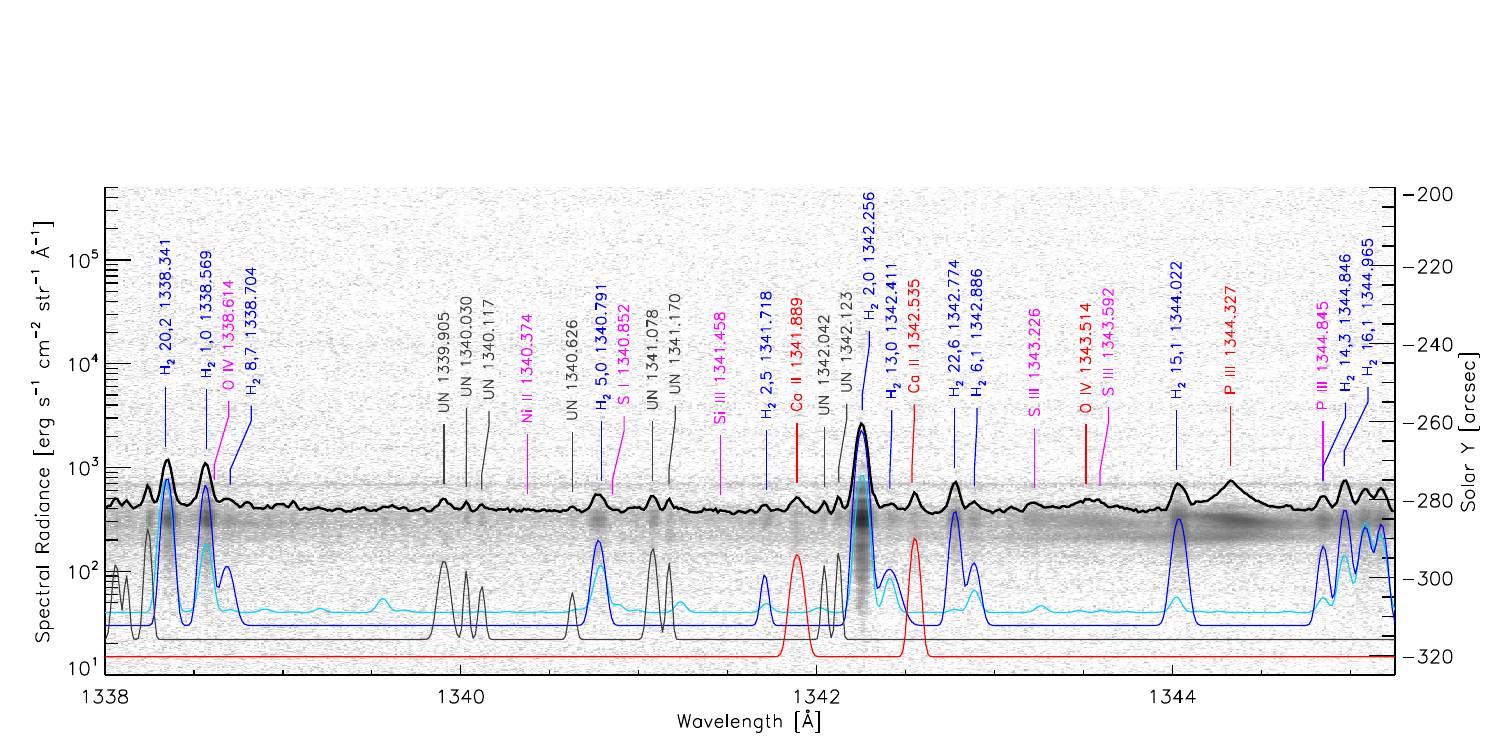}
	\end{center}
	\caption{A detailed view of the average flare spectrum (black line) from the IRIS FUV bandpasses with line identifications. Fitted \hh{} Lyman band lines are identified in blue with the ($J,v$) quantum numbers of the upper level.  Fitted atomic lines are noted in red, while atomic lines that were not fitted are labeled in magenta.  Fitted lines with unknown identifications are labeled in gray.  The composite fitted profiles for \hh{}, unknown, and weak atomic lines are shown by the dark blue, gray, and red lines respectively, where the continuum level of each is given an arbitrary offset from zero.  The FALC $\times$ 100 synthetic \hh{} spectrum is shown in cyan.  The background 2D spectrum has a reversed logarithmic color scale and comes from the time and slit position indicated in Figures \ref{fig:sji_frame} and \ref{fig:spec_map}.}
	\label{fig:FUV_spec}
\end{figure*}

\addtocounter{figure}{-1}
\begin{figure*}
	\begin{center}
		\includegraphics[width=7in]{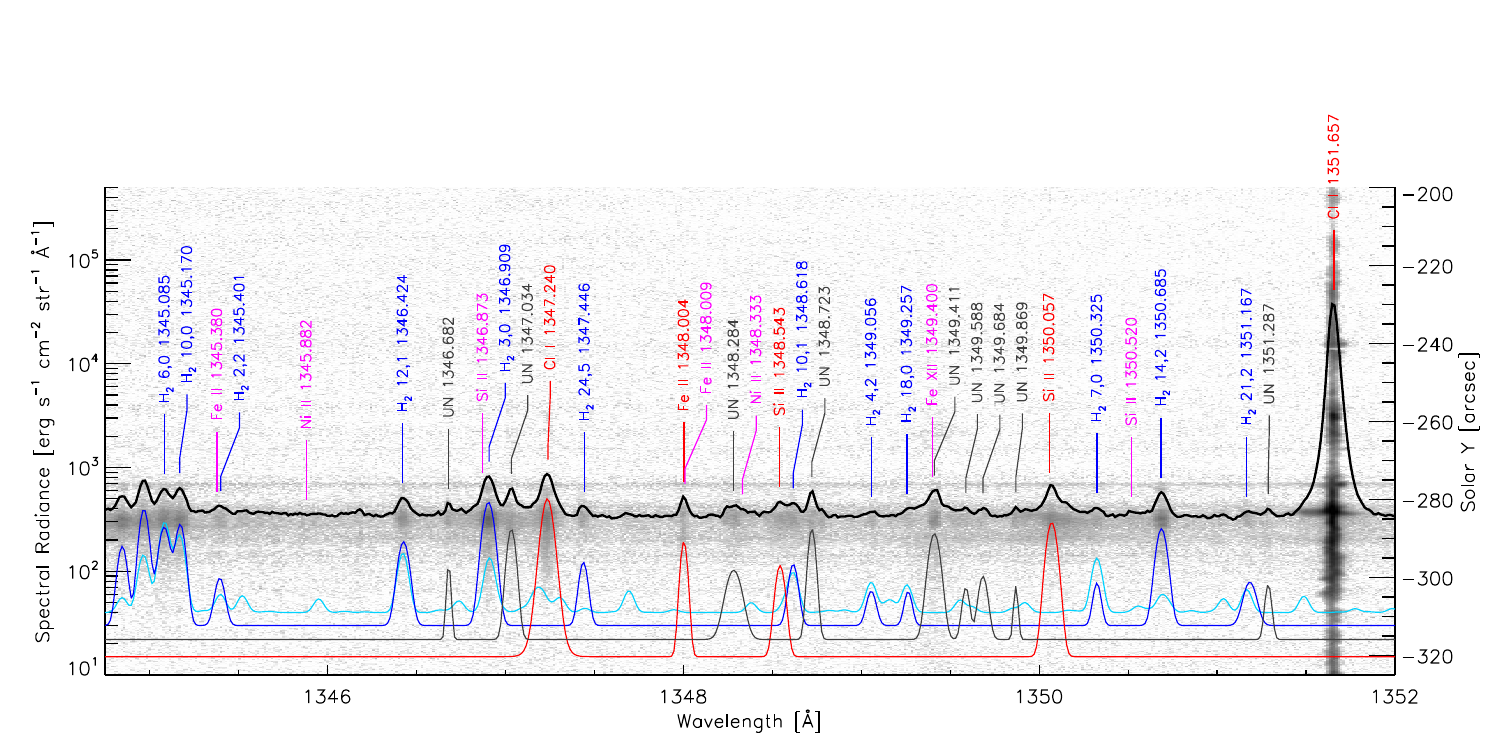}
		\includegraphics[width=7in]{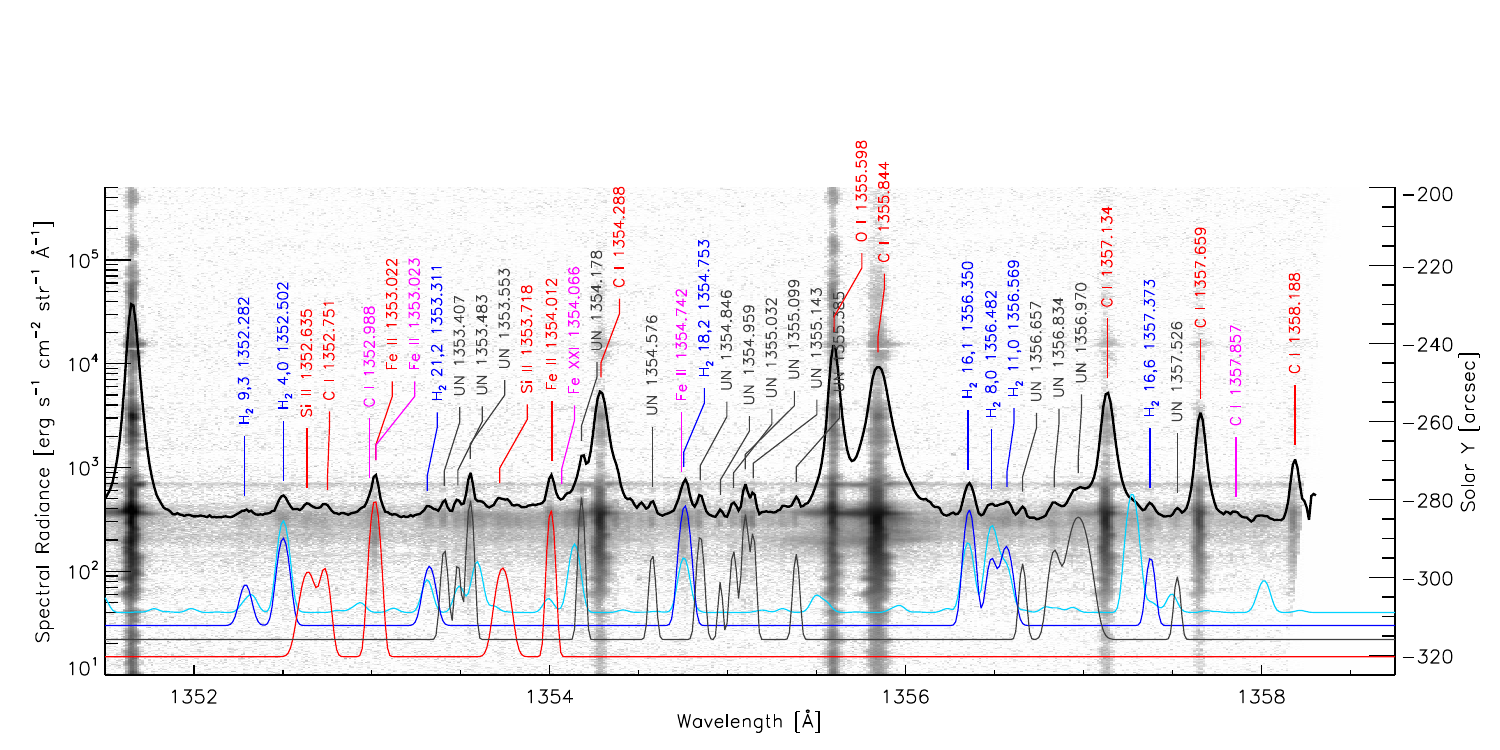}
	\end{center}
	\caption{{\it Continued.}}
\end{figure*}

\addtocounter{figure}{-1}
\begin{figure*}
	\begin{center}
		\includegraphics[width=7in]{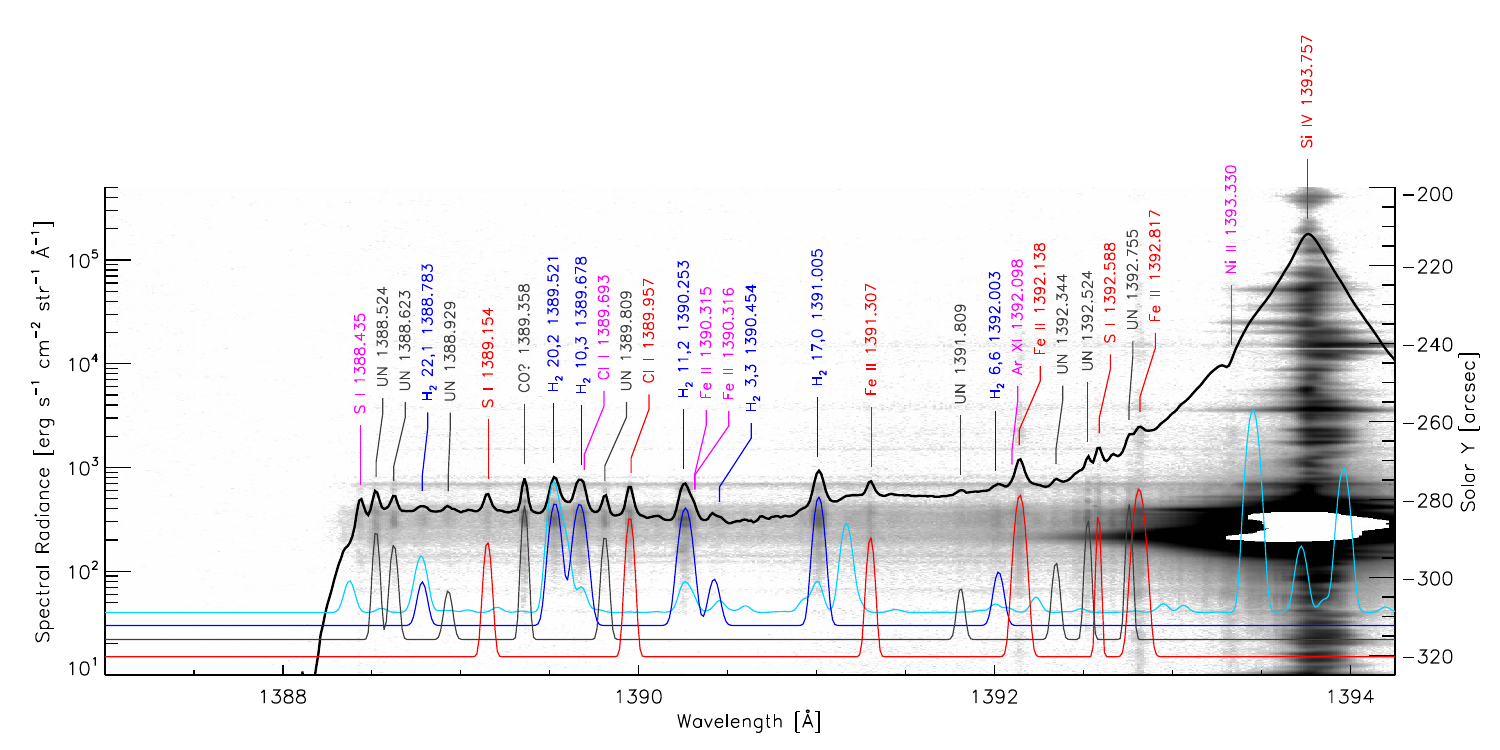}
		\includegraphics[width=7in]{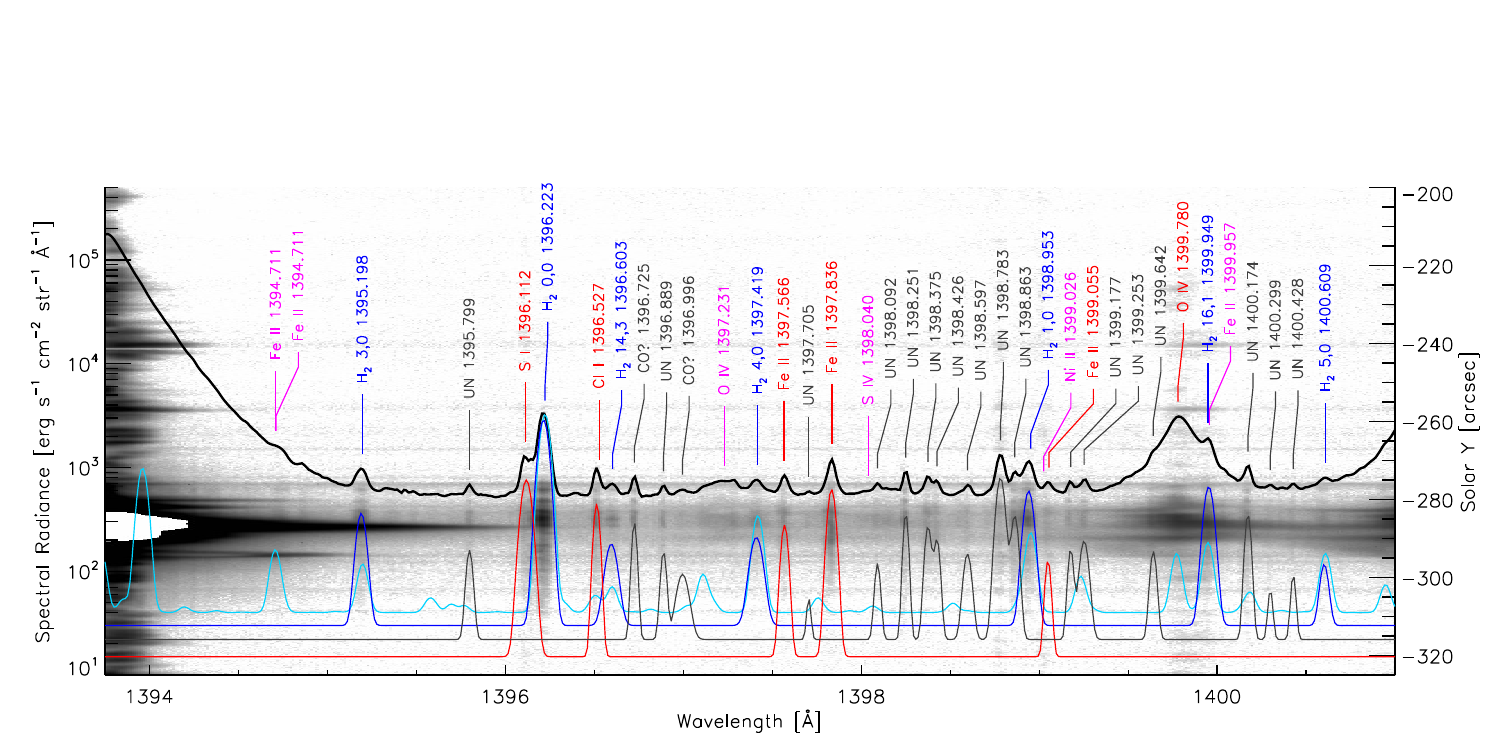}
	\end{center}
	\caption{{\it Continued.}}
\end{figure*}

\addtocounter{figure}{-1}
\begin{figure*}
	\begin{center}
		\includegraphics[width=7in]{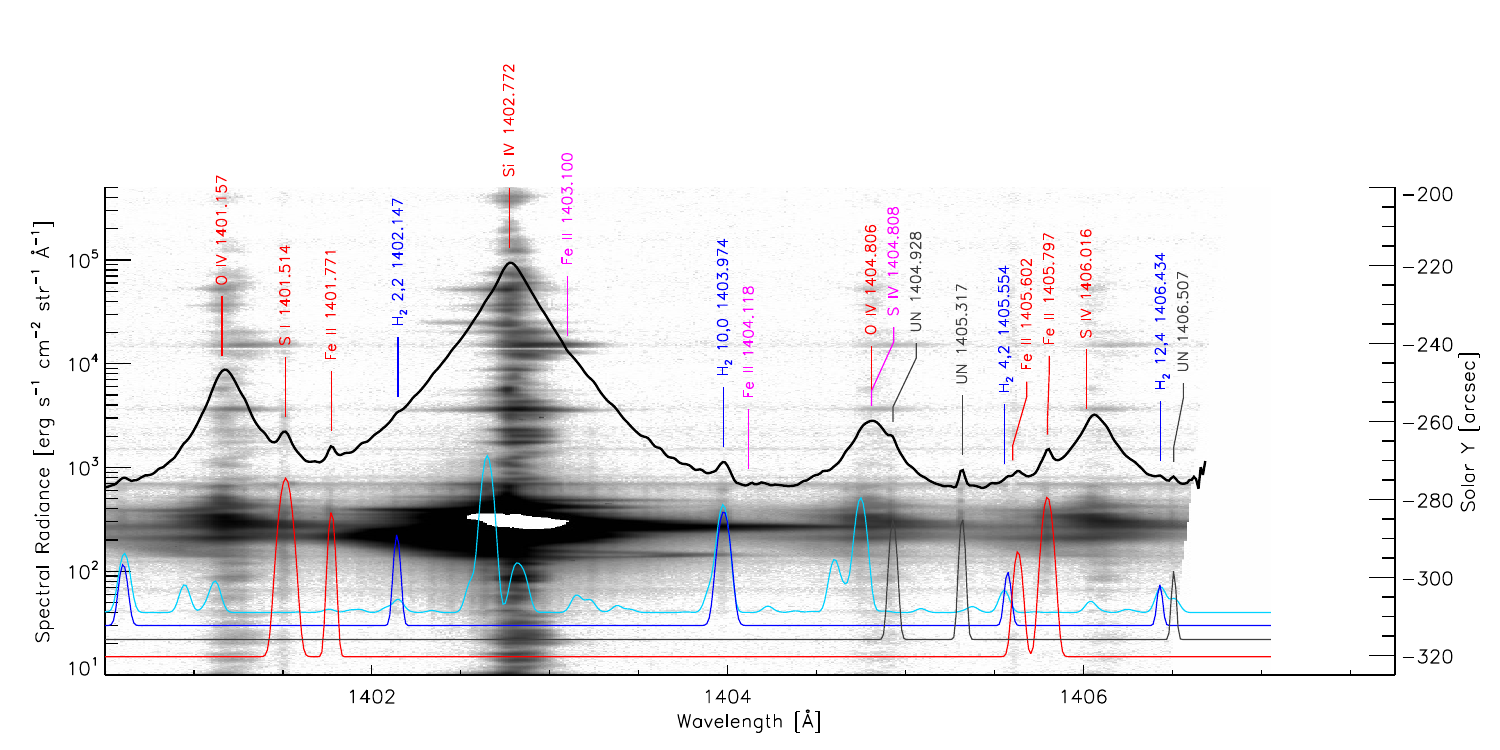}
	\end{center}
	\caption{{\it Continued.}}
\end{figure*}

We cross-referenced the atomic line lists from the \citet{sandlin86} HRTS/ATM atlas and the \citet{curdt01} SUMER atlas with the National Institute of Standards and Technology (NIST) Atomic Spectra Database \citep[][and references therein]{nist} and version 10.1 of the CHIANTI database \citep{chianti1, chianti2} to compile a complete list of wavelengths and atomic data.  
In cases where the wavelengths in multiple references did not agree, we adopted the more recent reference.  
For atoms greater than two times ionized, we adopted the CHIANTI wavelengths.  
Additional lines not identified by \citet{sandlin86} or \citet{curdt01} were suggested by Peter Young (\url{https://www.pyoung.org/iris/iris_line_list.pdf}).
Although \citet{curdt22} provides updated line identifications for SUMER, overlapping the IRIS FUV bandpasses, we do not include these lines in our list.
The newly added atomic line identifications are dim lines from \ion{O}{4} and \ion{S}{4} that cannot be seen in this spectrum.  

Table \ref{tbl:lines} includes all of the cross-referenced atomic lines, but only the \hh{} lines with positive identifications are listed.
For all lines from the compiled line lists, we provide the laboratory or predicted rest wavelength and the upper and lower level configurations for the transition.  
For atomic lines, the literature reference listed is for the wavelength, while for \hh{} lines it refers to the first identification in the solar spectrum.

\startlongtable
\input{linetable}

\subsection{Line Fitting}
The average flare spectrum was compared with our extended \hh{} linelist from spectral synthesis, and the previously identified \hh{} and atomic line lists.
We attempted to fit as many spectral lines as possible with the minimum requirement that the peak signal divided by the standard deviation of the data-minus-fit residual have a signal-to-noise ratio greater than five.
Weak, narrow lines were typically fit well with a single Gaussian profile.  
A linear or polynomial function was sometimes added to the fit to account for the shape of the underlying background (e.g. for \hh{} lines blended in the wings of the \ion{C}{2} or \ion{Si}{4} lines).  
In crowded regions of the spectrum it was necessary to fit up to three Gaussian profiles at once plus a constant term to get a reasonable match to the profiles.  
The transition region lines of \ion{Si}{4}, \ion{O}{4}, and \ion{S}{4} were fit with Lorentzian functions, while Moffat functions produced better fits to the bright atomic lines of \ion{Cl}{1}, \ion{C}{1}, and \ion{O}{1}.  
The \ion{C}{2} lines were fit with a composite Moffat and Gaussian function to account for the broad line wings and central depression of the line core, respectively.  
We ascribe no particular scientific significance to the type of profile used in each fit, as they depend both on the underlying line shape and the distribution of intensities and velocities along the line of sight and across the field of view during the flare.  
To compare the line width properties for all of these different kinds of peaked functions, we calculated the full width at half the maximum intensity(FWHM) of the profile or composite profile based on the fitted parameters.

Errors in the IRIS spectrograph data are a combination of photon counting, dark current, read noise, uncorrected bright pixels, and systematic errors (i.e. flat field residuals, stray light, etc.).  
Assuming an inverse gain of 4 photons DN$^{-1}$ for the FUV detector and a combined read and dark noise of 3 DN per pixel, estimated from the un-illuminated portion of the FUV detector, even small, 5 DN signals in the spectrum should achieve a signal-to-noise of over 100.
Due to the large number of individual spectra combined to make the average flare spectrum, poorly characterized systematic noise sources probably dominate over detector and photon counting noise sources at this level.
The errors in the fitted parameters were therefore estimated by running the fit twice.  
After the first fit, the standard deviation of the data-minus-fit residual was calculated.  
This estimate of the error in the intensity was provided during the second pass through the fitting procedure to obtain the 1-$\sigma$ errors in the fitted parameters.

The results of fitting lines in the 1D flare spectrum are shown in Figure \ref{fig:FUV_spec}, the dark blue line shows the composite profiles from fitting lines attributed to \hh{}, the red line shows the composite profiles from fitting lines attributed to weak atomic species, and the gray line includes the composite profiles of all unknown lines.
Arbitrary background values have been added to the composite line profiles so that they can be easily distinguished from each other.
The spectral line labels appear above the spectrum.  
The \hh{} lines are labeled in dark blue.  
Atomic lines that were fit are shown in red, but if a line was not able to be fit, it is shown in magenta.
Unknown lines are labeled ``UN'' and shown in gray.
The measured wavelength, peak spectral radiance ($R_{peak}$), and full-width at half-max are given in Table \ref{tbl:lines} for those lines that were fitted.

\begin{figure}
    \centering
    \includegraphics[width=3.4in]{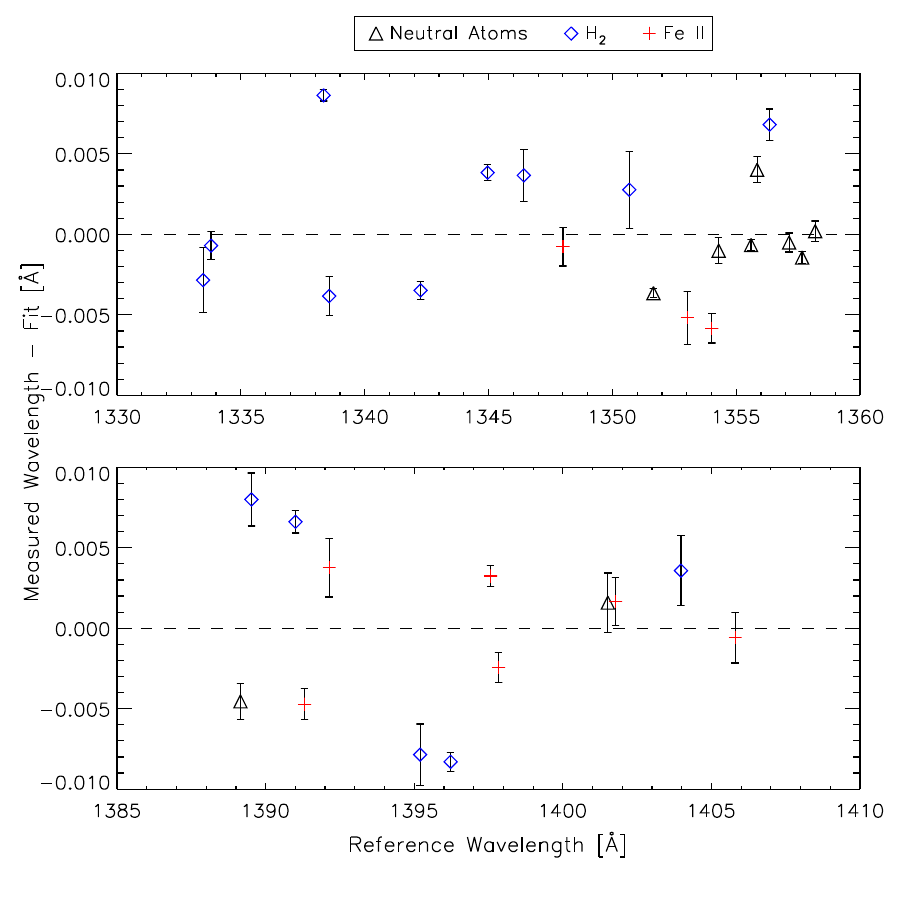}
    \caption{Selected lines of \hh{}, neutral atomic species, and \ion{Fe}{2} were used to determine the approximate local rest frame for the 1D flare spectrum in the IRIS FUV 1 (top) and FUV 2 (bottom) channels.  The error bars plotted for each point represent the 1-$\sigma$ errors in the fitted wavelength.}
    \label{fig:wavelength_fit}
\end{figure}

\subsection{Wavelength Correction}
A selection of lines were used to calibrate the local rest wavelength in the average flare spectrum separately for the FUV1 and FUV2 passbands.  
The lines selected are ones with unambiguous identifications, that could be easily fit from the spectrum, and primarily due to \hh{} or neutral atomic species, although some \ion{Fe}{2} lines were also included.
Figure \ref{fig:wavelength_fit} shows the residuals of a linear fit to the measured wavelength versus the reference wavelength.  
This fit provides a small correction to the linearization and slit-by-slit shift in wavelength applied to the spectra during the data reduction.  
The standard deviations of the residuals in each of the spectral bandpasses are about 5 m\AA, about half a pixel with the 12.8 m\AA\ sampling of the spectrograph, which corresponds to approximately 1 km s$^{-1}$ at these wavelengths.
This wavelength correction was applied to the observed line wavelengths listed in Table \ref{tbl:lines} and used in all further analysis.

\subsection{\hh{} Line Identification Criteria}\label{sec:idcrit}
Based on the compiled line lists, line fitting results, and using the FALC$\times100$ synthetic \hh\ spectrum shown in Figure \ref{fig:FUV_spec},
we applied a set of criteria for determining if it was likely that a line in the spectrum belonged to the \hh{} Lyman band.
The previously identified \hh{} lines were used as a template for identifying new lines.  
For all of the previous identifications we determined the maximum deviation from the expected wavelength and the minimum and maximum line width to set expectations for the new lines.
These criteria are as follows:
\begin{itemize}
    \item[(a)] The observed line matches the expected wavelength within $\pm$ 18 m\AA.
    \item[(b)] The observed line has a width between 48 and 80 m\AA.
    \item[(c)] Other lines from the same upper level are observed with the approximate relative intensities within a factor of 2 of the value, or within the measurement error, whichever is higher.
    \item[(d)] Other lines from the same upper level cannot be observed due to blends, because they are predicted to be weak, or they fall outside of the observed bandpasses.
    \item[(e)] The observed line intensity approximately matches the prediction by spectral synthesis.
    \item[(f)] The observed line does not match any existing line identifications.
\end{itemize}
Lines that fulfill multiple of these criteria, particularly (c), are considered to be more certain identifications.
Table \ref{tbl:ratios} in the appendix shows our evaluation of the criteria fulfilled by each line.
Further discussion and justification for line identifications in the case of ambiguous and blended lines is given in the appendix.

\section{Analysis of the Spatially Resolved Flare Spectra}\label{sec:spec2d}
To learn more about the variation in properties of \hh{} emission during the flare activity, we performed profile fitting for a small selection of the brightest \hh{} lines for each spatial pixel within the area shown in Figure \ref{fig:spec_map}.
As an estimate of the irradiation received by those \hh{} lines we determined the intensities at the corresponding wavelengths that excite their upper levels.
To place the properties of the \hh{} lines in context with other spectral diagnostics, we carried out profile fitting of other lines in the IRIS spectra representative of the photosphere and chromosphere.

\begin{figure}
    \centering
    \includegraphics[width=2.75in]{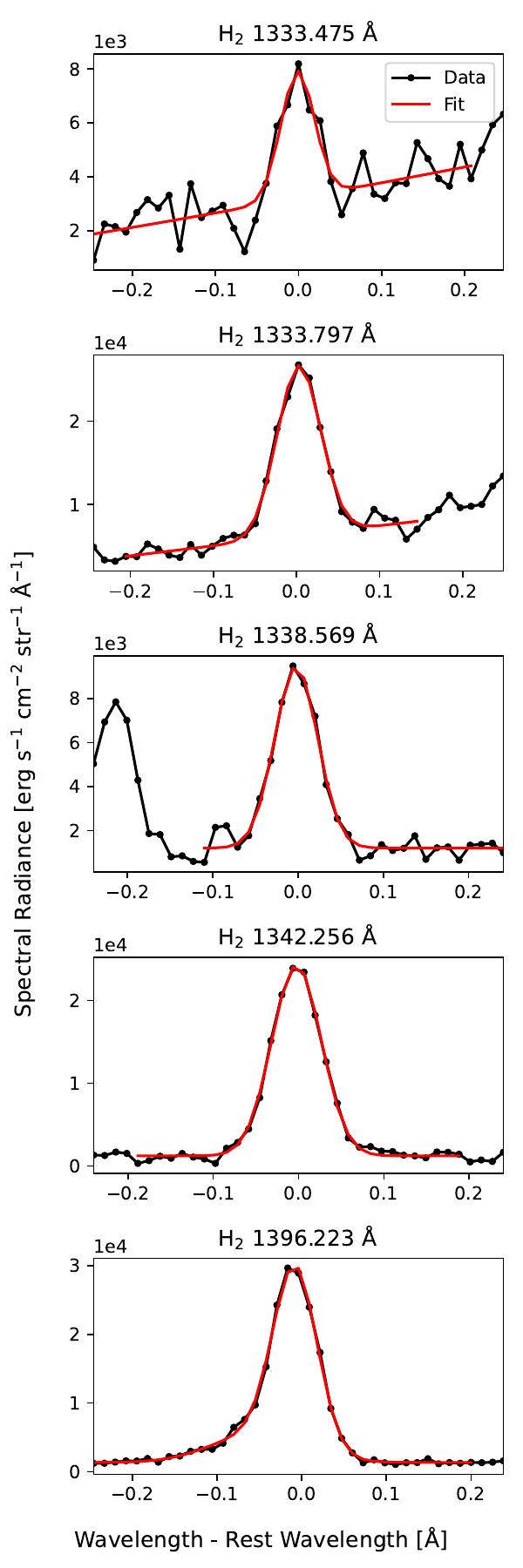}
    \caption{An example of \hh{} line fits performed on the spatially resolved flare spectrum.  The line profiles shown here come from the position indicated by the (+) symbol in Figure \ref{fig:spec_map}.}
    \label{fig:h2linefit}
\end{figure}

We fit one \hh{} line at 1396.223 \AA\ from the (0,0) upper level, two lines at 1333.474 and 1338.573 \AA\ from the (1,0) upper level, and two lines at 1333.796 and 1342.256 \AA\ from the (2,0) upper level.
As during the fitting of the average flare spectrum, these lines were fit with Gaussian profiles.  
For the line at 1396.223 \AA, a double Gaussian fit was performed to take into account the neighboring \ion{S}{1} line in the blue wing.
For the lines in the wing of the 1334.532 \AA\ \ion{C}{2} line (1333.474 and 1333.796 \AA), an additional quadratic function was used to account for the background which is often present during flares because the \ion{C}{2} lines can become very broad or Doppler shifted due to large velocities.
Figure \ref{fig:h2linefit} shows an example of the \hh{} line fitting for the position indicated by the (+) symbol in Figure \ref{fig:spec_map}.

\begin{figure}
    \centering
    \includegraphics[width=2.75in]{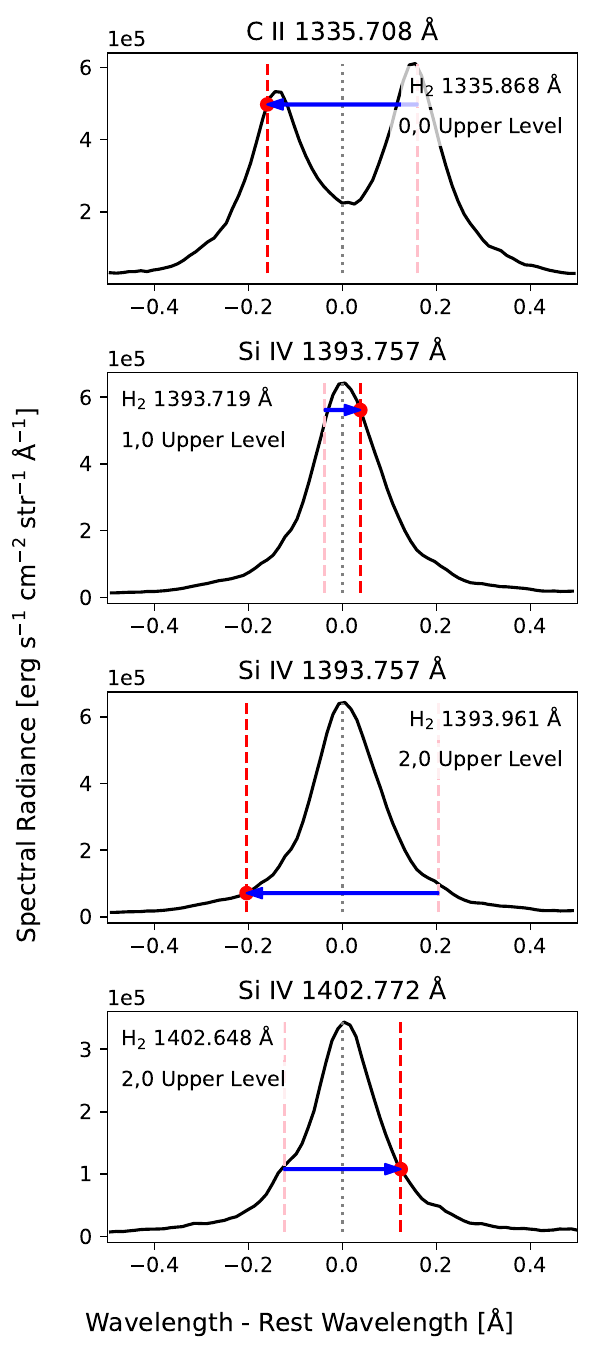}
    \caption{An example of how the spectral radiance for the \hh{} pumping line was derived from the spatially resolved flare spectra.  For each \hh{} pumping wavelength considered (pink dashed line), we reversed the \hh{} line wavelength with respect to the rest wavelength of the transition region line that provides the pumping photons and interpolated the spectral radiance at this location (red dashed line). The line profiles shown here come from the position indicated by the (+) symbol in Figure \ref{fig:spec_map}.}
    \label{fig:2demis}
\end{figure}

For each of the \hh{} upper levels referenced above, we determined the spectral radiance at the location of the strongest pumps identified based on the spectral synthesis results shown in Figure \ref{fig:pump_contrib}.
Because of the viewing geometry near disk center, the velocity shifts of the bright lines that excite \hh{} (\ion{C}{2} and \ion{Si}{4}) as seen by \hh{} in the low chromosphere are reversed from the observer's reference frame.
It is necessary to take the spectral radiance value from the opposite side of the exciter line profile relative to the exciter line center wavelength to get the spectral intensity as seen in the rest frame of \hh{}.
A demonstration of this technique is shown in Figure \ref{fig:2demis} for each of the transitions that populate the upper levels.
The spectra in this figure come from the same spatial pixel indicated by the (+) symbol in Figure \ref{fig:spec_map}.

\begin{figure}
    \centering
    \includegraphics[width=2.75in]{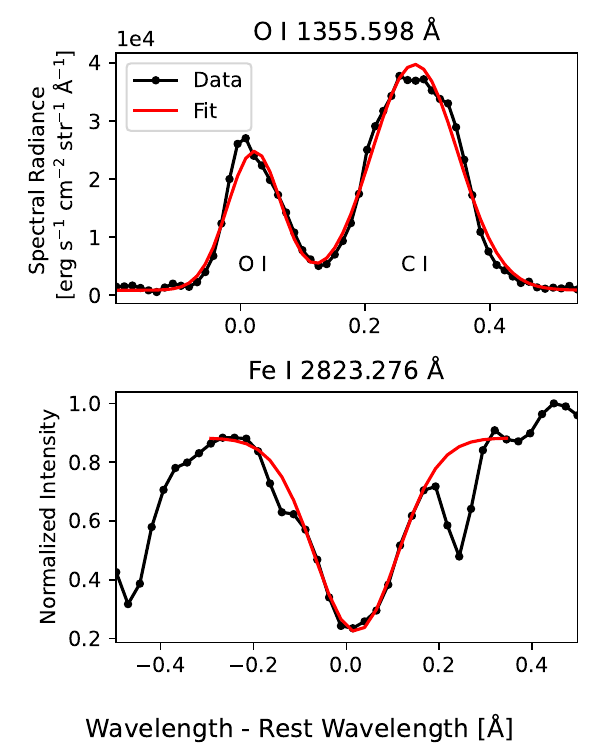}
    \caption{An example of the chromospheric \ion{O}{1} and photospheric \ion{Fe}{1} line fits performed on the spatially resolved flare spectrum. The line profiles shown here come from the position indicated by the (+) symbol in Figure \ref{fig:spec_map}.}
    \label{fig:atmlinefit}
\end{figure}

Although there are many neutral and singly ionized lines in the FUV spectrum that should form at similar heights to \hh{}, most of these lines are dimmer than the brightest \hh\ lines and difficult to fit over the full area of the observation.
Instead, we have chosen lines that are more easy to measure and that should bracket \hh\ in terms of their formation height in the upper photosphere/lower chromosphere.
As a reference for photospheric behaviour, we fit an \ion{Fe}{1} line at 2823.276 \AA\footnote{Following the standard convention, all wavelengths less than 2000 \AA\ are listed as vacuum wavelengths, and all wavelengths longer than 2000 \AA\ are listed as in-air wavelengths.} in the far red wing of \ion{Mg}{2} from the IRIS NUV spectrum.
The NUV spectrum is crowded by absorption.  This line is fairly isolated, and it is possible to fit the local continuum on each side of the line.
The \ion{Fe}{1} line profile was fit using a simple Gaussian profile plus a constant term to account for the variable amplitude of the continuum.
The wavelengths about weaker blending lines in the wing of the \ion{Fe}{1} line were excluded from the fit.
As a reference for the mid-chromosphere, we fit the \ion{O}{1} line at 1355.598 \AA\ in the FUV 1 spectrum.
The formation of this line has been investigated thoroughly by \citet{lin15}.
It forms at approximately 1.5 Mm in the chromosphere and is optically thin, which is rare for chromspheric lines.
During the flare, the \ion{O}{1} line becomes contaminated by the nearby \ion{C}{1} line at 1355.844 \AA, so it is necessary to fit them together.
To account for the narrow line core and broad wing that we see in both the \ion{O}{1} and \ion{C}{1} lines, we employ a Moffat function.
The Moffat function is based on a Lorentzian profile and is often used for characterization of the PSF of imaging systems.
It provides a better single-profile characterization of the lines than either a single Gaussian or Lorentzian profile because these lines very often have a compact peak with broad wings that extend far out to the red and blue.
The lines might be fit more accurately by multiple components for some locations during the flare activity, but such detailed fitting is beyond the scope of this work.
Examples of the fitted profiles of the \ion{Fe}{1} and \ion{O}{1} line are shown in Figure \ref{fig:atmlinefit}.

\section{Results}\label{sec:results}
In this section we present the results of our various analyses conducted with the \hh{} lines in this IRIS observation informed by synthesis results.
Because these results are based on a spectrograph raster scan, variation in the atmospheric parameters may be variations in space or time.  
Non-equilibrium effects are not considered, though they may be important for \hh{}, especially in dynamic events such as flares.
We interpret the spectra as if they come from an atmosphere that has reached a steady state.

\subsection{Line Identifications}
Using the average flare spectrum we verify the presence of many of the \hh{} lines previously identified by \citet{jordan77}, \citet{jordan78}, \citet{bartoe79}, \citet{sandlin86}, and \citet{jaeggli18} in the IRIS FUV bandpasses between 1332-1356 \AA\ and 1388-1406 \AA.  
Some previous line identifications could not be verified in the average flare spectrum for several lines too deeply embedded in regions with bright lines. 
There is a line at 1356.834 \AA\ in a heavily blended region; it may be the line 1356.859 \AA\ belonging to the (3,4) upper level identified by \citet{sandlin86}, but this line is not predicted to be very bright based on our radiation model, so we have labeled it as ``unidentified.''

Using the average flare spectrum we identify 37 new lines of \hh{} that are consistent with the width and spatial extent of the previously identified \hh{} lines, they appear alongside lines at wavelengths consistent with their upper level family, and/or they are a good match to the wavelengths of lines predicted to be bright by modeling.

Many of the \hh{} lines are brighter than the weak atomic species, such as \ion{Fe}{2}, \ion{S}{1}, and \ion{Ca}{2}.  
The \hh{} line widths are 60 m\AA\ on average, larger than these other species because \hh{} is lightweight, so the impact of thermal broadening is larger.
Some \hh{} lines are brighter than the spectral synthesis prediction from FALC $\times$ 100, while some lines are dimmer.

Many weak lines in the average flare spectrum remain unidentified.
A number of these lines are similar to each other in appearance, with very narrow widths smaller than \hh{} and the known atomic lines, indicating that they originate from a cool species with large mass.
A few of these lines at 1389.358, 1396.725, 1396.996 \AA\ have wavelengths consistent with CO lines previously identified in \citet{jordan79co}.
These lines originate from the 4th positive system transitions from the A$^{1}\Pi$ excited state to the X$^1\Sigma^+$ ground state.
We have tentatively labeled these in Table \ref{tbl:lines} and provided their upper and lower level rotational and vibrational quantum numbers using the same notation as used for \hh{}.
Other similar lines in the spectrum may be due to CO, but detailed modeling of the excitation, comparable to what we have done for \hh{}, is necessary to confirm their identities.

\begin{figure*}
    \begin{center}
        \includegraphics[width=6.0in]{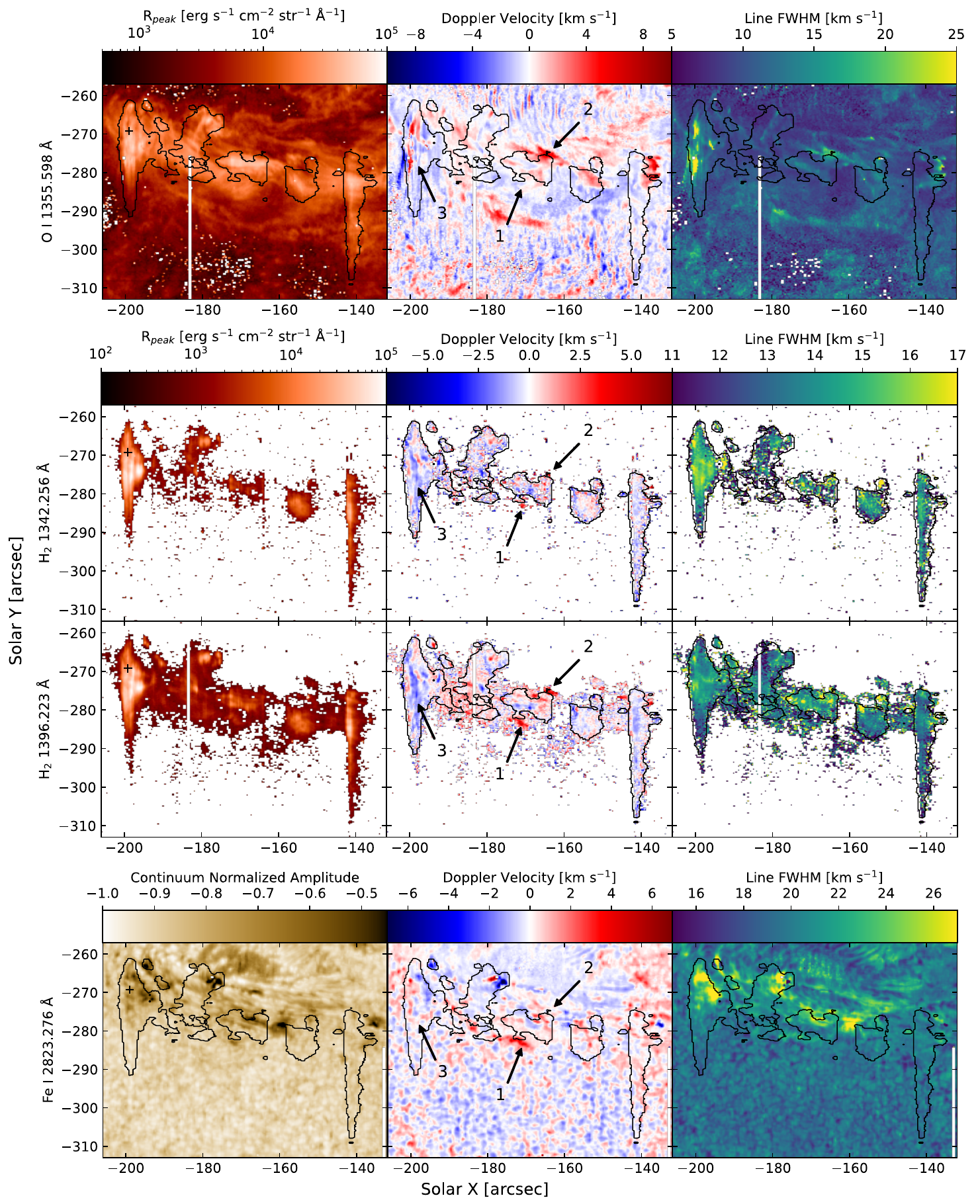}
    \end{center}
    \caption{Maps of line amplitude (left), Doppler velocity (center), and FWHM (right) derived from line-fitting analysis of the mid-chromospheric \ion{O}{1} line at 1355.598 \AA\ (top), the two brightest \hh{} lines in the IRIS FUV (middle panels), and the photospheric \ion{Fe}{1} line from the NUV (bottom).  The \hh{} velocity maps share some features in common with both the photospheric and chromospheric lines, but they also have unique features, indicating that \hh{} probes a unique height between the photosphere and choromosphere. The color scales are specific to each plot grouping.}
    \label{fig:fit_maps}
\end{figure*}

\subsection{Morphology of \hh{} Emission in Context}
Modeling places the formation height for the \hh{} emission lines in the IRIS FUV bandpass just above the temperature minimum in the upper photosphere or lower chromosphere, around 650 km above $\tau_{500 nm} = 1$ in the FALC model \citep[][Figure 8]{jaeggli18}.
It is possible to perform some validation of the formation height based on observations by comparing the properties seen in \hh{} to other diagnostics with well understood formation.
To that end, we compare maps of the line intensity, Doppler velocity, and width derived from the line-fitting analysis of the spatially resolved spectra for the two brightest \hh\ lines in the IRIS FUV bandpass at 1342.256 and 1396.223 \AA\ with the photospheric \ion{Fe}{1} line at 2823.276 \AA\ in the IRIS NUV bandpass and the chromospheric \ion{O}{1} line at 1355.598 \AA.

Figure \ref{fig:fit_maps} shows maps of the line amplitude, Doppler velocity, and line FWHM (from left to right) for the \ion{O}{1} 1355.598, \hh{} 1342.256, \hh{} 1396.223, and \ion{Fe}{1} 2823.276 \AA\ lines (from top to bottom).
The field of view has been reduced to the region where the majority of \hh\ emission is seen so that it can be shown in greater detail.
In the case of the NUV \ion{Fe}{1} line, the amplitude of the absorption line has been normalized by the nearby continuum value and is expressed in relative units, while the amplitudes of the FUV lines are in calibrated radiance units.
The Doppler velocity and line FWHM are expressed in velocity units (km s$^{-1}$).
We use slightly different color scaling for the quantities of each line, the color bars are specific to each plot grouping.
The data has been masked to exclude regions where the intensity, Doppler velocity, and line FWHM fall outside of specified values, so the results for \hh{} are mainly restricted to regions with high line intensity.
For context we show the \hh{} intensity contour (as in Figures \ref{fig:sji_frame} and \ref{fig:spec_map}) on all other plots to provide a common reference.

The two \hh{} lines have different exciters (\ion{C}{2} and \ion{Si}{4}), but their intensity is very similar.
They also have similar velocity shifts and line widths, which we would expect because they form at roughly the same height.
The brightest \hh{} regions correspond well with bright regions seen in \ion{O}{1}.
\ion{O}{1} 1355.598 \AA\ is populated through a cascade following recombination from \ion{O}{2}.  Although the line excitation mechanism is very different it ultimately comes from the same driver as \hh{}, heating of the chromosphere during the flare.

The \ion{Fe}{1} line shows expected behavior of a photospheric line, namely it reproduces the pattern of granulation, penumbral fibrils, and dark sunspot umbrae and pores, but the line amplitude is reversed with respect to the continuum intensity.
The line amplitude is smaller when continuum intensity is low, and larger when continuum intensity is high.
The \ion{O}{1} line shows the irregular structure of the active region chromosphere.
In velocity, it is possible to see the pattern of umbral oscillations in the sunspot.

For all lines, strong redshifts and blueshifts are seen in the flare regions along with enhanced line width.
The \hh{} Doppler velocity shares some features in common with the photosphere and chromosphere, but also shows some features that are not shown by the other lines.
The magnitude of Doppler velocities seen in the \hh{} lines is consistent with the photosphere $\pm$5 km s$^{-1}$.
A strong red-shifted patch is present in the \hh{} velocity map, indicated by arrow 1, and a similar patch appears in the \ion{Fe}{1} velocity map but does not appear to have a component in the chromosphere.
Just to the northwest of this patch there is another red-shifted patch in \hh{}, indicated by arrow 2, that has a corresponding patch in the chromosphere, but no photospheric component.
Intermediate to these two red-shifted patches, the \hh{} map shows two small blue-shifted patches side by side that do not appear in either the photospheric or chromospheric line velocities.
In addition, there is a blue-shifted patch that runs north-south, indicated by arrow 3, that does not seem to correspond to features in either the photospheric or chromospheric velocity map.

We conclude from the \hh{} Doppler velocity that we are seeing material intermediate between the photosphere and the chromosphere.
Some of the velocity flows span the height between the \hh{} and the photosphere and between \hh{} and the mid-chromosphere, but \hh{} also shows the unique behavior of material that falls in between these two heights.

\subsection{\hh{} Line Intensity Ratios}

\begin{figure*}
	\begin{center}
		\includegraphics[width=7.0in]{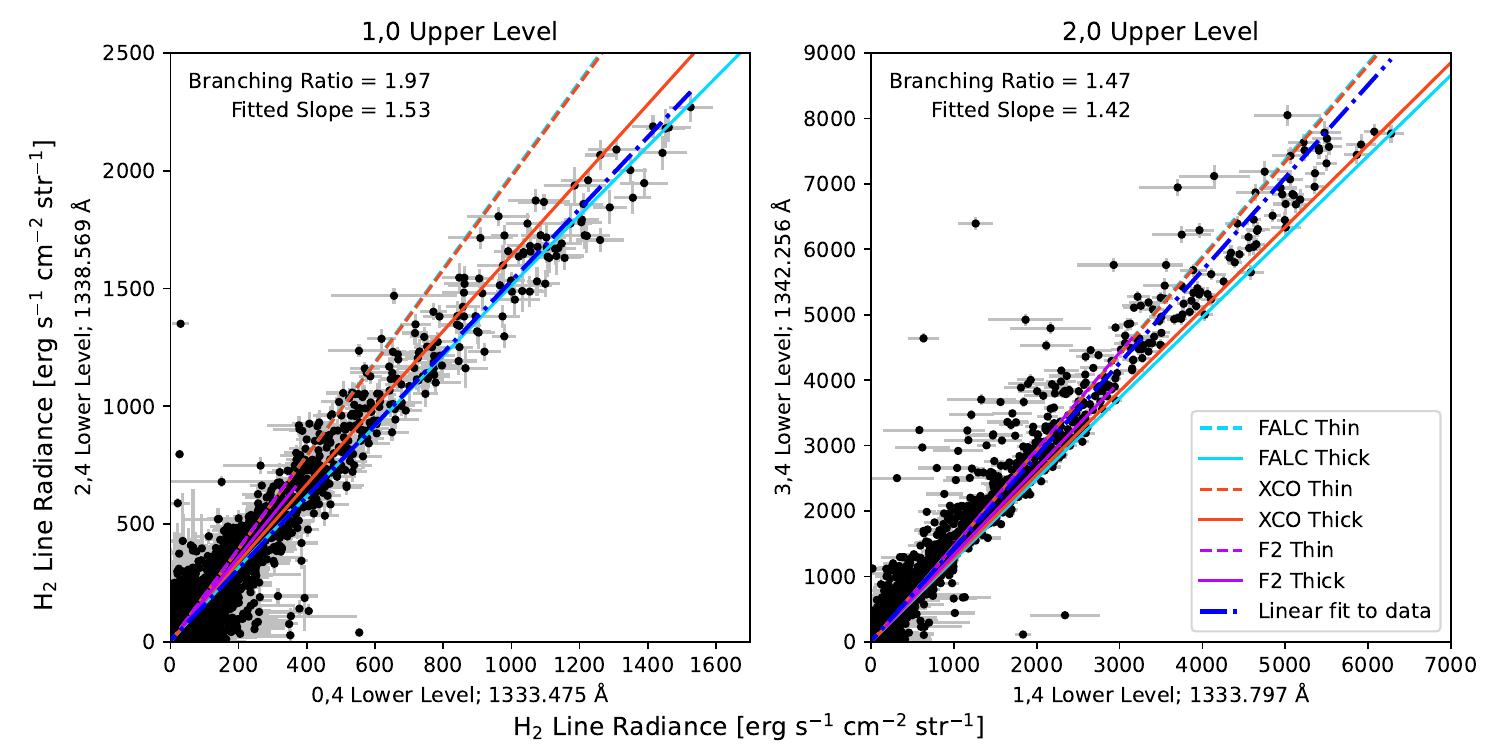}
	\end{center}
	\caption{Scatter plots of the intensity from bright \hh{} lines of two different upper levels measured from the spatially resolved spectrum.  In the left panel the intensity of 1338.572 \AA\ is plotted on the x-axis, and the intensity of 1333.474 \AA\ is plotted on the y-axis ($J=1, v=0$ upper level).  In the right panel the intensity of 1342.256 \AA\ is plotted on the x-axis, and the intensity of 1333.796 \AA\ is plotted on the y-axis ($J=2, v=0$ upper level).  The black data points are plotted with gray 1-$\sigma$ error bars in the total radiance for each line. The line ratio resulting from spectral synthesis is plotted for each pair of lines using solid lines for the optically thick case and dashed lines for the optically thin case of synthesis calculation from the three model atmospheres. The dark blue dot-dashed line shows a linear fit to the data points.}
	\label{fig:line_ratios}
\end{figure*}

The ratio of intensities for \hh{} lines sharing a common upper level should depend on the transition probability, the amount of reabsorption and redistribution that happens due to the optical density of \hh{} in the line formation region, and the amount of background absorption from other line and continuum sources that occurs between the \hh{} line formation region and the observer.
The ratio of \hh{} line intensities therefore has some dependence on the atmospheric parameters, they allow us to distinguish between the optically thin and optically thick cases of line formation.
In this section we discuss the results of fitting the brightest \hh{} lines in the spatially resolved spectra and how those line intensities compare within the same upper level family.
The line intensity ratios that result from the average flare spectrum are more difficult to compare with the spectral synthesis results and are given in Table \ref{tbl:ratios} in the Appendix.

Figure \ref{fig:line_ratios} shows the comparison of the total line radiance determined from the fitted Gaussian profiles of the 1333.475 and 1338.569 \AA\ lines from the (1,0) upper level (left panel), and the 1333.797 and 1342.256 \AA\ lines from the (2,0) upper level (right panel).
The total line radiance is shown along with the associated 1-$\sigma$ errors for each data point (black points with gray error bars).
Data points with errors that were too high, i.e. due to low signal levels, were rejected and not plotted here.
A linear fit was applied to the plotted data points (dark blue dot-dashed line) and the slope of this fitted line is printed in the top left corner of each panel along with the expected branching ratio between the two transitions.
The line intensity relations from the spectral synthesis results are shown for the optically thin case (dashed lines) and optically thick (solid lines) calculations for line formation for each model as indicated in the figure legend.

For the chosen \hh{} lines and upper levels, the optically thin relations shown by the overlapping red, cyan, and purple dashed lines are consistent with the branching ratio, but this is not always the case.
If there is a significant difference in the background opacity between the two wavelengths the slope of the intensity relation will be different from the branching ratio.
The hotter F2 model does not produce very large \hh{} line intensities, and the synthetic results do not fully cover the measured line radiances even with a multiplier of 10,000, but the cooler FALC and XCO models do encompass the measurements.

In the case of the (1,0) upper level, the data points seem to show behavior which is more consistent with the optically thick approximation where the XCO model provides the best match to the measured slope.
For the (2,0) upper level the results are not so clear.  
The difference between the optically thin and optically thick calculations is not large, and the majority of the data points at the upper end of the relation seem to fall between the optically thick and optically thin synthesis results.
A scattering of points with larger errors appear to the upper left of the main relation, and the contribution from these points results in a steeper slope in the linear fit to the data points that is more consistent with the optically thin calculation.
Considering the potential for systematic errors in the line fits, the 1342.256 \AA\ line is isolated, fairly strong, and has no known blends, hence the small errors that can be seen in the y-direction.
The 1333.797 \AA\ line does show larger fitting errors.
Although it is nearly as strong as the 1342.256 \AA\ line, it is in the blue wing of the \ion{C}{2} line, so the background underlying the \hh{} line can be difficult to characterize when \ion{C}{2} is bright and the wing becomes steeply sloped or curved.
The 1333.797 \AA\ line is also potentially in an unresolved blend with \ion{S}{1} at 1333.792 \AA.
Contamination by \ion{S}{1} would tend to make the measured line brighter with respect to the 1342.256 \AA\ line, while failure to fit the curved background of the \ion{C}{2} line wing could underestimate the line intensity.

\begin{figure}
	\begin{center}
		\includegraphics[width=3.5in]{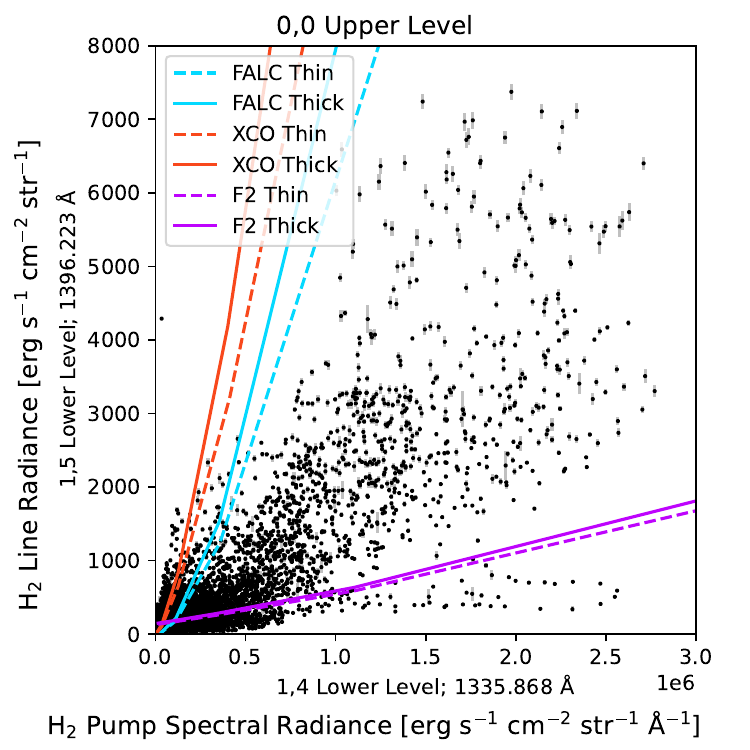}
	\end{center}
	\caption{The total \hh{} line radiance versus the spectral radiance observed at the wavelength of the pumped \hh{} line for the (0,0) upper level.  The black data points are plotted with the 1-$\sigma$ error bars in the total line radiance.  The same quantity was derived from the spectral synthesis results for the three atmospheric models for both optically thin (dashed lines) and optically thick calculations (solid lines).}
	\label{fig:pump_ratio_00}
\end{figure}

\subsection{\hh{} Emission Versus Excitation}
The strength of emission from \hh{} is strongly related to the strength of emission from the bright lines that excite the transition to the upper level, but there are several other factors related to the atmospheric parameters that can impact the fluorescent emission produced by \hh{}.
Continuum opacity, mainly due to photo-ionization of Si, and opacity of the exciting line itself determine how far down the exciting radiation can travel to reach \hh{}.
The \hh{} population is dependent on atmospheric temperature and density.
The \hh{} transition probability, optical density in the \hh{} line formation region, and the background opacity determine the number of photons that exit the atmosphere from each transition from the upper level. 
The relation between the intensity of an \hh{} line and its excitation source therefore probes the atmospheric structure.
We cannot know exactly what the downward radiation from the exciter looks like, but for optically thin transition region lines like \ion{Si}{4}, it should be similar to the radiation exiting the atmosphere upward.

\begin{figure}
	\begin{center}
		\includegraphics[width=3.5in]{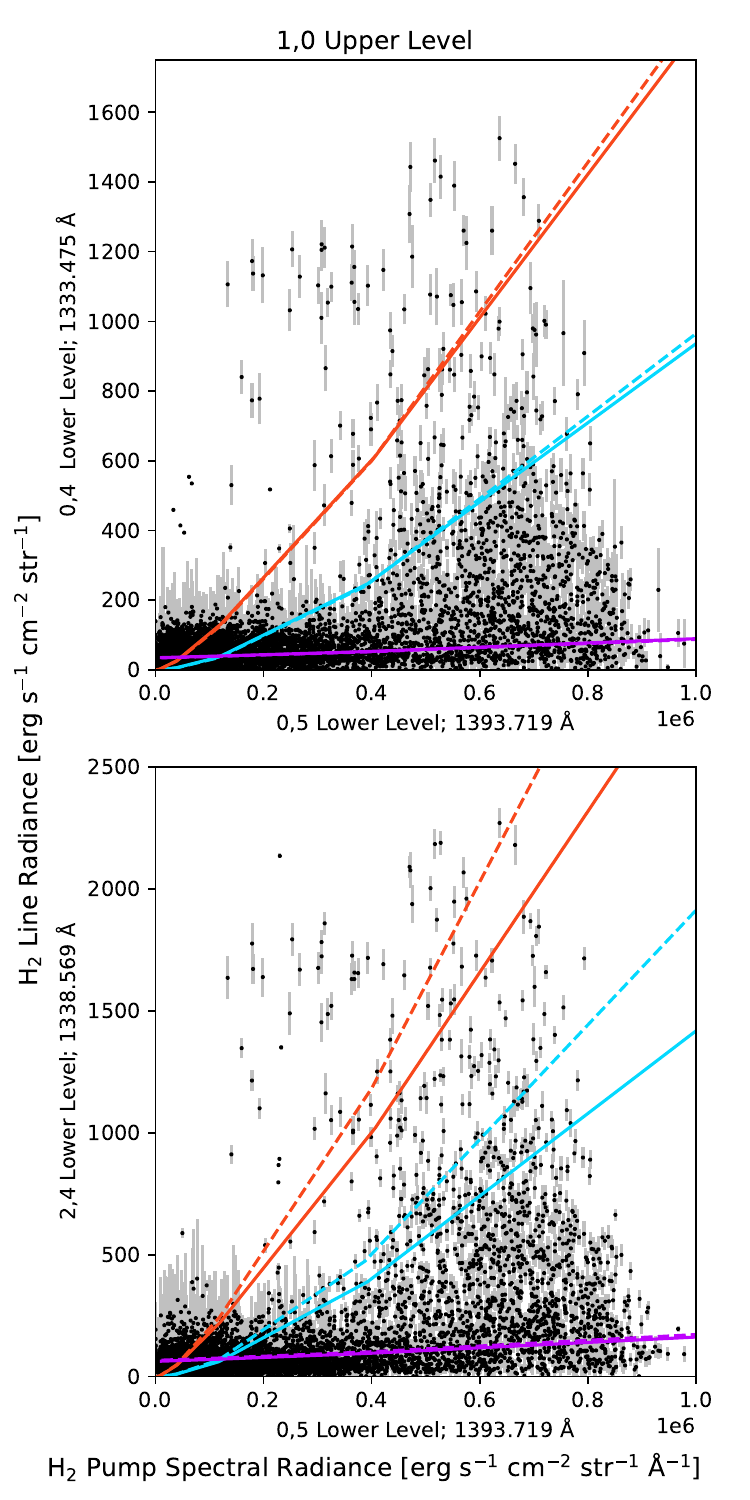}
	\end{center}
	\caption{The total \hh{} line radiance versus the spectral radiance observed at the wavelength of the pumped \hh{} line for the (1,0) upper level.  Plot symbols are the same as in Figure \ref{fig:pump_ratio_00}.}
	\label{fig:pump_ratio_10}
\end{figure}

\begin{figure*}
	\begin{center}
		\includegraphics[width=7.0in]{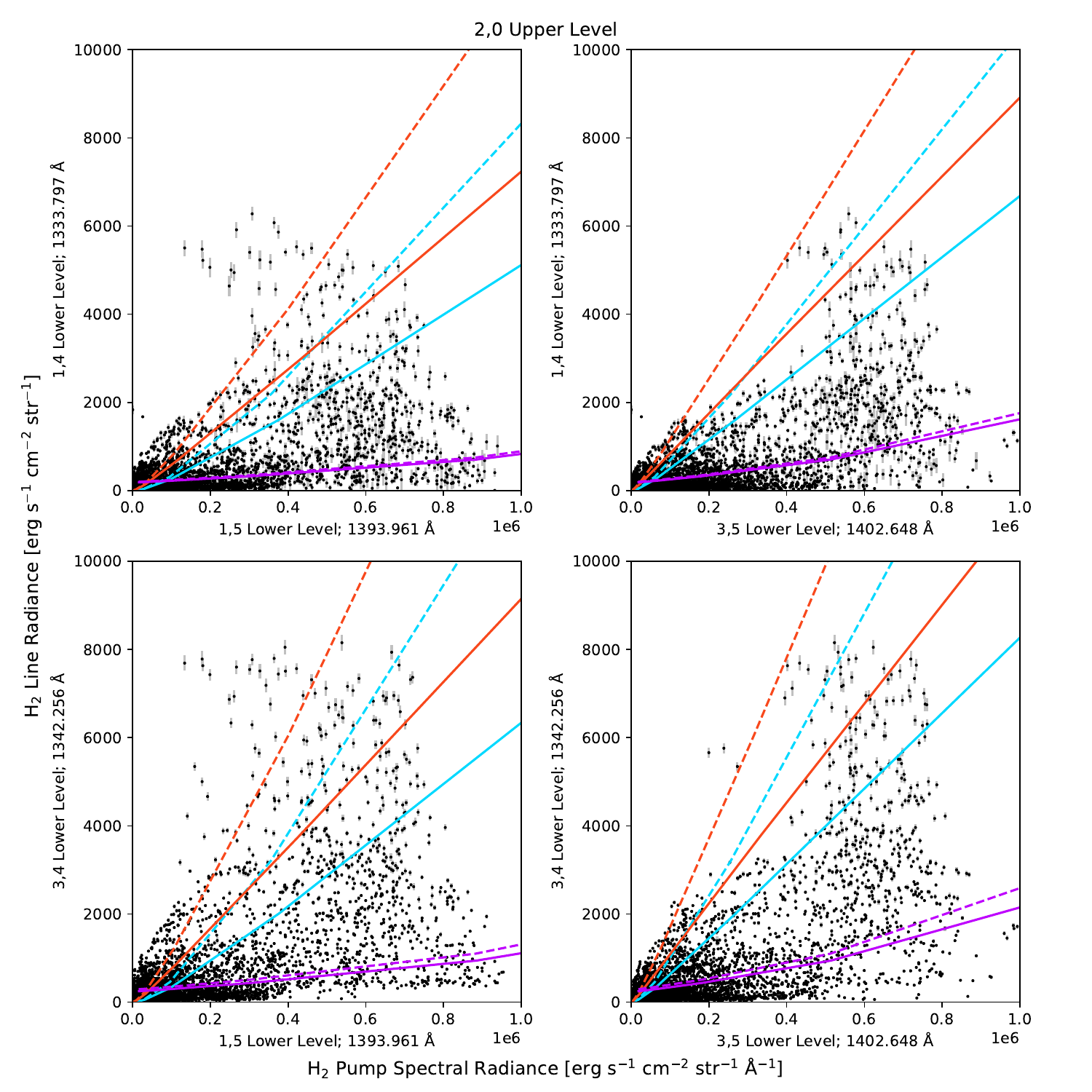}
	\end{center}
	\caption{The total \hh{} line radiance versus the spectral radiance observed at the wavelength of the pumped \hh{} lines for the (2,0) upper level.  Plot symbols are the same as in Figure \ref{fig:pump_ratio_00}.}
	\label{fig:pump_ratio_20}
\end{figure*}

The IRIS FUV bandpasses allow us to see both the excitation source(s) and the \hh{} lines they excite for the (0,0), (1,0), and (2,0) upper levels which produce the brightest \hh{} lines observed by IRIS.
Figure \ref{fig:pump_ratio_00} shows the relationship between the total radiance measured for the 1396.223 \AA\ \hh{} line and the spectral radiance at the wavelength corresponding to the \hh{} pumping line at 1335.868 \AA, excited by \ion{C}{2}, which both belong to the (0,0) upper level.
Figure \ref{fig:pump_ratio_10} shows the relation between the (1,0) upper level lines at 1333.475 and 1338.569 \AA\ with respect to their pumping line at 1393.719 \AA\ excited by the 1393 \AA\ \ion{Si}{4} line.
Figure \ref{fig:pump_ratio_20} shows the relation between the (2,0) upper level lines at 1333.797 and 1342.256 \AA\ with respect to both pumping wavelengths at 1393.961 and 1402.648 \AA\ in each of the \ion{Si}{4} lines.
In each of these figures we show the \hh{} line total radiance versus the pumping spectral radiance for each fitted spatial pixel (black points) with error bars based on the line fitting (gray).
We show the same relations derived from the spectral synthesis calculations for each atmosphere, for both optically thin (dashed lines) and optically thick (solid lines) cases.
As discussed in Section \ref{sec:spec2d}, the line-of-sight velocities we observe from above the atmosphere are reversed for \hh{} since it is located below the transition region, so we take the spectral radiance from the wavelength on the opposite side of the exciter line profile.

The location of a particular data point in this parameter space says something about the physical parameters of the atmosphere in that spatial pixel.
From the variation in slope of the relations from the models, we can tell that the temperature of the chromosphere is a key ingredient.
The cool XCO model produces the most \hh{} emission per input excitation, i.e. a larger \hh{} population is exposed to exciting photons.  
The slightly warmer FALC model shows a slightly shallower slope, and the F2 model, with its hot chromosphere has a much shallower slope and is widely separated from the other two models.
For the (0,0) upper level, the datapoints are bounded by the spectral synthesis results from the various models, implying that a model with intermediate chromospheric temperature might better describe these locations.
The lines of the (1,0) upper level show similar behavior, although there is more scatter and the error bars are larger since these lines are weaker than the 1396.223 \AA\ \hh{} line.
The intensity relations for the (2,0) upper level lines essentially fill the full parameter space between the models.

Each plot shows a grouping of points extending vertically from the main relation at the highest spectral radiance levels for the pumped line.
Although we explicitly excluded pixels which showed saturated values from these relations, pixels near to saturated areas may still be impacted by charge overflow on the IRIS FUV spectrograph detector.
The \ion{Si}{4} lines, especially the 1393 \AA\ line, have higher detector count rates than \ion{C}{2}, and the line that stimulates the (1,0) upper level sits very close to the peak of the 1393.757 \AA\ \ion{Si}{4} line (see the second panel in Figure \ref{fig:2demis}), so the relations for that upper level should be impacted the most by saturation effects.

\begin{figure*}
    \begin{center}
        \includegraphics[width=7.0in]{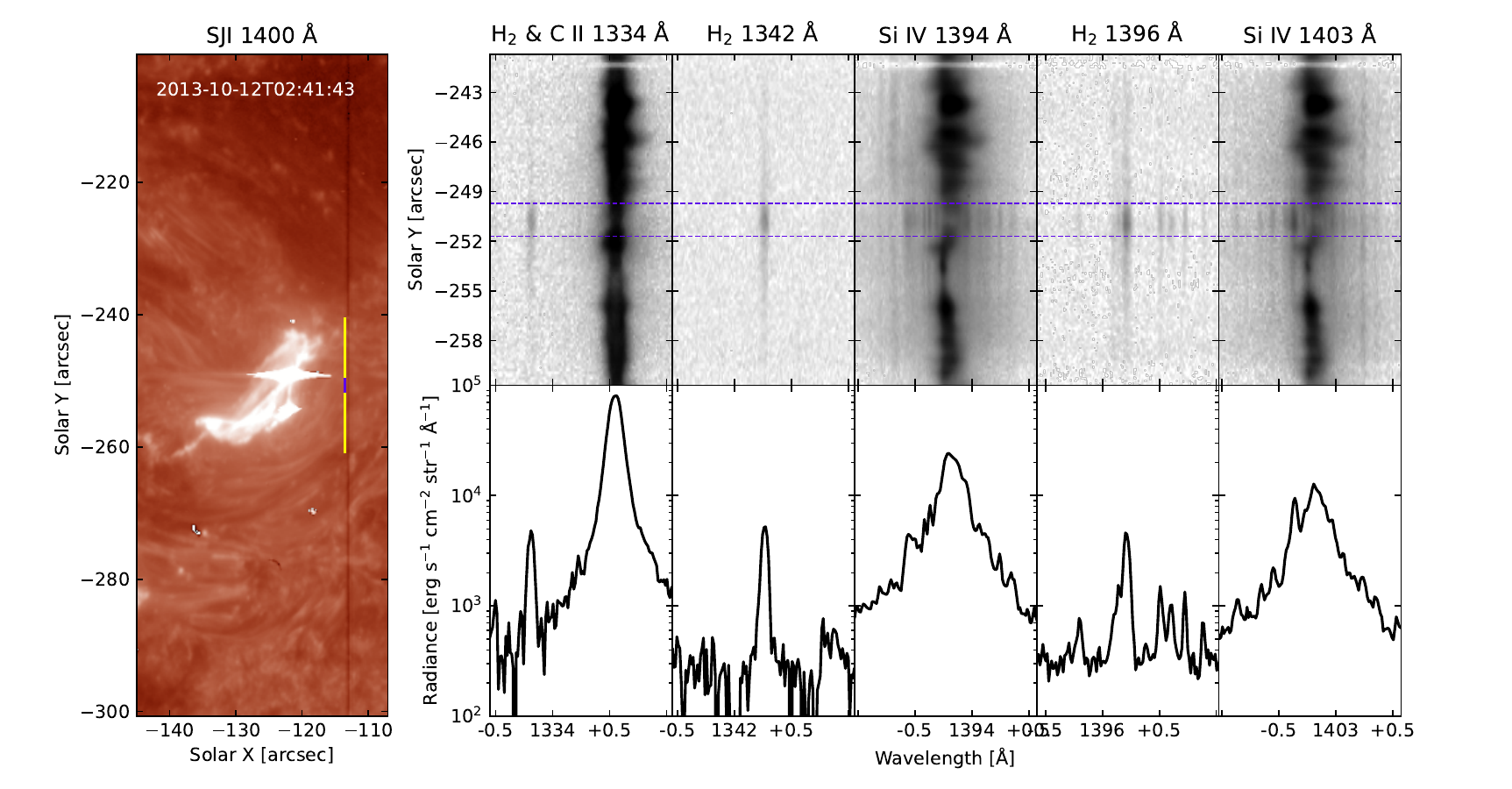}
    \end{center}
    \caption{An example of \hh{} excitation by a source that is not directly above the region sampled by the spectrograph slit of IRIS.  The SJI 1400 \AA\ context image (left panel) shows bright emission from a loop-like structure.  The brightest part of the structure shows a saturation spike in the image extending left-right along the detector pixels.  Segments of the FUV spectrum from the region covered by the spectrograph slit are shown in the right set of panels.  The top row shows the spectrum from the region covered by the yellow line in the SJI image.  The brightest \hh{} emission occurs in the region enclosed by the dashed lines, and the average spectral profile from this region is shown in the lower set of panels.  The \hh{} emission from this region is very strong relative to the local emission from the bright lines of \ion{C}{2} and \ion{Si}{4}, with a peak radiance only about an order of magnitude smaller.}
    \label{fig:nonlocal}
\end{figure*}

\subsection{The Impact of Non-Local Excitation Sources}\label{sec:nonlocal}
For the results presented in the previous section, we compared the \hh{} line intensities to the transition region emission sources in the same spatial pixel.
One effect not yet accounted for is the impact of radiation sources that are not directly above the observed \hh{} emission.
In the 1D radiation model used for calculating the \hh{} emission, it is assumed that the radiation source shines down isotropically from the top of the atmosphere.
In the observations, some instances during the flare activity are a good match this situation, when a broad region around the spectrograph slit has fairly uniform illumination, but there are also times when \hh{} excitation occurs in response to distant sources.

One good example of this is seen at 2013-10-12T02:41:43.
In the spectroheliogram map of \ion{Si}{4} shown by the left panel of Figure \ref{fig:spec_map}, this time corresponds to the heliocentric position (X:-114$\arcsec$,Y:-250$\arcsec$).
The contour indicating bright \hh{} emission does not correspond to bright emission in \ion{Si}{4} as seen by the spectrograph, but the SJI reveals that bright flare activity from a loop-like structure occurs to the left of the spectrograph slit position.
Figure \ref{fig:nonlocal} shows the SJI context image and selected sections of the FUV spectra.
In this situation, the \hh{} lines are extremely bright relative to the \ion{C}{2} and \ion{Si}{4} lines at the same location seen by the slit.
The \hh{} lines have a peak radiance of only about a factor of 10 lower than the co-spatial transition region lines, and \hh{} lines located near to the \ion{Si}{4} lines are visible when they could not be seen at other locations during the flare activity.
It is obvious that the bright region to the left of the slit provides the excitation for \hh{} in this region, although it is about 8$\arcsec$ or 6 Mm away.

For any given position in the observation, \hh{} may receive more radiation from sources that are not directly above the location of emission.
Due to the isotropic illumination assumed in the radiation model, the spectral synthesis results can only provide a lower bound on the observed relation between the \hh{} line intensity and the exciting radiation seen by the spectrograph slit shown in Figures \ref{fig:pump_ratio_00}, \ref{fig:pump_ratio_10}, and \ref{fig:pump_ratio_20}.

\section{Discussion}\label{sec:discussion}
We have presented diverse results from our investigation of \hh{} emission during flares in an active region.  The following subsections serve as a guide of the discussion of these diverse results, how our results relate to previous work, and what might be the next steps in the investigation of \hh{} diagnostics.

\subsection{Identification of New \hh{} Lines}
We have derived new identifications for lines of the \hh{} Lyman band using FUV flare spectra from IRIS.
We have also confirmed the identity of \hh{} lines within the IRIS FUV bandpasses previously identified by \citet{jordan77,jordan78}, \citet{bartoe79}, and \citet{jaeggli18} based on HRTS and SUMER spectra.
The \hh{} line identifications presented in \citet{jaeggli18} and those compiled in this work are based on comparison of the observed spectra with spectral synthesis based on 1D model atmospheres and making use of the expanded \hh\ line list and an up-to-date partition function not available when previous work was carried out.
Some of the new lines listed in this work can be seen in the SUMER atlas spectrum used in \citet{jaeggli18}, but they did not meet the intensity threshold for being included in that list.
IRIS has superior spectral resolution with respect to SUMER and HRTS, so some of the new lines were in blends that could not be resolved by those instruments.
More progress on \hh{} line identifications for the Lyman and Werner bands may be possible with the development of instruments with greater sensitivity and spectral range than what IRIS currently achieves, for example, the spectrograph for the Extreme Ultraviolet Solar Telescope \citep[EUVST,][]{suematsu21}.

\subsection{Is \hh{} line formation optically thin or thick?}\label{sec:thick_or_thin}
It remains unclear if \hh{} lines are optically thin or if there is significant absorption and redistribution to transitions with the highest photon escape probabilities.
Considering the \hh{} lines excited by Lyman-$\alpha$ emission, \citet{jordan78} expected that \hh{} emission from quiet-Sun regions would be impacted more by opacity effects than emission from sunspots, which could be assumed to be optically thin.
They confirmed that the relative \hh{} intensities observed in sunspot spectra matched the optically thin assumption within photometric errors for their observation over a sunspot.
Based on this expectation, the calculation of \hh{} line intensities, developed in \citet{jaeggli18} and used in this work, was carried out for two cases, one treating the lines as optically thin and the other using escape probabilities to approximate the optically thick case.

Using these synthetic results we have compared the intensity relation for the 1333.481 and 1338.565 \AA\ lines of the (1,0) upper level and the 1333.797 and 1342.257 \AA\ \hh{} lines of the (2,0) upper level with both the optically thin and optically thick calculations for many different locations throughout flare activity occurring mainly over granulation and penumbra in an active region.
We have found that the (1,0) upper level is in better agreement with the optically thick approximation using escape probabilities, while the (2,0) upper level lines favor the optically thin calculation, although systematic effects in the fitting have not been ruled out.

For the observations shown in \cite{mulay23}, \hh{} emission comes from beneath a flare ribbon which occurs over granulation between two sunspots in an active region.
For the same lines that we have compared, \citet{mulay23} found that the line ratios for the (1,0) and (2,0) upper levels were in agreement with the expected ratio (the branching ratio based on the transition probabilities) under optically thin assumption, however, the error in their relation (Figure 8 in that paper) is fairly large and likely would not distinguish between the optically thin and optically thick cases that we have shown in Figure \ref{fig:line_ratios}.
In addition, the intensities shown are in detector counts rather than physical units and do not appear to have been corrected for the wavelength-dependent sensitivity of the IRIS spectrograph.
This correction is especially important for taking the ratio of lines between the FUV1 and FUV2 channels, but is probably not as critical for lines nearby in wavelength such as these.

Although optically thin treatment may be ``good enough'' in most cases, the question of optically thick vs. optically thin remains open and may deserve more detailed treatment in radiative transfer calculations.

\subsection{Collisions Revisited}
Another effect to consider is that of possible collisions with excited \hh{}. 
In \citet{jordan78}, \citet{bartoe79}, \citet{jaeggli18}, and in this work, the effect of collisions on the upper level is assumed to be negligible. 
Noting that the effect of collisions will be greater for cases where \hh{} emission occurs from deeper in the atmosphere, we briefly revisit this assumption at a height of 650 km above the photosphere in the FALC model, where the majority of \hh{} fluorescence occurs, the temperature is 4750~K, and the H and He densities are $6\times 10^{14}$ and $6\times 10^{13}$~cm$^{-3}$, respectively.

Unfortunately, information regarding collisional rates for electronically excited \hh{} with the dominant species (H and He) is scarce, so we provide some examples using what data is available for the $X^1\Sigma^+_g$ ground state. 
\cite{vargas24} provide a reactive (dissociative) collision rate coefficient for \hh{}+H at 4750~K of $5\times 10^{-13}$~cm$^3$~s$^{-1}$, and a factor of 3 less for collisions with He, implying a rate of $\sim 300$~s$^{-1}$ for collisions with H at the typical height of \hh{} florescence, and about 10~s$^{-1}$ for He. 
These collision frequencies are insignificant compared to the total probability for spontaneous emission of the upper levels ($\sim 10^9$~s$^{-1}$).

For non-reactive collisions altering the state of \hh{}, we estimate from \cite{mandy99} that the total rate coefficient to other levels from a given level of \hh{} is less than $10^{-9}$~cm$^3$ s$^{-1}$, implying state-to-state collision frequencies less than $6\times 10^{4}$~s$^{-1}$, still substantially less than the transition probabilities. 
This implies that the probability of an upper level being de-populated before it radiatively decays, and corresponding effect on intensities from that level, are all well under 1\%. 
Nevertheless, in order to fully evaluate the possibility of collisional de-population of the upper levels, collision rate calculations are needed for excited electron configurations of \hh{}. 
Note that collisions will not affect the line ratios from a given upper level, and will therefore not affect the discussion in the previous Subsection \ref{sec:thick_or_thin}.

\subsection{Do regions with cooler photospheric temperatures produce brighter \hh{} emission?}
Based on the observations from the HRTS first and second flight \citet{jordan78} and \citet{bartoe79} saw enhancement of \hh{} in sunspot spectra.
Observations presented in \citet{innes08} and \citet{schuhle99} show enhancement of \hh{} over sunspots.  The Werner-band lines used in their analyses are excited by \ion{O}{6} 1031.94 \AA\ which has enhanced emission over sunspots, although resonance with the \hh{} emission from the sunspot may further enhance this line \citep{morgan05, labrosse07}.

For the observation shown in this work, bright \hh{} emission is seen in regions that are mostly above regions of granulation and sunspot penumbra, as can be seen in the NUV continuum map on the right side of Figure \ref{fig:spec_map}.
We do not see particular enhancement of \hh{} above the sunspot, indeed, emission from \hh{} cannot be seen in regions without strong transition region emission, unless the illumination comes from outside the spectrograph slit as in the example in Figure \ref{fig:nonlocal}.
The IRIS observations used in \citet{mulay21} and \citet{mulay23} are of similar character, where bright emission from \hh{} occurs over granulation between sunspots in an active region and is very closely associated with the \ion{Si}{4} emission from the flare ribbon.

The generalization that \hh{} emission comes from sunspots is misleading.
\citet{jordan78} notes that the \hh{} emission seen during the first flight of HRTS comes from particular locations in the sunspot umbra where H Lyman $\alpha$ shows broad wings, implying that there is activity occurring.
Likewise the \hh{} emission seen during the second flight of HRTS \citep{bartoe79} is enhanced above a sunspot light bridge in a flare productive active region (see discussion in \citet{jaeggli18}).
For the Lyman-band \hh{} lines in the IRIS bandpass, the main ingredient to produce \hh{} emission is a sufficiently bright excitation source.
However, it is worth investigating the difference between flare emission over sunspot umbrae and over granulation to see if the temperature of the lower solar atmosphere has a significant impact of the level of \hh{} emission.

\subsection{Height of \hh{} Emission Line Formation}
The UV emission from \hh{} originates at a critical height in the Sun's atmosphere at the transition from the photosphere to the chromosphere.
The modeling of \hh{} emission carried out here and in \citet{jaeggli18} places the formation height for the lines in the IRIS bandpass around 650 km above the photosphere (defined as where $\tau=1$ for the 500 nm continuum) in the FALC and XCO atmospheres, but for the hot F2 flare atmosphere they form around 450 km.
This is around the same height where the infrared lines of the CO fundamental band have been shown to form \citep[e.g.,][]{stauffer22}, and seem to show extremely cool material \citep{ayres81}.

Comparing the Doppler velocities measured for strong \hh{} lines with photospheric \ion{Fe}{1} and chromospheric \ion{O}{1},
some features match behavior in the photospheric diagnostic, some features match behavior in the chromospheric diagnostic, and some behavior is unique to \hh{}.
The matching features may indicate that plasma motions span the formation heights of the diagnostics and \hh{}, but they might also indicate a change in the formation height of the diagnostics due to atmospheric changes during the flare activity.
In particular, heating of the atmosphere during a flare can cause changes in the atmospheric parameters, including the continuum opacity through ionization of the neutral species that are responsible (Si, C, and Fe).
As demonstrated by the F2 atmosphere, the formation height of \hh{} is driven downward in a hot atmosphere, although our simulation results represent the steady-state solution.
Abrupt heating might allow UV photons to reach deeper, denser material where \hh{} is abundant before dissociation of the molecules can occur.
Changes in the formation height might be diagnosed through time-dependent changes in the \hh{} emission with respect to its exciter, or changes in the line width.
It should also be possible to constrain the formation height of \hh{} using simultaneously observed diagnostics from different heights throughout the photosphere and chromosphere.
IRIS's NUV \ion{Mg}{2} lines can provide height-dependent information through inversion of the lines \citep{delacruz19, sainzdalda24}.

\subsection{What are the limitations of using 1D models to understand \hh{} emission?}
It is not necessary to invoke a very cool or highly thermally structured atmosphere to explain the presence of \hh{} emission in the UV spectrum of solar flares.  
Our comparison of \hh{} line intensities to their excitation wavelengths alongside the spectral synthesis shows that atmospheres warmer than the FALC model can account for emission during the flare activity observed by IRIS.
At the same time, there are obvious challenges to using 1D models to interpret the 3D radiative transfer problem of transition region radiation shining down on the lower solar atmosphere.

The radiation that excites \hh{} may come from diverse heights.
\citet{mulay23} estimates the depth of the \hh{} emission from the compact flare ribbon based on geometric arguments due to the fact that \hh{} emission is more extended than its source, and place the source flare ribbon 1.9 and 1.2 Mm above for two different regions.
Illumination of \hh{} can be complicated, coming from distant regions and illuminating from the side, as shown by the example in Section \ref{sec:nonlocal}.
One location in the very same IRIS observation used in this study shows \hh{} in absorption on top of the \ion{Si}{4} transition region lines \citep{schmit14}, indicating an apparent reversal in the atmospheric temperature gradient.

Despite these complications, 1D modeling still remains an attractive option because the lower solar atmosphere is stratified on average.
Improvements to 1D modeling can be made with more accurate treatment of the exciting radiation and extending the range atmospheric models with simulations of flares.
Grids of flare simulations \citep[e.g.][]{fchroma23} can provide a broader range of physical situations than what has been done with the current set of atmospheres and provide insight into how the signature of \hh{} might change in time.
Rather than injecting ad hoc radiation through the top of the atmosphere, the radiation should be generated self-consistently by the model atmosphere.
CHIANTI might be used to calculate emission from truly optically thin lines that excite \hh{}, while RH can be used to properly account for the optical density of lines that become optically thick, such as \ion{Si}{4} and \ion{C}{2}.

\section{Conclusions}\label{sec:conclusions}
The bipolar active region NOAA 11861 produced two C-class flares early on October 12, 2013.
IRIS observed the flares using a full spectrum raster with deep, 30 sec exposures while the automatic exposure control was off.
The resulting spectra from the FUV channels of IRIS contain a rich forest of lines produced by molecular hydrogen and other cool species due excitation of the lower solar atmosphere by the flare activity.

Working from the raw IRIS data, we reduced the FUV spectra to mitigate noise and avoid loss of information due to multiple resamplings of the data.
The spectra were corrected for the instrumental throughput and converted into physical units for comparison with spectral synthesis results produced using the modeling framework developed in \citet{jaeggli18}.
Selected spectra from the flare were averaged together to produce a high signal-to-noise spectrum, and lists of previous identifications of molecular and atomic lines were compiled for carrying out identification of new \hh{} lines.
Extensive line fitting was performed on the average flare spectrum to derive the wavelength, peak intensity, and line width for as many lines as possible.
Selected line fitting was performed for the strongest \hh{} lines, and lines representative of the photosphere and chromosphere, over the individual spectra from the IRIS raster of the active region.

We have produced what is probably the most complete list of atomic and molecular transitions for the IRIS FUV bandpasses, covering the wavelengths 1330 - 1358 and 1389 - 1407 \AA.
We have confirmed many \hh{} line identifications made in previous work.
We propose 37 new line identifications of \hh{} with a high level of certainty based on wavelength conincidences, comparison with synthetic \hh{} spectra, and the presence of other \hh{} lines with the same upper level.

Looking at the strongest \hh{} lines in context with other diagnostics, the Doppler velocity of \hh{} has some features in common with photospheric \ion{Fe}{1} and with the chromospheric \ion{O}{1} line at various locations during the flare, indicating that both photospheric and chromospheric activity can span heights where \hh{} forms.
The question of whether \hh{} line formation occurs in an optically thin or thick regime still appears to be an open one because we see preference for either based on different line pairs.
Spectral synthesis results reveal that the relationship between \hh{} emission and the intensity of its excitation source can be used to differentiate between atmospheric models with different temperatures, but the 3-dimensional nature of illumination coming down from the transition region into the lower solar atmosphere, and incomplete coverage of the source and excitation region with a slit spectrograph, remain a challenge.

We have demonstrated the diagnostic potential of \hh{} fluorescent emission in the UV.
During flares, \hh{} outshines other weak neutral species and shows the response of the lower solar atmosphere to flare radiation.
While there are significant challenges, \hh{} emission provides information about a critical height in the Sun's low chromosphere where other molecular diagnostics show the presence of extremely cool gas.
Further work modeling \hh{} emission using state-of-the-art simulations and more accurate treatment of the radiation for the excitation sources will provide more insight for the interpretation of \hh{} observations.
There exists a wealth of IRIS observations to investigate for molecular studies, and although the IRIS sensitivity in the FUV channels has degraded, making it more difficult to study these weak lines, future prospects for spectral observations of the \hh{} Lyman band seem bright with the anticipated launch of EUVST.

\acknowledgements
The work of S. Jaeggli on this project from 2015-2016 was supported by the NASA postdoctoral program and from 2011-2015 under the NASA/IRIS subcontract from Lockheed Martin to Montana State University.
IRIS is a NASA small explorer mission developed and operated by LMSAL with mission operations executed at NASA Ames Research center and major contributions to downlink communications funded by ESA and the Norwegian Space Center.
The authors would like to thank Graham Kerr, Sargram Mulay, and Daniel N\'obrega Siverio for encouragement on completing this work.

\appendix

\section{Notes on Line Identifications}\label{sec:lineids_notes}
This section is intended to accompany Figure \ref{fig:FUV_spec} and Table \ref{tbl:lines} to provide some justification for the line identification decisions that were made and provide an evaluation of blended lines.  Each item below provides notes for a particular blend or issue.

\begin{description}
    \item[\ion{S}{1} 1333.792 and \hh{} 1333.797 \AA]
    These lines are close enough in wavelength that the blend would not be resolved.  As in \citet{mulay21}, we have determined that in these observations, the \hh{} line is the dominant one in this blend.  The measured line intensity is well correlated with the other line of the (2,0) upper level.  Additional emission from \ion{S}{1} would tend to decrease the slope of the relation shown in the right panel of Figure \ref{fig:line_ratios}.

    \item[\hh{} 1338.569 and \ion{O}{4} 1338.704 \AA]
    The \ion{O}{4} line may be present at a low level, but it should match the broad appearance of other \ion{O}{4} lines and cannot be easily confused with the much more narrow \hh{} lines in this region which show intensities in good agreement with the synthetic spectrum.  The broad \ion{O}{4} line may contaminate fits of the 1338.569 \AA\ \hh{} line, leading to excess intensity over the other line in the (1,0) upper level and causing an increase in the slope of the relation in the left panel of Figure \ref{fig:line_ratios}, but more likely it boosts the background level under the line, which is easily accounted for in the fit.

    \item[\ion{P}{3} 1344.845 and \hh{} 1344.846 \AA]
    The line fitted at this wavelength more closely resembles other \hh{} lines.  The nearby \ion{P}{3} line at 1344.327 \AA\ provides a template for what this \ion{P}{3} might look like, and if this line is present, it is weaker and may contribute to background emission which is difficult to see under the tight grouping of \hh{} lines.

    \item[\ion{Si}{2} 1346.873 and \hh{} 1346.909 \AA]
    The center wavelength at the peak of this line is a closer match to \hh{}, and the line width and appearance are similar to other \hh{} lines.  The \ion{Si}{2} line at 1350.057 \AA\ seems to show a narrow peak with a broad wings.  That could be consistent with the appearance of the spectrum at 1346.9 \AA\ if the \ion{Si}{2} line here is weaker than the one at 1350.1 \AA.  This line may be a blend.

    \item[\ion{Fe}{12} 1349.400 and Unknown Line at 1349.411 \AA]
    Although the coronal \ion{Fe}{12} line may get bright enough during major flares to be seen on the disk, the line fitted at this wavelength is much too narrow to be a coronal line, and we attribute it to another unknown species.  Several of the other unknown lines share similar appearance, e.g. the line at 1341.078 \AA.
    There may be a diffuse contribution from \ion{Fe}{12} in the background.

    \item[\ion{Fe}{2} 1354.742 and \hh{} 1354.753 \AA]
    The line fitted here has a narrow width more consistent with other \ion{Fe}{2} lines that can be seen in the spectrum (e.g. nearby at 1353.022 and 1354.012 \AA).  In addition, the spectral synthesis calculation predicts a relatively small intensity for the \hh{} line, so at best this is a blend where the contribution from \ion{Fe}{2} is dominant. 
    
    \item[\hh{} 1389.678 \AA\ and \ion{Cl}{1} 1389.693 \AA]
    The measured line wavelength at 1389.668 \AA\ is a better match to \hh{} at 1389.678 \AA. The line at this wavelength is broader than the nearby \ion{Cl}{1} line at 1389.957 \AA\ and is a good match in appearance to other \hh{} lines.  The upper level for this line is probably excited by the line at 1336.123 \AA, in the wing of \ion{C}{2}, which cannot be seen due to the high intensity there.  The \hh{} line amplitude is somewhat under predicted by spectral synthesis, but the observations have much broader line widths than used in the \hh\ radiation model.
    
    \item[\hh{} 1390.253 and \ion{Fe}{2} 1390.315 \AA]
    The major component to this line appears to be \hh{} due to the center wavelength of the peak and the line width, although it does appear that the \ion{Fe}{2} line pair contributes a weaker component, skewing the red wing of the \hh{} line.

    \item[\ion{O}{4} 1397.231 \AA\ and Unknown Lines]
    Although the diffuse \ion{O}{4} line appears to be present at this wavelength, it seems to be blended with unidentified narrow lines with similar intensity, making it difficult to fit any of them.
    
    \item[\ion{Ni}{2} 1399.026 and \ion{Fe}{2} 1399.055 \AA]
    There appear to be two lines at this wavelength.  A broader line close to the \ion{Ni}{2} wavelength appears in the regions outside of the flare.  In the flare spectrum, there is a narrower line slightly shifted to the red with respect to the \ion{Ni}{2} rest wavelength.  This line appears to be a previously missed \ion{Fe}{2} line.
    
    \item[\hh{} 1399.949 \AA\ and \ion{Fe}{2} 1399.957 \AA]
    This might be an unresolved blend between \hh{} and \ion{Fe}{2}, but if so, the \ion{Fe}{2} component is minor.  The fitted line width is consistent with other \hh{} lines and broader than other \ion{Fe}{2} lines seen in the spectrum.  Although the broad \ion{O}{4} line may be interfering with the fitting, two other lines from the (16,1) upper level are visible in the spectrum and have the correct approximate intensities compared to this line, and all are in good agreement with the predicted intensities from spectral synthesis.
    
\end{description}

\section{Identification of \hh{} Lines and Their Shared Upper Levels}\label{sec:ush}
\hh{} lines should appear alongside the other lines in their common upper level and should have intensity ratios determined approximately by the branching ratio.
Non-detection of other lines from the same upper level that are predicted to be weak, or masked by blends with brighter lines also strengthens our case for identification.
In this section we present a list of all \hh{} lines identified in this work grouped by upper level, listing any other lines from the upper level that might be seen.
The table provides the ratio of line intensities based on the spectral synthesis results from the FALC$\times100$ atmosphere using the optically thick calculation.
Upper levels that only have one line that occurs in the IRIS FUV bandpasses are also included for completeness, but we do not list any line ratio data for them.
For each line that is identified, we provide the criteria that are fulfilled as defined in Section \ref{sec:idcrit}.

\clearpage
\startlongtable
\input{H2_ratio_table}
\clearpage

\bibliography{references}{}
\bibliographystyle{aasjournal}

\end{document}

%% file: linetable.tex
\begin{deluxetable*}{llcllccl}
\tablecaption{List of Atomic and Molecular Lines}\label{tbl:lines}
\tablewidth{0 pt}
\tablehead{\colhead{$\lambda_{obs}$} & \colhead{$\lambda_{ref}$} & \colhead{Species} & \multicolumn{2}{c}{Transition} & \colhead{R$_{peak}$} & \colhead{FWHM} & \colhead{Reference} \\
\cline{4-5}
\colhead{[\AA]} & \colhead{[\AA]} & & \colhead{Lower} & \colhead{Upper}  & \colhead{[erg s$^{-1}$ cm$^{-2}$ str$^{-1}$ \AA$^{-1}$]} & \colhead{[m\AA]} & }
\startdata
1331.962 & 1331.9540 & \hh\ & 11,  3  X & 12,  0  B & 7.3E+01 &  48.4 & new \\
1332.340 & 1332.3410 & \hh\ & 22,  1  X & 21,  3  B & 2.3E+02 &  80.3 & \citet{jaeggli18} \\
1332.560 & 1332.5510 & \hh\ & 23,  1  X & 24,  3  B & 1.1E+02 &  52.6 & new \\
1333.339 &  & UN &  &  & 6.0E+01 &  35.7 &  \\
1333.472 & 1333.4750 & \hh\ &  0,  4  X &  1,  0  B & 3.6E+02 &  49.5 & \citet{jordan77} \\
 & 1333.7920 & S I & 3s$^{2}$ 3p$^{4}$ \ $^{3}$P$_{2}$ & 3s$^{2}$ 3p$^{3}$ 4d \ $^{5}$D$^\circ_{3}$ &  &  & NIST \\
1333.796 & 1333.7970 & \hh\ &  1,  4  X &  2,  0  B & 1.6E+03 &  56.3 & \citet{jordan77} \\
1334.541 & 1334.5323 & C II & 2s$^{2}$ 2p \ $^{2}$P$^\circ_{1/2}$ & 2s 2p$^{2}$ \ $^{2}$D$_{3/2}$ & 2.6E+05 & 337.5 & NIST \\
 & 1335.2030 & Ni II & 3d$^{9}$ \ $^{2}$D$_{3/2}$ & 3d$^{8}$ 4p \ $^{2}$F$^\circ_{5/2}$ &  &  & \citet{shenstone70} \\
1335.710 & 1335.7080 & C II & 2s$^{2}$ 2p \ $^{2}$P$^\circ_{3/2}$ & 2s 2p$^{2}$ \ $^{2}$D$_{5/2}$ & 2.8E+05 & 415.4 & NIST \\
1336.661 &  & UN &  &  & 1.4E+02 &  33.0 &  \\
1337.355 & 1337.3570 & \hh\ & 15,  2  X & 14,  0  B & 1.2E+02 &  66.1 & new \\
1337.427 &  & UN &  &  & 1.4E+02 &  36.5 &  \\
1337.472 & 1337.4670 & \hh\ &  3,  4  X &  4,  0  B & 1.6E+02 &  52.9 & \citet{bartoe79} \\
1337.628 & 1337.6220 & \hh\ &  9,  4  X &  8,  2  B & 2.1E+02 &  54.8 & new \\
1337.728 &  & UN &  &  & 1.6E+02 &  52.6 &  \\
1337.825 & 1337.8310 & \hh\ & 13,  4  X & 12,  4  B & 1.1E+02 &  57.5 & new \\
1338.061 &  & UN &  &  & 9.6E+01 &  39.6 &  \\
1338.121 &  & UN &  &  & 6.9E+01 &  29.0 &  \\
1338.241 &  & UN &  &  & 2.4E+02 &  36.9 &  \\
1338.350 & 1338.3410 & \hh\ & 19,  2  X & 20,  2  B & 7.6E+02 &  56.5 & \citet{jaeggli18} \\
1338.565 & 1338.5690 & \hh\ &  2,  4  X &  1,  0  B & 6.4E+02 &  57.1 & \citet{jordan77} \\
 & 1338.6140 & O IV & 2s 2p$^{2}$ \ $^{2}$P$_{1/2}$ & 2p$^{3}$ \ $^{2}$D$^\circ_{3/2}$ &  &  & CHIANTI \\
1338.685 & 1338.7040 & \hh\ &  9,  6  X &  8,  7  B & 8.2E+01 &  78.8 & new \\
1339.905 &  & UN &  &  & 1.0E+02 &  63.8 &  \\
1340.030 &  & UN &  &  & 7.9E+01 &  27.5 &  \\
1340.117 &  & UN &  &  & 5.1E+01 &  34.1 &  \\
 & 1340.3740 & Ni II & 3d$^{8}$ 4p \ $^{2}$F$^\circ_{7/2}$ & 3d$^{8}$ 6d \ $^{2}$H$_{9/2}$ &  &  & \citet{shenstone70} \\
1340.626 &  & UN &  &  & 4.0E+01 &  35.2 &  \\
1340.773 & 1340.7910 & \hh\ &  4,  4  X &  5,  0  B & 1.7E+02 &  66.8 & \citet{bartoe79} \\
 & 1340.8520 & S I & 3s$^{2}$ 3p$^{4}$ \ $^{3}$P$_{1}$ & 3s$^{2}$ 3p$^{3}$ 4d \ $^{5}$D$^\circ_{2}$ &  &  & NIST \\
1341.078 &  & UN &  &  & 1.4E+02 &  46.3 &  \\
1341.170 &  & UN &  &  & 1.0E+02 &  27.9 &  \\
 & 1341.4580 & Si III & 3s 3d \ $^{3}$D$_{3}$ & 3p 3d \ $^{3}$D$^\circ_{3}$ &  &  & NIST \\
1341.708 & 1341.7180 & \hh\ &  3,  6  X &  2,  5  B & 6.3E+01 &  43.8 & new \\
1341.890 & 1341.8890 & Ca II & 4s \ $^{2}$S$_{1/2}$ & 6p \ $^{2}$P$^\circ_{3/2}$ & 1.3E+02 &  70.4 & \citet{edlen56} \\
1342.042 &  & UN &  &  & 9.2E+01 &  28.3 &  \\
1342.123 &  & UN &  &  & 1.4E+02 &  27.3 &  \\
1342.253 & 1342.2560 & \hh\ &  3,  4  X &  2,  0  B & 2.2E+03 &  63.5 & \citet{jordan77} \\
1342.409 & 1342.4110 & \hh\ & 12,  3  X & 13,  0  B & 7.4E+01 & 113.9 & new \\
1342.552 & 1342.5350 & Ca II & 4s \ $^{2}$S$_{1/2}$ & 6p \ $^{2}$P$^\circ_{1/2}$ & 2.0E+02 &  52.4 & \citet{edlen56} \\
1342.777 & 1342.7740 & \hh\ & 21,  3  X & 22,  6  B & 3.5E+02 &  56.6 & new \\
1342.883 & 1342.8860 & \hh\ &  7,  4  X &  6,  1  B & 9.0E+01 &  67.8 & \citet{jordan78} \\
 & 1343.2260 & S III & 3s 3p$^{3}$ \ $^{3}$P$^\circ_{2}$ & 3s$^{2}$ 3p 4p \ $^{3}$P$_{2}$ &  &  & CHIANTI \\
1343.531 & 1343.5140 & O IV & 2s 2p$^{2}$ \ $^{2}$P$_{3/2}$ & 2p$^{3}$ \ $^{2}$D$^\circ_{5/2}$ & 1.1E+02 & 194.7 & CHIANTI \\
 & 1343.5920 & S III & 3s 3p$^{3}$ \ $^{3}$P$^\circ_{1}$ & 3s$^{2}$ 3p 4p \ $^{3}$P$_{2}$ &  &  & NIST \\
1344.036 & 1344.0220 & \hh\ & 14,  3  X & 15,  1  B & 2.9E+02 &  73.7 & new \\
1344.329 & 1344.3270 & P III & 3s$^{2}$ 3p \ $^{2}$P$^\circ_{3/2}$ & 3s 3p$^{2}$ \ $^{2}$D$_{5/2}$ & 4.2E+02 & 191.6 & NIST \\
 & 1344.8450 & P III & 3s$^{2}$ 3p \ $^{2}$P$^\circ_{3/2}$ & 3s 3p$^{2}$ \ $^{2}$D$_{3/2}$ &  &  & NIST \\
1344.847 & 1344.8460 & \hh\ & 13,  4  X & 14,  3  B & 1.4E+02 &  58.2 & new \\
1344.969 & 1344.9650 & \hh\ & 17,  2  X & 16,  1  B & 3.6E+02 &  59.7 & new \\
1345.083 & 1345.0850 & \hh\ &  5,  4  X &  6,  0  B & 2.4E+02 &  68.5 & \citet{bartoe79} \\
1345.174 & 1345.1700 & \hh\ & 11,  3  X & 10,  0  B & 2.5E+02 &  59.2 & \citet{jaeggli18} \\
 & 1345.3800 & Fe II & 3d$^{6}$ 4s \ $^{4}$G$_{11/2}$ & 3d$^{5}$ 4s 4p \ $^{2}$G$^\circ_{9/2}$ &  &  & NIST \\
1345.395 & 1345.4010 & \hh\ &  1,  5  X &  2,  2  B & 5.6E+01 &  60.4 & new \\
 & 1345.8820 & Ni II & 3d$^{9}$ \ $^{2}$D$_{5/2}$ & 3d$^{8}$ 4p \ $^{4}$S$^\circ_{3/2}$ &  &  & \citet{shenstone70} \\
1346.428 & 1346.4240 & \hh\ & 13,  3  X & 12,  1  B & 1.6E+02 &  68.9 & new \\
1346.682 &  & UN &  &  & 9.4E+01 &  26.1 &  \\
 & 1346.8730 & Si II & 3s 3p$^{2}$ \ $^{4}$P$_{3/2}$ & 3s 3p 4s \ $^{4}$P$^\circ_{5/2}$ &  &  & \citet{moore65} \\
1346.905 & 1346.9090 & \hh\ &  4,  4  X &  3,  0  B & 4.3E+02 &  72.0 & \citet{bartoe79} \\
1347.034 &  & UN &  &  & 2.3E+02 &  55.2 &  \\
1347.236 & 1347.2400 & Cl I & 3s$^{2}$ 3p$^{5}$ \ $^{2}$P$^\circ_{3/2}$ & 3s$^{2}$ 3p$^{4}$ 4s \ $^{2}$P$_{3/2}$ & 4.8E+02 &  72.8 & NIST \\
1347.439 & 1347.4460 & \hh\ & 25,  1  X & 24,  5  B & 9.2E+01 &  53.8 & new \\
1348.003 & 1348.0037 & Fe II & 3d$^{6}$ 4s \ $^{4}$P$_{3/2}$ & 3d$^{5}$ 4s 4p \ $^{4}$D$^\circ_{5/2}$ & 1.8E+02 &  43.6 & NIST \\
 & 1348.0094 & Fe II & 3d$^{7}$ \ $^{2}$D2$_{5/2}$ & 3d$^{6}$ 4p \ $^{2}$D$^\circ_{5/2}$ &  &  & NIST \\
1348.284 &  & UN &  &  & 8.0E+01 &  95.6 &  \\
 & 1348.3330 & Ni II & 3d$^{8}$ 4s $^2$D$_{5/2}$ & 3d$^{7}$ 4s 4p $^{4}$D$^\circ_{5/2}$ &  &  & \citet{shenstone70} \\
1348.544 & 1348.5430 & Si II & 3s 3p$^{2}$ \ $^{4}$P$_{1/2}$ & 3s 3p 4s \ $^{4}$P$^\circ_{3/2}$ & 9.9E+01 &  58.1 & \citet{moore65} \\
1348.618 & 1348.6180 & \hh\ &  9,  4  X & 10,  1  B & 8.7E+01 &  56.7 & new \\
1348.723 &  & UN &  &  & 2.3E+02 &  45.9 &  \\
1349.057 & 1349.0560 & \hh\ &  3,  5  X &  4,  2  B & 3.4E+01 &  57.8 & new \\
1349.264 & 1349.2570 & \hh\ & 17,  2  X & 18,  0  B & 3.4E+01 &  48.0 & new \\
 & 1349.4000 & Fe XII & 3s$^{2}$ 3p$^{3}$ \ $^{4}$S$^\circ_{3/2}$ & 3s$^{2}$ 3p$^{3}$ \ $^{2}$P$^\circ_{1/2}$ &  &  & CHIANTI \\
1349.411 &  & UN &  &  & 2.1E+02 &  70.8 &  \\
1349.588 &  & UN &  &  & 4.7E+01 &  45.4 &  \\
1349.684 &  & UN &  &  & 6.7E+01 &  56.8 &  \\
1349.869 &  & UN &  &  & 5.0E+01 &  25.4 &  \\
1350.071 & 1350.0570 & Si II & 3s 3p$^{2}$ \ $^{4}$P$_{5/2}$ & 3s 3p 4s \ $^{4}$P$^\circ_{5/2}$ & 2.8E+02 &  68.5 & \citet{moore65} \\
1350.325 & 1350.3250 & \hh\ &  6,  4  X &  7,  0  B & 4.7E+01 &  48.0 & \citet{jaeggli18} \\
 & 1350.5200 & Si II & 3s 3p$^{2}$ \ $^{4}$P$_{3/2}$ & 3s 3p 4s \ $^{4}$P$^\circ_{3/2}$ &  &  & NIST \\
1350.688 & 1350.6850 & \hh\ & 15,  3  X & 14,  2  B & 2.3E+02 &  68.5 & new \\
1351.184 & 1351.1670 & \hh\ & 22,  1  X & 21,  2  B & 4.9E+01 &  78.2 & new \\
1351.287 &  & UN &  &  & 5.3E+01 &  39.2 &  \\
1351.653 & 1351.6568 & Cl I & 3s$^{2}$ 3p$^{5}$ \ $^{2}$P$^\circ_{1/2}$ & 3s$^{2}$ 3p$^{4}$ 4s \ $^{2}$P$_{1/2}$ & 3.9E+04 &  49.3 & NIST \\
1352.290 & 1352.2820 & \hh\ &  8,  5  X &  9,  3  B & 4.4E+01 &  70.4 & new \\
1352.501 & 1352.5020 & \hh\ &  5,  4  X &  4,  0  B & 1.8E+02 &  66.2 & \citet{bartoe79} \\
1352.641 & 1352.6350 & Si II & 3s 3p$^{2}$ \ $^{4}$P$_{3/2}$ & 3s 3p 4s \ $^{4}$P$^\circ_{1/2}$ & 8.4E+01 &  90.5 & \citet{moore65} \\
1352.735 & 1352.7510 & C I & 2s$^{2}$ 2p$^{2}$ \ $^{1}$D$_{2}$ & 2s$^{2}$ 2p 4d \ $^{3}$P$^\circ_{1}$ & 8.8E+01 &  59.1 & \citet{johansson66} \\
 & 1352.9880 & C I & 2s$^{2}$ 2p$^{2}$ \ $^{1}$D$_{2}$ & 2s$^{2}$ 2p 4d \ $^{3}$P$^\circ_{2}$ &  &  & \citet{johansson66} \\
1353.017 & 1353.0218 & Fe II & 3d$^{6}$ 4s \ $^{2}$D$_{5/2}$ & 3d$^{6}$ 5p \ $^{4}$G$^\circ_{7/2}$ & 4.7E+02 &  52.9 & NIST \\
 & 1353.0228 & Fe II & 3d$^{6}$ 4s \ $^{4}$P$_{5/2}$ & 3d$^{5}$ 4s 4p \ $^{4}$P$^\circ_{3/2}$ &  &  & NIST \\
1353.323 & 1353.3110 & \hh\ & 20,  2  X & 21,  2  B & 8.2E+01 &  70.7 & new \\
1353.407 &  & UN &  &  & 1.4E+02 &  38.2 &  \\
1353.483 &  & UN &  &  & 9.6E+01 &  33.8 &  \\
1353.553 &  & UN &  &  & 4.6E+02 &  35.1 &  \\
1353.735 & 1353.7180 & Si II & 3s 3p$^{2}$ \ $^{4}$P$_{5/2}$ & 3s 3p 4s \ $^{4}$P$^\circ_{3/2}$ & 9.2E+01 &  81.0 & \citet{moore65} \\
1354.006 & 1354.0120 & Fe II & 3d$^{6}$ 4s \ $^{4}$P$_{5/2}$ & 3d$^{5}$ 4s 4p \ $^{6}$P$^\circ_{5/2}$ & 3.8E+02 &  39.0 & NIST \\
 & 1354.0665 & Fe XXI & 2s$^{2}$ 2p$^{2}$ \ $^{3}$P$_{0}$ & 2s$^{2}$ 2p$^{2}$ \ $^{3}$P$_{1}$ &  &  & CHIANTI \\
1354.178 &  & UN &  &  & 4.9E+02 &  30.8 &  \\
1354.287 & 1354.2880 & C I & 2s$^{2}$ 2p$^{2}$ \ $^{1}$D$_{2}$ & 2s$^{2}$ 2p 4d \ $^{1}$P$^\circ_{1}$ & 5.3E+03 &  67.0 & \citet{johansson66} \\
1354.576 &  & UN &  &  & 1.2E+02 &  34.5 &  \\
 & 1354.7420 & Fe II & 3d$^{6}$ 4s \ $^{2}$I$_{13/2}$ & 3d$^{5}$ 4s 4p \ $^{2}$H$^\circ_{11/2}$ &  &  & NIST \\
1354.759 & 1354.7530 & \hh\ & 19,  2  X & 18,  2  B & 4.0E+02 &  58.7 & new \\
1354.846 &  & UN &  &  & 2.0E+02 &  38.2 &  \\
1354.959 &  & UN &  &  & 5.7E+01 &  24.8 &  \\
1355.032 &  & UN &  &  & 1.3E+02 &  36.9 &  \\
1355.099 &  & UN &  &  & 3.3E+02 &  32.7 &  \\
1355.143 &  & UN &  &  & 2.2E+02 &  28.7 &  \\
1355.385 &  & UN &  &  & 1.2E+02 &  34.0 &  \\
1355.597 & 1355.5977 & O I & 2s$^{2}$ 2p$^{4}$ \ $^{3}$P$_{2}$ & 2s$^{2}$ 2p$^{3}$ 3s \ $^{5}$S$^\circ_{2}$ & 1.5E+04 &  48.2 & \citet{eriksson68} \\
1355.848 & 1355.8440 & C I & 2s$^{2}$ 2p$^{2}$ \ $^{1}$D$_{2}$ & 2s$^{2}$ 2p 4d \ $^{1}$F$^\circ_{3}$ & 9.2E+03 &  99.2 & \citet{johansson66} \\
1356.357 & 1356.3500 & \hh\ & 15,  3  X & 16,  1  B & 3.6E+02 &  56.8 & new \\
1356.482 & 1356.4820 & \hh\ &  7,  4  X &  8,  0  B & 1.0E+02 &  58.3 & \citet{bartoe79} \\
1356.566 & 1356.5690 & \hh\ & 12,  3  X & 11,  0  B & 1.4E+02 &  70.4 & new \\
1356.657 &  & UN &  &  & 9.6E+01 &  36.3 &  \\
1356.834 &  & UN &  &  & 1.3E+02 &  64.8 &  \\
1356.970 &  & UN &  &  & 3.1E+02 & 114.8 &  \\
1357.133 & 1357.1340 & C I & 2s$^{2}$ 2p$^{2}$ \ $^{1}$D$_{2}$ & 2s$^{2}$ 2p 5s \ $^{1}$P$^\circ_{1}$ & 5.0E+03 &  62.0 & \citet{johansson66} \\
1357.376 & 1357.3730 & \hh\ & 15,  5  X & 16,  6  B & 1.0E+02 &  48.5 & new \\
1357.526 &  & UN &  &  & 6.7E+01 &  36.3 &  \\
1357.658 & 1357.6590 & C I & 2s$^{2}$ 2p$^{2}$ \ $^{1}$D$_{2}$ & 2s$^{2}$ 2p 4d \ $^{3}$D$^\circ_{3}$ & 3.0E+03 &  49.9 & \citet{johansson66} \\
 & 1357.8570 & C I & 2s$^{2}$ 2p$^{2}$ \ $^{1}$D$_{2}$ & 2s$^{2}$ 2p 4d \ $^{3}$D$^\circ_{2}$ &  &  & \citet{johansson66} \\
1358.188 & 1358.1880 & C I & 2s$^{2}$ 2p$^{2}$ \ $^{1}$D$_{2}$ & 2s$^{2}$ 2p 4d \ $^{3}$D$^\circ_{1}$ & 8.7E+02 &  49.6 & \citet{johansson66} \\
 & 1388.4347 & S I & 3s$^{2}$ 3p$^{4}$ \ $^{3}$P$_{2}$ & 3s 3p$^{5}$ \ $^{3}$P$^\circ_{2}$ &  &  & NIST \\
1388.524 &  & UN &  &  & 2.2E+02 &  36.5 &  \\
1388.623 &  & UN &  &  & 1.6E+02 &  42.7 &  \\
1388.783 & 1388.7830 & \hh\ & 21,  2  X & 22,  1  B & 4.9E+01 &  57.6 & new \\
1388.929 &  & UN &  &  & 4.3E+01 &  47.9 &  \\
1389.149 & 1389.1538 & S I & 3s$^{2}$ 3p$^{4}$ \ $^{3}$P$_{1}$ & 3s 3p$^{5}$ \ $^{3}$P$^\circ_{1}$ & 1.8E+02 &  43.4 & NIST \\
1389.358 &  & CO? & 27, 7 X & 28, 1 A & 4.1E+02 &  34.7 & \citet{jordan79co} \\
1389.529 & 1389.5210 & \hh\ & 19,  3  X & 20,  2  B & 4.2E+02 &  65.2 & \citet{jaeggli18} \\
1389.668 & 1389.6780 & \hh\ & 11,  5  X & 10,  3  B & 4.1E+02 &  73.0 & new \\
 & 1389.6930 & Cl I & 3s$^{2}$ 3p$^{5}$ \ $^{2}$P$^\circ_{3/2}$ & 3s$^{2}$ 3p$^{4}$ 4s \ $^{4}$P$_{5/2}$ &  &  & NIST \\
1389.809 &  & UN &  &  & 2.0E+02 &  39.3 &  \\
1389.950 & 1389.9570 & Cl I & 3s$^{2}$ 3p$^{5}$ \ $^{2}$P$^\circ_{1/2}$ & 3s$^{2}$ 3p$^{4}$ 4s \ $^{4}$P$_{1/2}$ & 3.2E+02 &  43.9 & NIST \\
1390.262 & 1390.2530 & \hh\ & 10,  5  X & 11,  2  B & 3.8E+02 &  73.2 & new \\
 & 1390.3146 & Fe II & 3d$^{6}$ 4p \ $^{6}$P$^\circ_{7/2}$ & 3d$^{6}$ 6d \ 114584$_{5/2}$ &  &  & NIST \\
 & 1390.3162 & Fe II & 3d$^{6}$ 4s \ $^{2}$H$_{9/2}$ & 3d$^{6}$ 4p \ $^{2}$H$^\circ_{11/2}$ &  &  & NIST \\
1390.423 & 1390.4540 & \hh\ &  4,  6  X &  3,  3  B & 5.4E+01 &  64.8 & new \\
1391.012 & 1391.0050 & \hh\ & 16,  3  X & 17,  0  B & 4.9E+02 &  58.9 & new \\
1391.302 & 1391.3070 & Fe II & 3d$^{6}$ 4s \ $^{4}$F$_{7/2}$ & 3d$^{5}$ 4s 4p \ $^{6}$P$^\circ_{5/2}$ & 2.0E+02 &  42.5 & NIST \\
1391.809 &  & UN &  &  & 4.7E+01 &  42.7 &  \\
1392.022 & 1392.0030 & \hh\ &  7,  7  X &  6,  6  B & 6.9E+01 &  59.0 & new \\
 & 1392.0980 & Ar XI & 2s$^{2}$ 2p$^{4}$ \ $^{3}$P$_{2}$ & 2s$^{2}$ 2p$^{4}$ \ $^{1}$D$_{2}$ &  &  & CHIANTI \\
1392.142 & 1392.1380 & Fe II & 3d$^{7}$ \ $^{2}$H$_{11/2}$ & 3d$^{6}$ 4p \ $^{2}$G$^\circ_{9/2}$ & 5.3E+02 &  57.4 & NIST \\
1392.344 &  & UN &  &  & 1.0E+02 &  42.8 &  \\
1392.524 &  & UN &  &  & 2.9E+02 &  32.7 &  \\
1392.583 & 1392.5878 & S I & 3s$^{2}$ 3p$^{4}$ \ $^{3}$P$_{0}$ & 3s 3p$^{5}$ \ $^{3}$P$^\circ_{1}$ & 3.6E+02 &  23.7 & NIST \\
1392.755 &  & UN &  &  & 4.2E+02 &  32.1 &  \\
1392.811 & 1392.8170 & Fe II & 3d$^{7}$ \ $^{2}$H$_{9/2}$ & 3d$^{6}$ 4p \ $^{2}$G$^\circ_{7/2}$ & 6.0E+02 &  62.9 & \citet{johansson78} \\
 & 1393.3300 & Ni II & 3d$^{9}$ \ $^{2}$D$_{5/2}$ & 3d$^{8}$ 4p \ $^{2}$D$^\circ_{5/2}$ &  &  & \citet{shenstone70} \\
1393.769 & 1393.7570 & Si IV & 3s \ $^{2}$S$_{1/2}$ & 3p \ $^{2}$P$^\circ_{3/2}$ & 1.6E+05 & 279.0 & CHIANTI \\
 & 1394.7105 & Fe II & 3d$^{7}$ \ $^{2}$F$_{7/2}$ & 3d$^{6}$ 4f \ $^{2}$[3]$^\circ_{5/2}$ &  &  & NIST \\
 & 1394.7111 & Fe II & 3d$^{7}$ \ $^{2}$D2$_{5/2}$ & 3d$^{6}$ 4p \ $^{2}$D$^\circ_{3/2}$ &  &  & NIST \\
1395.190 & 1395.1980 & \hh\ &  2,  5  X &  3,  0  B & 3.3E+02 &  59.8 & \citet{bartoe79} \\
1395.799 &  & UN &  &  & 1.4E+02 &  40.2 &  \\
1396.118 & 1396.1122 & S I & 3s$^{2}$ 3p$^{4}$ \ $^{3}$P$_{1}$ & 3s 3p$^{5}$ \ $^{3}$P$^\circ_{2}$ & 7.4E+02 &  67.0 & NIST \\
1396.215 & 1396.2230 & \hh\ &  1,  5  X &  0,  0  B & 2.8E+03 &  64.8 & \citet{jordan77} \\
1396.513 & 1396.5270 & Cl I & 3s$^{2}$ 3p$^{5}$ \ $^{2}$P$^\circ_{1/2}$ & 3s$^{2}$ 3p$^{4}$ 4s \ $^{4}$P$_{3/2}$ & 4.3E+02 &  39.4 & NIST \\
1396.596 & 1396.6030 & \hh\ & 13,  5  X & 14,  3  B & 1.5E+02 &  72.3 & new \\
1396.725 &  & CO? & 28, 5 X & 27, 0 A & 2.7E+02 &  31.8 & \citet{jordan79co} \\
1396.889 &  & UN &  &  & 1.3E+02 &  35.1 &  \\
1396.996 &  & CO? & 40, 7 X & 41, 1 A & 7.1E+01 &  73.9 & \citet{jordan79co} \\
 & 1397.2310 & O IV & 2s$^{2}$ 2p \ $^{2}$P$^\circ_{1/2}$ & 2s 2p$^{2}$ \ $^{4}$P$_{3/2}$ &  &  & CHIANTI \\
1397.411 & 1397.4190 & \hh\ &  3,  5  X &  4,  0  B & 1.8E+02 &  76.4 & \citet{bartoe79} \\
1397.569 & 1397.5660 & Fe II & 3d$^{7}$ \ $^{2}$F$_{5/2}$ & 3d$^{6}$ 4f \ $^{2}$[3]$^\circ_{5/2}$ & 2.6E+02 &  44.2 & NIST \\
1397.705 &  & UN &  &  & 3.3E+01 &  25.2 &  \\
1397.834 & 1397.8360 & Fe II & 3d$^{6}$ 4s \ $^{2}$P$_{3/2}$ & 3d$^{5}$ 4s 4p \ $^{2}$P$^\circ_{3/2}$ & 6.0E+02 &  50.7 & NIST \\
 & 1398.0400 & S IV & 3s$^{2}$ 3p \ $^{2}$P$^\circ_{1/2}$ & 3s 3p$^{2}$ \ $^{4}$P$_{3/2}$ &  &  & CHIANTI \\
1398.092 &  & UN &  &  & 9.7E+01 &  33.8 &  \\
1398.251 &  & UN &  &  & 3.4E+02 &  31.7 &  \\
1398.375 &  & UN &  &  & 2.5E+02 &  39.8 &  \\
1398.426 &  & UN &  &  & 1.8E+02 &  36.6 &  \\
1398.597 &  & UN &  &  & 1.3E+02 &  49.6 &  \\
1398.783 &  & UN &  &  & 7.7E+02 &  53.0 &  \\
1398.863 &  & UN &  &  & 3.2E+02 &  44.8 &  \\
1398.941 & 1398.9530 & \hh\ &  2,  5  X &  1,  0  B & 5.6E+02 &  61.5 & \citet{jordan77} \\
 & 1399.0260 & Ni II & 3d$^{9}$ \ $^{2}$D$_{3/2}$ & 3d$^{8}$ 4p \ $^{2}$P$^\circ_{3/2}$ &  &  & \citet{shenstone70} \\
1399.050 & 1399.0550 & Fe II & 3d$^{7}$ \ $^{2}$H$_{9/2}$ & 3d$^{5}$ 4s 4p \ $^{4}$F$^\circ_{7/2}$ & 1.1E+02 &  36.1 & NIST \\
1399.177 &  & UN &  &  & 1.4E+02 &  34.1 &  \\
1399.253 &  & UN &  &  & 1.7E+02 &  53.5 &  \\
1399.642 &  & UN &  &  & 1.3E+02 &  39.4 &  \\
1399.792 & 1399.7800 & O IV & 2s$^{2}$ 2p \ $^{2}$P$^\circ_{1/2}$ & 2s 2p$^{2}$ \ $^{4}$P$_{1/2}$ & 2.5E+03 & 209.4 & CHIANTI \\
1399.953 & 1399.9490 & \hh\ & 17,  3  X & 16,  1  B & 6.2E+02 &  62.4 & new \\
 & 1399.9570 & Fe II & 3d$^{6}$ 4s \ $^{4}$P$_{1/2}$ & 3d$^{5}$ 4s 4p \ $^{6}$D$^\circ_{3/2}$ &  &  & NIST \\
1400.174 &  & UN &  &  & 3.3E+02 &  37.1 &  \\
1400.299 &  & UN &  &  & 4.4E+01 &  28.7 &  \\
1400.428 &  & UN &  &  & 7.0E+01 &  29.9 &  \\
1400.602 & 1400.6090 & \hh\ &  4,  5  X &  5,  0  B & 8.7E+01 &  50.6 & \citet{sandlin86} \\
1401.180 & 1401.1570 & O IV & 2s$^{2}$ 2p \ $^{2}$P$^\circ_{3/2}$ & 2s 2p$^{2}$ \ $^{4}$P$_{5/2}$ & 8.1E+03 & 189.4 & CHIANTI \\
1401.516 & 1401.5136 & S I & 3s$^{2}$ 3p$^{4}$ \ $^{3}$P$_{2}$ & 3s$^{2}$ 3p$^{3}$ 5s \ $^{3}$S$^\circ_{1}$ & 7.7E+02 &  71.4 & NIST \\
1401.773 & 1401.7710 & Fe II & 3d$^{6}$ 4s \ $^{4}$F$_{7/2}$ & 3d$^{5}$ 4s 4p \ $^{4}$G$^\circ_{9/2}$ & 3.6E+02 &  38.6 & NIST \\
1402.141 & 1402.1470 & \hh\ &  1,  6  X &  2,  2  B & 2.0E+02 &  36.9 & new \\
1402.783 & 1402.7720 & Si IV & 2p$^{6}$ 3s \ $^{2}$S$_{1/2}$ & 2p$^{6}$ 3p \ $^{2}$P$^\circ_{1/2}$ & 9.3E+04 & 258.4 & CHIANTI \\
 & 1403.1000 & Fe II & 3d$^{7}$ \ $^{4}$F$_{9/2}$ & 3d$^{6}$ 4p \ $^{2}$G$^\circ_{7/2}$ &  &  & NIST \\
1403.978 & 1403.9740 & \hh\ & 11,  4  X & 10,  0  B & 3.5E+02 &  66.9 & \citet{bartoe79} \\
 & 1404.1180 & Fe II & 3d$^{7}$ \ $^{4}$F$_{9/2}$ & 3d$^{6}$ 4p \ $^{2}$G$^\circ_{9/2}$ &  &  & NIST \\
1404.810 & 1404.8060 & O IV & 2s$^{2}$ 2p \ $^{2}$P$_{3/2}$ & 2s 2p$^{2}$ \ $^{4}$P$_{3/2}$ & 2.3E+03 & 231.5 & CHIANTI \\
 & 1404.8080 & S IV & 3s$^{2}$ 3p \ $^{2}$P$^\circ_{1/2}$ & 3s 3p$^{2}$ \ $^{4}$P$_{1/2}$ &  &  & CHIANTI \\
1404.928 &  & UN &  &  & 2.9E+02 &  40.8 &  \\
1405.317 &  & UN &  &  & 3.1E+02 &  32.7 &  \\
1405.574 & 1405.5540 & \hh\ &  3,  6  X &  4,  2  B & 6.8E+01 &  38.1 & new \\
1405.631 & 1405.6020 & Fe II & 3d$^{7}$ \ $^{4}$F$_{9/2}$ & 3d$^{6}$ 4p \ $^{2}$F$^\circ_{7/2}$ & 1.4E+02 &  50.2 & NIST \\
1405.796 & 1405.7970 & Fe II & 3d$^{6}$ 4s \ $^{4}$F$_{5/2}$ & 3d$^{5}$ 4s 4p \ $^{4}$G$^\circ_{7/2}$ & 5.1E+02 &  59.8 & NIST \\
1406.067 & 1406.0160 & S IV & 3s$^{2}$ 3p \ $^{2}$P$^\circ_{3/2}$ & 3s 3p$^{2}$ \ $^{4}$P$_{5/2}$ & 2.5E+03 & 202.7 & CHIANTI \\
1406.428 & 1406.4340 & \hh\ & 11,  6  X & 12,  4  B & 4.4E+01 &  31.6 & new \\
1406.507 &  & UN &  &  & 7.9E+01 &  25.5 &  \\
\enddata
\tablecomments{All \hh{} theoretical wavelengths come from \citet{abgrall00}.  The references provided for \hh{} lines indicate their first identification in the solar spectrum.  The references for the atomic species provide the laboratory or theoretical rest wavelength for the transition.  Lines labeled ``NIST'' come from the NIST Atomic Spectral Database and references therein \citep{nist}, while lines labeled ``CHIANTI'' come from the CHIANTI database version 10.1 \citep{chianti1,chianti2}.}
\end{deluxetable*}

%% file: H2_ratio_table.tex
\begin{deluxetable*}{rrccll}
\tablecaption{Relative Intensities of \hh\ Lines with Shared Upper Levels}\label{tbl:ratios}
\tablewidth{0pt}
\tablehead{\multicolumn{2}{c}{Transition} & \colhead{Wavelength} & \multicolumn{2}{c}{Line Intensity Ratio} & Criteria \\
\cmidrule(lr){1-2}\cmidrule(lr){4-5}
\colhead{Upper} & \colhead{Lower} & \colhead{[\AA]} & \colhead{Model\tablenotemark{a}} & \colhead{Observed} & Fulfilled}
\startdata
 0, 0 B &  1, 4 X & 1335.868 &  0.77 &  \\
 0, 0 B &  1, 5 X & 1396.223 &  1.00 &  1.00$\pm$ 0.01 & abdef \\
\hline
 1, 0 B &  0, 4 X & 1333.475 &  0.57 &  0.55$\pm$ 0.02 & abcdef \\
 1, 0 B &  0, 5 X & 1393.719 &  0.70 &  \\
 1, 0 B &  2, 4 X & 1338.569 &  1.00 &  1.00$\pm$ 0.02 & abcdef \\
 1, 0 B &  2, 5 X & 1398.953 &  1.27 &  0.87$\pm$ 0.19 & abcdef \\
\hline
 2, 0 B &  1, 4 X & 1333.797 &  0.76 &  0.72$\pm$ 0.01 & abcde \\
 2, 0 B &  1, 5 X & 1393.961 &  1.16 &  \\
 2, 0 B &  3, 4 X & 1342.256 &  1.00 &  1.00$\pm$ 0.01 & abcdef \\
 2, 0 B &  3, 5 X & 1402.648 &  1.55 &  \\
\hline
 2, 2 B &  1, 5 X & 1345.401 &  1.43 &  0.29$\pm$ 0.91 & abce \\
 2, 2 B &  1, 6 X & 1402.147 &  1.00 &  1.00$\pm$ 1.28 & acef \\
 2, 2 B &  3, 5 X & 1353.491 &  2.56 &  \\
\hline
 2, 5 B &  1, 6 X & 1334.275 &  0.85 &  \\
 2, 5 B &  3, 6 X & 1341.718 &  1.00 &  1.00$\pm$ 3.63 & adef \\
 2, 5 B &  3, 7 X & 1393.060 &  0.74 &  \\
\hline
 3, 0 B &  2, 4 X & 1335.131 &  0.62 &  \\
 3, 0 B &  2, 5 X & 1395.198 &  0.92 &  0.77$\pm$ 0.03 & abcdef \\
 3, 0 B &  4, 4 X & 1346.909 &  1.00 &  1.00$\pm$ 0.41 & abcde \\
\hline
 3, 3 B &  4, 5 X & 1335.545 &  1.14 &  \\
 3, 3 B &  4, 6 X & 1390.454 &  1.00 &  1.00$\pm$ 1.35 & bdef \\
\hline
 4, 0 B &  3, 4 X & 1337.467 &  0.56 &  0.91$\pm$ 1.69 & abcef \\
 4, 0 B &  3, 5 X & 1397.419 &  1.00 &  1.00$\pm$ 0.57 & abcef \\
 4, 0 B &  5, 4 X & 1352.502 &  0.88 &  0.98$\pm$ 0.87 & abcef \\
\hline
 4, 2 B &  3, 5 X & 1349.056 &  1.51 &  0.50$\pm$ 0.05 & abcef \\
 4, 2 B &  3, 6 X & 1405.554 &  1.00 &  1.00$\pm$ 1.93 & cef \\
\hline
 5, 0 B &  4, 4 X & 1340.791 &  1.00 &  1.00$\pm$ 1.16 & abcef \\
 5, 0 B &  4, 5 X & 1400.609 &  1.62 &  0.51$\pm$ 0.65 & abcef \\
\hline
 6, 0 B &  5, 4 X & 1345.085 &  1.00 &  1.00$\pm$ 0.82 & abdef \\
 6, 0 B &  5, 5 X & 1404.747 &  1.84 &  \\
\hline
 6, 1 B &  7, 4 X & 1342.886 &  1.00 &  1.00$\pm$ 2.11 & abdef \\
 6, 1 B &  7, 5 X & 1400.947 &  1.31 &  \\
\hline
 6, 6 B &  7, 6 X & 1343.033 &  0.00 &  \\
 6, 6 B &  7, 7 X & 1392.003 &  1.00 &  1.00$\pm$ 1.41 & bdf \\
\hline
 7, 0 B &  6, 4 X & 1350.325 & & & abdef \\
\hline
 8, 0 B &  7, 4 X & 1356.482 & & & abdef \\
\hline
 8, 2 B &  9, 4 X & 1337.622 &  1.00 &  1.00$\pm$ 1.48 & abdef \\
 8, 2 B &  9, 5 X & 1393.441 &  4.08 &  \\
\hline
 8, 7 B &  9, 6 X & 1338.704 & & & bdef \\
\hline
 9, 3 B &  8, 5 X & 1352.282 &  1.00 &  1.00$\pm$ 3.26 & abdef \\
 9, 3 B &  8, 6 X & 1405.732 &  0.05 &  \\
\hline
10, 0 B & 11, 3 X & 1345.170 &  0.45 &  0.73$\pm$ 0.56 & abcef \\
10, 0 B & 11, 4 X & 1403.974 &  1.00 &  1.00$\pm$ 0.03 & abcef \\
\hline
10, 1 B &  9, 4 X & 1348.618 &  1.00 &  1.00$\pm$ 2.07 & abdef \\
10, 1 B &  9, 5 X & 1405.377 &  0.02 &  \\
\hline
10, 3 B & 11, 4 X & 1336.123 &  1.87 &  \\
10, 3 B & 11, 5 X & 1389.678 &  1.00 &  1.00$\pm$ 0.19 & abde \\
\hline
11, 0 B & 12, 3 X & 1356.569 & & & abdef \\
\hline
11, 2 B & 10, 4 X & 1335.628 &  0.03 &  \\
11, 2 B & 10, 5 X & 1390.253 &  1.00 &  1.00$\pm$ 0.20 & abdef \\
\hline
12, 0 B & 11, 3 X & 1331.954 &  1.00 &  1.00$\pm$ 4.23 & abdef \\
12, 0 B & 11, 4 X & 1389.583 &  2.40 &  \\
\hline
12, 1 B & 13, 3 X & 1346.424 &  1.00 &  1.00$\pm$ 0.02 & abdef \\
12, 1 B & 13, 4 X & 1402.854 &  0.41 &  \\
\hline
12, 4 B & 11, 5 X & 1355.920 &  0.11 &  \\
12, 4 B & 11, 6 X & 1406.434 &  1.22 &  0.42$\pm$ 1.35 & acdef \\
12, 4 B & 13, 4 X & 1337.831 &  1.00 &  1.00$\pm$ 2.26 & abcdef \\
12, 4 B & 13, 5 X & 1389.039 &  0.10 &  \\
\hline
13, 0 B & 12, 3 X & 1342.411 &  1.00 &  1.00$\pm$ 2.71 & adef \\
13, 0 B & 12, 4 X & 1399.773 &  2.24 &  \\
\hline
14, 0 B & 13, 3 X & 1353.591 &  2.04 &  \\
14, 0 B & 15, 2 X & 1337.357 &  1.00 &  1.00$\pm$ 1.98 & abdef \\
14, 0 B & 15, 3 X & 1394.706 &  2.90 &  \\
\hline
14, 2 B & 15, 3 X & 1350.685 &  1.00 &  1.00$\pm$ 0.04 & abdef \\
14, 2 B & 15, 4 X & 1404.670 &  0.82 &  \\
\hline
14, 3 B & 13, 4 X & 1344.846 &  0.56 &  0.94$\pm$ 1.28 & abce \\
14, 3 B & 13, 5 X & 1396.603 &  1.00 &  1.00$\pm$ 0.62 & abcef \\
\hline
15, 1 B & 14, 3 X & 1344.022 &  1.00 &  1.00$\pm$ 0.61 & abdef \\
15, 1 B & 14, 4 X & 1398.894 &  0.88 &  \\
\hline
16, 1 B & 15, 3 X & 1356.350 &  1.00 &  0.58$\pm$ 0.01 & abcef \\
16, 1 B & 17, 2 X & 1344.965 &  0.67 &  0.59$\pm$ 0.01 & abcef \\
16, 1 B & 17, 3 X & 1399.949 &  1.00 &  1.00$\pm$ 0.20 & abce \\
\hline
16, 6 B & 15, 5 X & 1357.373 &  1.00 &  1.00$\pm$ 1.57 & abdef \\
16, 6 B & 15, 6 X & 1402.337 &  0.29 &  \\
16, 6 B & 17, 4 X & 1348.749 &  0.05 &  \\
16, 6 B & 17, 5 X & 1394.745 &  1.11 &  \\
\hline
17, 0 B & 16, 2 X & 1335.298 &  0.30 &  \\
17, 0 B & 16, 3 X & 1391.005 &  1.00 &  1.00$\pm$ 0.01 & abdef \\
\hline
18, 0 B & 17, 2 X & 1349.257 &  1.00 &  1.00$\pm$ 8.07 & abdef \\
18, 0 B & 17, 3 X & 1404.600 &  2.52 &  \\
18, 0 B & 19, 1 X & 1341.234 &  0.32 &  \\
18, 0 B & 19, 2 X & 1397.110 &  1.56 &  \\
\hline
18, 2 B & 19, 2 X & 1354.753 & & & abdef \\
\hline
20, 2 B & 19, 2 X & 1338.341 &  1.00 &  1.00$\pm$ 0.01 & abcdef \\
20, 2 B & 19, 3 X & 1389.521 &  0.93 &  0.55$\pm$ 0.02 & abcdef \\
20, 2 B & 21, 1 X & 1334.424 &  0.66 &  \\
\hline
21, 2 B & 20, 2 X & 1353.311 &  1.00 &  1.00$\pm$ 1.83 & abcdef \\
21, 2 B & 20, 3 X & 1403.903 &  0.69 &  \\
21, 2 B & 22, 1 X & 1351.167 &  0.62 &  0.60$\pm$ 1.94 & abcdef \\
21, 2 B & 22, 2 X & 1402.557 &  0.91 &  \\
\hline
21, 3 B & 20, 2 X & 1334.426 &  1.16 &  \\
21, 3 B & 22, 1 X & 1332.341 &  1.00 &  1.00$\pm$ 1.08 & abdef \\
\hline
22, 1 B & 21, 1 X & 1336.714 &  0.25 &  \\
22, 1 B & 21, 2 X & 1388.783 &  1.00 &  1.00$\pm$ 1.52 & abdef \\
22, 1 B & 23, 0 X & 1335.580 &  0.06 &  \\
22, 1 B & 23, 1 X & 1388.374 &  0.50 &  \\
\hline
22, 6 B & 21, 3 X & 1342.774 &  1.00 &  1.00$\pm$ 0.61 & abdf \\
22, 6 B & 23, 2 X & 1343.213 &  0.15 &  \\
\hline
24, 3 B & 23, 1 X & 1332.551 &  1.00 &  1.00$\pm$ 2.56 & abdef \\
24, 3 B & 25, 0 X & 1334.656 &  0.20 &  \\
\hline
24, 5 B & 23, 2 X & 1344.645 &  0.27 &  \\
24, 5 B & 23, 3 X & 1388.944 &  0.43 &  \\
24, 5 B & 25, 1 X & 1347.446 &  1.00 &  1.00$\pm$ 1.98 & abdef \\
24, 5 B & 25, 2 X & 1392.712 &  0.00 &  \\
\hline
\enddata
\tablenotetext{a}{FALC$\times100$ optically thick calculation.}
\end{deluxetable*}

%% file: manuscript.bbl
\begin{thebibliography}{}
\expandafter\ifx\csname natexlab\endcsname\relax\def\natexlab#1{#1}\fi
\providecommand{\url}[1]{\href{#1}{#1}}

\bibitem[{{Abgrall} {et~al.}(2000){Abgrall}, {Roueff}, \& {Drira}}]{abgrall00}
{Abgrall}, H., {Roueff}, E., \& {Drira}, I. 2000, \aaps, 141, 297

\bibitem[{{Avrett}(1995)}]{avrett95}
{Avrett}, E.~H. 1995, in Infrared tools for solar astrophysics: What's next?,
  ed. J.~R. {Kuhn} \& M.~J. {Penn}, 303--311

\bibitem[{{Ayres} \& {Testerman}(1981)}]{ayres81}
{Ayres}, T.~R., \& {Testerman}, L. 1981, \apj, 245, 1124

\bibitem[{{Barklem} \& {Collet}(2016)}]{barklem16}
{Barklem}, P.~S., \& {Collet}, R. 2016, \aap, 588, A96

\bibitem[{{Bartoe} {et~al.}(1979){Bartoe}, {Brueckner}, {Nicolas}, {Sandlin},
  {Vanhoosier}, \& {Jordan}}]{bartoe79}
{Bartoe}, J. D.~F., {Brueckner}, G.~E., {Nicolas}, K.~R., {et~al.} 1979,
  \mnras, 187, 463

\bibitem[{{Carlsson} {et~al.}(2023){Carlsson}, {Fletcher}, {Allred}, {Heinzel},
  {Ka{\v{s}}parov{\'a}}, {Kowalski}, {Mathioudakis}, {Reid}, \&
  {Sim{\~o}es}}]{fchroma23}
{Carlsson}, M., {Fletcher}, L., {Allred}, J., {et~al.} 2023, \aap, 673, A150

\bibitem[{{Curdt} {et~al.}(2001){Curdt}, {Brekke}, {Feldman}, {Wilhelm},
  {Dwivedi}, {Sch{\"u}hle}, \& {Lemaire}}]{curdt01}
{Curdt}, W., {Brekke}, P., {Feldman}, U., {et~al.} 2001, \aap, 375, 591

\bibitem[{{Curdt} {et~al.}(2022){Curdt}, {Wilhelm}, {Sch{\"u}hle}, {Vial},
  {Lemaire}, \& {Bocchialini}}]{curdt22}
{Curdt}, W., {Wilhelm}, K., {Sch{\"u}hle}, U., {et~al.} 2022, \solphys, 297,
  145

\bibitem[{{de la Cruz Rodr{\'\i}guez} {et~al.}(2019){de la Cruz
  Rodr{\'\i}guez}, {Leenaarts}, {Danilovic}, \& {Uitenbroek}}]{delacruz19}
{de la Cruz Rodr{\'\i}guez}, J., {Leenaarts}, J., {Danilovic}, S., \&
  {Uitenbroek}, H. 2019, \aap, 623, A74

\bibitem[{{De Pontieu} {et~al.}(2014){De Pontieu}, {Title}, {Lemen}, {Kushner},
  {Akin}, {Allard}, {Berger}, {Boerner}, {Cheung}, {Chou}, {Drake}, {Duncan},
  {Freeland}, {Heyman}, {Hoffman}, {Hurlburt}, {Lindgren}, {Mathur}, {Rehse},
  {Sabolish}, {Seguin}, {Schrijver}, {Tarbell}, {W{\"u}lser}, {Wolfson},
  {Yanari}, {Mudge}, {Nguyen-Phuc}, {Timmons}, {van Bezooijen}, {Weingrod},
  {Brookner}, {Butcher}, {Dougherty}, {Eder}, {Knagenhjelm}, {Larsen},
  {Mansir}, {Phan}, {Boyle}, {Cheimets}, {DeLuca}, {Golub}, {Gates}, {Hertz},
  {McKillop}, {Park}, {Perry}, {Podgorski}, {Reeves}, {Saar}, {Testa}, {Tian},
  {Weber}, {Dunn}, {Eccles}, {Jaeggli}, {Kankelborg}, {Mashburn}, {Pust},
  {Springer}, {Carvalho}, {Kleint}, {Marmie}, {Mazmanian}, {Pereira}, {Sawyer},
  {Strong}, {Worden}, {Carlsson}, {Hansteen}, {Leenaarts}, {Wiesmann},
  {Aloise}, {Chu}, {Bush}, {Scherrer}, {Brekke}, {Martinez-Sykora}, {Lites},
  {McIntosh}, {Uitenbroek}, {Okamoto}, {Gummin}, {Auker}, {Jerram}, {Pool}, \&
  {Waltham}}]{depontieu14}
{De Pontieu}, B., {Title}, A.~M., {Lemen}, J.~R., {et~al.} 2014, \solphys, 289,
  2733

\bibitem[{{Dere} {et~al.}(2023){Dere}, {Del Zanna}, {Young}, \&
  {Landi}}]{chianti2}
{Dere}, K.~P., {Del Zanna}, G., {Young}, P.~R., \& {Landi}, E. 2023, \apjs,
  268, 52

\bibitem[{{Dere} {et~al.}(1997){Dere}, {Landi}, {Mason}, {Monsignori Fossi}, \&
  {Young}}]{chianti1}
{Dere}, K.~P., {Landi}, E., {Mason}, H.~E., {Monsignori Fossi}, B.~C., \&
  {Young}, P.~R. 1997, \aaps, 125, 149

\bibitem[{{Edl\'en} \& {Risberg}(1956)}]{edlen56}
{Edl\'en}, B., \& {Risberg}, P. 1956, Ark. Fys., 10, 553

\bibitem[{{Eriksson} \& {Isberg}(1968)}]{eriksson68}
{Eriksson}, K., \& {Isberg}, H. 1968, Ark. Fys., 37, 221

\bibitem[{{Fontenla} {et~al.}(1993){Fontenla}, {Avrett}, \&
  {Loeser}}]{fontenla93}
{Fontenla}, J.~M., {Avrett}, E.~H., \& {Loeser}, R. 1993, \apj, 406, 319

\bibitem[{{Innes}(2008)}]{innes08}
{Innes}, D.~E. 2008, \aap, 481, L41

\bibitem[{{Jaeggli} {et~al.}(2018){Jaeggli}, {Judge}, \& {Daw}}]{jaeggli18}
{Jaeggli}, S.~A., {Judge}, P.~G., \& {Daw}, A.~N. 2018, \apj, 855, 134

\bibitem[{{Johansson}(1966)}]{johansson66}
{Johansson}, L. 1966, Ark. Fys., 31, 201

\bibitem[{{Johansson}(1978)}]{johansson78}
{Johansson}, S. 1978, \physscr, 18, 217

\bibitem[{{Jordan} {et~al.}(1979){Jordan}, {Bartoe}, {Brueckner}, {Nicolas},
  {Sandlin}, \& {Vanhoosier}}]{jordan79co}
{Jordan}, C., {Bartoe}, J. D.~F., {Brueckner}, G.~E., {et~al.} 1979, \mnras,
  187, 473

\bibitem[{{Jordan} {et~al.}(1977){Jordan}, {Brueckner}, {Bartoe}, {Sandlin}, \&
  {van Hoosier}}]{jordan77}
{Jordan}, C., {Brueckner}, G.~E., {Bartoe}, J. D.~F., {Sandlin}, G.~D., \& {van
  Hoosier}, M.~E. 1977, \nat, 270, 326

\bibitem[{{Jordan} {et~al.}(1978){Jordan}, {Brueckner}, {Bartoe}, {Sandlin}, \&
  {Vanhoosier}}]{jordan78}
{Jordan}, C., {Brueckner}, G.~E., {Bartoe}, J. D.~F., {Sandlin}, G.~D., \&
  {Vanhoosier}, M.~E. 1978, \apj, 226, 687

\bibitem[{{Kramida} {et~al.}(2018){Kramida}, {Ralchenko}, {Reader}, \& the NIST
  ASD~Team}]{nist}
{Kramida}, A., {Ralchenko}, Y., {Reader}, J., \& the NIST ASD~Team. 2018, NIST
  Atomic Spectra Database (version 5.6.1), , ,
  doi:https://doi.org/10.18434/T4W30F.
\newblock \url{http://physics.nist.gov/asd}

\bibitem[{{Kuhn} \& {Morgan}(2006)}]{kuhn06}
{Kuhn}, J.~R., \& {Morgan}, H. 2006, in Astronomical Society of the Pacific
  Conference Series, Vol. 354, Solar MHD Theory and Observations: A High
  Spatial Resolution Perspective, ed. J.~{Leibacher}, R.~F. {Stein}, \&
  H.~{Uitenbroek}, 230

\bibitem[{{Labrosse} {et~al.}(2007){Labrosse}, {Morgan}, {Habbal}, \&
  {Brown}}]{labrosse07}
{Labrosse}, N., {Morgan}, H., {Habbal}, S.~R., \& {Brown}, D. 2007, in
  Astronomical Society of the Pacific Conference Series, Vol. 368, The Physics
  of Chromospheric Plasmas, ed. P.~{Heinzel}, I.~{Dorotovi{\v{c}}}, \& R.~J.
  {Rutten}, 247

\bibitem[{{Leenaarts} {et~al.}(2011){Leenaarts}, {Carlsson}, {Hansteen}, \&
  {Gudiksen}}]{leenaarts11}
{Leenaarts}, J., {Carlsson}, M., {Hansteen}, V., \& {Gudiksen}, B.~V. 2011,
  \aap, 530, A124

\bibitem[{{Lin} \& {Carlsson}(2015)}]{lin15}
{Lin}, H.-H., \& {Carlsson}, M. 2015, \apj, 813, 34

\bibitem[{{Machado} {et~al.}(1980){Machado}, {Avrett}, {Vernazza}, \&
  {Noyes}}]{machado80}
{Machado}, M.~E., {Avrett}, E.~H., {Vernazza}, J.~E., \& {Noyes}, R.~W. 1980,
  \apj, 242, 336

\bibitem[{{Maltby} {et~al.}(1986){Maltby}, {Avrett}, {Carlsson},
  {Kjeldseth-Moe}, {Kurucz}, \& {Loeser}}]{maltby86}
{Maltby}, P., {Avrett}, E.~H., {Carlsson}, M., {et~al.} 1986, \apj, 306, 284

\bibitem[{Mandy \& Martin(1999)}]{mandy99}
Mandy, M.~E., \& Martin, P.~G. 1999, The Journal of Chemical Physics, 110,
  7811.
\newblock \url{https://doi.org/10.1063/1.478731}

\bibitem[{{Moore}(1965)}]{moore65}
{Moore}, C. 1965, {Selected tables of atomic spectra - A atomic energy
  levels-second edition, B multiplet tables : Si II, Si III, Si IV - data
  derived from the analyses of optical spectra}, Tech. rep., National Bureau of
  Standards, Gaithersburg, MD, doi:10.6028/NBS.NSRDS.3sec1.
\newblock \url{https://doi.org/10.6028/NBS.NSRDS.3sec1}

\bibitem[{{Morgan} \& {Habbal}(2005)}]{morgan05}
{Morgan}, H., \& {Habbal}, S.~R. 2005, \apjl, 630, L189

\bibitem[{{Mulay} \& {Fletcher}(2021)}]{mulay21}
{Mulay}, S.~M., \& {Fletcher}, L. 2021, \mnras, 504, 2842

\bibitem[{{Mulay} {et~al.}(2023){Mulay}, {Fletcher}, {Hudson}, \&
  {Labrosse}}]{mulay23}
{Mulay}, S.~M., {Fletcher}, L., {Hudson}, H., \& {Labrosse}, N. 2023, \mnras,
  526, 4755

\bibitem[{{Sainz Dalda} {et~al.}(2024){Sainz Dalda}, {Agrawal}, {De Pontieu},
  \& {Go{\v{s}}i{\'c}}}]{sainzdalda24}
{Sainz Dalda}, A., {Agrawal}, A., {De Pontieu}, B., \& {Go{\v{s}}i{\'c}}, M.
  2024, \apjs, 271, 24

\bibitem[{{Sandlin} {et~al.}(1986){Sandlin}, {Bartoe}, {Brueckner}, {Tousey},
  \& {Vanhoosier}}]{sandlin86}
{Sandlin}, G.~D., {Bartoe}, J. D.~F., {Brueckner}, G.~E., {Tousey}, R., \&
  {Vanhoosier}, M.~E. 1986, \apjs, 61, 801

\bibitem[{{Schmit} {et~al.}(2014){Schmit}, {Innes}, {Ayres}, {Peter}, {Curdt},
  \& {Jaeggli}}]{schmit14}
{Schmit}, D.~J., {Innes}, D., {Ayres}, T., {et~al.} 2014, \aap, 569, L7

\bibitem[{{Sch{\"u}ehle} {et~al.}(1999){Sch{\"u}ehle}, {Brown}, {Curdt}, \&
  {Feldman}}]{schuhle99}
{Sch{\"u}ehle}, U., {Brown}, C.~M., {Curdt}, W., \& {Feldman}, U. 1999, in ESA
  Special Publication, Vol. 446, 8th SOHO Workshop: Plasma Dynamics and
  Diagnostics in the Solar Transition Region and Corona, ed. J.~C. {Vial} \&
  B.~{Kaldeich-Sch{\"u}}, 617

\bibitem[{{Shenstone}(1970)}]{shenstone70}
{Shenstone}, A. 1970, J. Res. Nat. Bur. Stand., 74, 801

\bibitem[{{Socas-Navarro} {et~al.}(2015){Socas-Navarro}, {de la Cruz
  Rodr{\'\i}guez}, {Asensio Ramos}, {Trujillo Bueno}, \& {Ruiz
  Cobo}}]{socasnavarro15}
{Socas-Navarro}, H., {de la Cruz Rodr{\'\i}guez}, J., {Asensio Ramos}, A.,
  {Trujillo Bueno}, J., \& {Ruiz Cobo}, B. 2015, \aap, 577, A7

\bibitem[{{Solanki} {et~al.}(1994){Solanki}, {Livingston}, \&
  {Ayres}}]{solanki94}
{Solanki}, S.~K., {Livingston}, W., \& {Ayres}, T. 1994, Science, 263, 64

\bibitem[{{Spinrad}(1966)}]{spinrad66}
{Spinrad}, H. 1966, \apj, 145, 195

\bibitem[{{Stauffer} {et~al.}(2022){Stauffer}, {Reardon}, \&
  {Penn}}]{stauffer22}
{Stauffer}, J.~R., {Reardon}, K.~P., \& {Penn}, M. 2022, \apj, 930, 87

\bibitem[{{Suematsu} {et~al.}(2021){Suematsu}, {Shimizu}, {Hara}, {Kawate},
  {Katsukawa}, {Ichimoto}, \& {Imada}}]{suematsu21}
{Suematsu}, Y., {Shimizu}, T., {Hara}, H., {et~al.} 2021, in Society of
  Photo-Optical Instrumentation Engineers (SPIE) Conference Series, Vol. 11852,
  International Conference on Space Optics \&mdash; ICSO 2020, ed. B.~{Cugny},
  Z.~{Sodnik}, \& N.~{Karafolas}, 118523K

\bibitem[{{Swamy}(1975)}]{krishna75}
{Swamy}, K.~S.~K. 1975, \solphys, 41, 301

\bibitem[{{Uitenbroek}(2000)}]{uitenbroek00}
{Uitenbroek}, H. 2000, \apj, 531, 571

\bibitem[{Vargas {et~al.}(2024)Vargas, Monge-Palacios, \& Lacoste}]{vargas24}
Vargas, J.~a., Monge-Palacios, M., \& Lacoste, D.~A. 2024, Journal of
  Thermophysics and Heat Transfer, 38, 210.
\newblock \url{https://doi.org/10.2514/1.T6878}

\bibitem[{{W{\"u}lser} {et~al.}(2018){W{\"u}lser}, {Jaeggli}, {De Pontieu},
  {Tarbell}, {Boerner}, {Freeland}, {Liu}, {Timmons}, {Brannon}, {Kankelborg},
  {Madsen}, {McKillop}, {Prchlik}, {Saar}, {Schanche}, {Testa}, {Bryans}, \&
  {Wiesmann}}]{wuelser18}
{W{\"u}lser}, J.~P., {Jaeggli}, S., {De Pontieu}, B., {et~al.} 2018, \solphys,
  293, 149

\end{thebibliography}
